\documentclass{acmtrans2e}

\newtheorem{theorem}{Theorem}[section]
\newtheorem{lemma}[theorem]{Lemma}
\newtheorem{proposition}[theorem]{Proposition}
\newtheorem{corollary}[theorem]{Corollary}

\newdef{definition}[theorem]{Definition}
\newdef{example}[theorem]{Example}
\newdef{remark}[theorem]{Remark}
\newdef{claim}{Claim}

\def \comment#1{}

\def\parentalphi{\renewcommand{\theenumi}{\alph{enumi}}
\renewcommand{\labelenumi}{(\theenumi)}}

\newcounter{symbol}
\newcommand{\indexsyma}[1]%
{\stepcounter{symbol}\index{zzz1 \thesymbol @\protect#1}}
\newcommand{\indexsymb}[1]%
{\stepcounter{symbol}\index{zzz2 \thesymbol @\protect#1}}
\newcommand{\indexsymc}[1]%
{\stepcounter{symbol}\index{zzz3 \thesymbol @\protect#1}}
\newcommand{\indexsymd}[1]%
{\stepcounter{symbol}\index{zzz4 \thesymbol @\protect#1}}
\newcommand{\indexsyme}[1]%
{\stepcounter{symbol}\index{zzz5 \thesymbol @\protect#1}}

\newcommand{\startclauses}{ \vspace{3mm} \\ \begin{math}
\begin{array}{llllll}}
\newcommand{\stopclauses}{\end{array} \end{math}
\vspace{3mm} \\}

\newcommand{\TEXC}[1]{\mbox{$ \exists X C $}}

\newcommand{\TEYD}[1]{\mbox{$ \exists Y D $}}

\newcommand{\la}{\mbox{$\:\leftarrow\:$}}
\newcommand{\ra}{\mbox{$\:\rightarrow\:$}}

\newcommand{\lra}{\mbox{$\:\leftrightarrow\:$}}

\newcommand{\And}{\mbox{$\ \wedge\ $}}
\newcommand{\A}{\mbox{$\ \wedge\ $}}
\newcommand{\U}{\mbox{$\:\cup\:$}}
\newcommand{\I}{\mbox{$\:\cap\:$}}

\newcommand{\NI}{\noindent}

\newcommand{\II}{\vspace{2 mm}}

\newcommand{\ol}[1]{\mbox{$\tilde{#1}$}}

\newcommand{\Var}{{\it Var}}
\newcommand{\Dom}{{\it Dom}}
\newcommand{\Ran}{{\it Ran}}

\newcommand{\tuple}[1]{\tilde{{#1}}}
\newcommand{\wrt}{{with respect to }}

\renewcommand{\ol}[1]{\tilde{#1}}

\newtheorem{obs}{\it Observation}

\newtheorem{ass}{\it Assumption}

\newtheorem{nottheorem}{\it Note}

\newenvironment{programs}{\sf \begin{tabbing}prop\= {\sf prop}\= {\sf
propro}\=\hspace{9cm}\= \kill}{\end{tabbing}}

\newenvironment{programss}{\small\sf \begin{tabbing}
ask\=
{\sf ask}\=
{\sf ask}\=
{\sf ask}\=
{\sf ask}\=
{\sf ask}\= \kill}{\end{tabbing}}

\renewcommand{\la}{\mbox{$\:\leftarrow\:$}}
\renewcommand{\ra}{\mbox{$\:\rightarrow\:$}}

\newcommand{\Par}{\mbox{${\sf \;\|\;}$}}

\renewenvironment{programss}{\small
\sf \begin{tabbing} ask\= {\sf ask \ \ }\= {\sf ask \ \ }\= {\sf
ask \ \ }\= {\sf ask \ \ }\= {\sf ask \ \ }\=
\kill}{\end{tabbing}}

\newcommand{\HS}{\hspace{3.8mm}}
\newcommand{\PS}{(\hspace{2.4mm}}

\title{Transformations of CCP programs}
\author{SANDRO ETALLE\\University of Twente and CWI\\
MAURIZIO GABBRIELLI\\ Universit\`a di Udine \and MARIA CHIARA
MEO\\ Universit\`{a} di L'Aquila}

\begin{abstract}
\noindent We introduce a transformation system for concurrent
constraint programming (CCP).  We define suitable applicability
conditions for the transformations which guarantee that the
input/output CCP semantics is preserved also when distinguishing
deadlocked computations from successful ones and when considering
intermediate results of (possibly) non-terminating computations.

The system allows us to optimize CCP programs while preserving their
intended meaning: In addition to the usual benefits that one has for
sequential declarative languages, the transformation of concurrent
programs can also lead to the elimination of communication channels
and of synchronization points, to the transformation of
non-deterministic computations into deterministic ones, and to the
crucial saving of computational space. Furthermore, since the
transformation system preserves the deadlock behavior of programs, it
can be used for proving deadlock freeness of a given program \wrt a
class of queries. To this aim it is sometimes sufficient to apply our
transformations and to specialize the resulting program \wrt the given
queries in such a way that the obtained program is trivially deadlock
free.
\end{abstract}


\category{F.3.2 }{Logics and Meanings of Programs}{Semantics of
  Programming Languages}[algebraic approaches to semantics]





\begin{document}

\setcounter{page}{1}

\begin{bottomstuff}
  Author's address: S. Etalle, Distributed and Embedded Systems group,
  Faculty of Computer Science, University of Twente, P.O. Box 217
  7500AE Enschede - The Netherlands. \newline M. Gabbrielli,
  Dipartimento di Matematica e Informatica, Universit\`a di Udine, Via
  delle Scienze 206, 33100 Udine, Italy, gabbri@dimi.uniud.it.
  \newline Maria Chiara Meo, Dipartimento di Matematica Pura e
  Applicata, Universit\`{a} di L'Aquila, Loc.~Coppito, 67010 L'Aquila,
  Italy, meo@univaq.it \permission \copyright\ 1999 ACM
  0164-0925/99/0100-0111 \$00.75
\end{bottomstuff}
\maketitle

\section{Introduction}
Optimization techniques, in the case of logic-based languages, fall
into two main categories: on one hand, there exist methods for
compile-time and low-level optimizations such as the ones presented
for constraint logic programs by \cite{JMM91}, which are usually based
on program analysis methodologies (e.g.  abstract interpretation).  On
the other hand, we find source to source transformation techniques
such as \emph{partial evaluation} \cite{MS97}
and more general techniques based on the \emph{unfold}
and \emph{fold} or on the \emph{replacement} operation.

Unfold/fold transformation techniques were first introduced for
functional programs in \cite{BD77}, and then adapted to logic
programming (LP) both for program synthesis \cite{CS77,Hog81}, and for
program specialization and optimization \cite{Kom82}.  Tamaki and Sato
\citeyear{TS84} proposed a general framework for the unfold/fold
transformation of logic programs, which has remained in the years the
main historical reference of the field, and has later been extended to
constraint logic programming (CLP) in \cite{Mah93,EG96-tcs,BG98} (for
an overview of the subject, see the survey by Pettorossi and Proietti
\citeyear{PP94}).  As shown by a number of applications, these
techniques provide a powerful methodology for the development and
optimization of large programs, and can be regarded as the
\emph{basic} transformations techniques, which might be further
adapted to be used for partial evaluation.

Despite a large literature in the field of declarative sequential
languages, unfold/fold transformation sequences have hardly been
applied to concurrent languages. Notable exceptions are the papers of
Ueda and Fukurawa \citeyear{UF88}, Sahlin \citeyear{Sah95}, and of de
Francesco and Santone \citeyear{DFS96} (their relations with this
paper are discussed in Section \ref{sec:related}).  Also when
considering partial evaluation we find only very few recent attempts
\cite{HKY96,MG97,GM97} to apply it in the field of concurrent
languages.

This situation is partially due to the fact that the
non-determinism and the synchronization mechanisms present in
concurrent languages substantially complicate their semantics,
thus complicating also the definition of \emph{correct}
transformation systems. Nevertheless these transformation
techniques can be very useful also for concurrent languages,
since they allow further optimizations related to the
simplification of synchronization and communication mechanisms.

In this paper we introduce a transformation system for concurrent
constraint programming (CCP) \cite{Sa89a,SR90,SRP91}.  This paradigm
derives from replacing the \emph{store-as-valuation} concept of von
Neumann computing by the \emph{store-as-constraint} model: Its
computational model is based on a global \emph{store}, which consists
of the conjunction of all constraints established until that moment
and expresses some partial information on the values of the variables
involved in the computation. Concurrent processes synchronize and
communicate asynchronously via the store by using elementary actions
(ask and tell) which can be expressed in a logical form (essentially
implication and conjunction \cite{BGMP97}).  On one hand, CCP enjoys a
clean logical semantics,
avoiding many of the complications arising in the concurrent
imperative setting; as argued in the position paper \cite{EG98} this
aspect is of great help in the development of effective transformation
tools. On the other hand, differently from the case of other
theoretical models for concurrency (e.g. the $\pi$-calculus), there
exist ``real'' implementations of concurrent constraint languages
(notably, the Oz language \cite{Smo95} and the related ongoing Mozart
project \texttt{http://www.mozart-oz.org/}); thus, in contrast to
other models for concurrency, in this framework transformation
techniques can be readily applied to practical problems.

The transformation system we are going to introduce is originally
inspired by the system of Tamaki and Sato \citeyear{TS84}. Compared to its
predecessors, it improves in three ways: Firstly, we managed to
eliminate the limitation that in a folding operation the \emph{folding
  clause} has to be non-recursive, a limitation which is present in
many other unfold/fold transformation systems, this improvement
possibly leads to the use of new more sophisticated transformation
strategies. Secondly, the applicability conditions we propose for the
folding operation are now independent from the \emph{transformation
  history}, making the operation much easier to understand and to
implement. In fact, following \cite{DFS96}, our applicability
conditions are based on the notion of ``guardedness'' and can be
checked locally on the program to be folded.
Finally, we introduced several new transformation operations.
It is also worth mentioning that the declarative nature of CCP
allows us to define reasonably simple applicability conditions
which ensure the correctness of our system.

We will illustrate with a practical example how our transformation
system for CCP can be even more useful than its predecessors for
sequential logic languages. Indeed, in addition to the usual benefits,
in this context the transformations can also lead to the elimination
of communication channels and of synchronization points, to the
transformation of non-deterministic computations into deterministic
ones, and to the crucial saving of computational \emph{space}.
These improvements
were possible already in the context of
GHC programs by using the system defined in \cite{UF88}.

Our results show that the original and the transformed program
have the same input/output semantics in a rather strong sense,
which distinguishes successful, deadlocked and failed
derivations. As a corollary, we obtain that the original program
is deadlock free iff the transformed one is and this allows us
to employ the transformation system as an effective tool for
proving deadlock-freeness: if, after the transformation, we can
prove or see that the process we are considering never deadlocks
(in some cases the transformation simplifies the program's
behavior so that this can be immediately checked), then we are
also sure that the original process does not deadlock either. We
also consider non-terminating computations by proving three
further correctness results. The first one shows that the
intermediate results of (possibly non-terminating) computations
are preserved up to logical implication, while the second one
ensures full preservation of (traces of) intermediate results,
provided we slightly restrict the applicability conditions for
our transformations. The third result shows that this restricted
transformation system preserves a certain kind of infinite
computations (active ones). We discuss the extension of this
result to the general case, claiming that our system does not
introduce any new infinite computation.
\II

This paper is organized as follows: in the next section we
present the notation and the necessary preliminary definitions,
most of them regarding the CCP paradigm. In Section
\ref{sec:transformation} we define the transformation system,
which consists of various different operations (for this reason
the section is divided in a number of subsections). We will also
use a working example to illustrate the application of our
methodology. Section \ref{sec:correctness} states the first main
result, concerning the correctness of the transformation system,
while Section \ref{sec:intermediate} contains the results for
non-terminating computations. Further examples are contained in
Section \ref{sec:example}.  Section \ref{sec:related} compares
this paper to related work in the literature and Section
\ref{sec:conclusions} concludes. For the sake of readability we
include in this paper only proof sketches of several results, the
(rather long) technical details being deferred to the (on-line
only) Appendix.

A preliminary version of this paper appeared in \cite{EGM98}.

\section{Preliminaries}\label{section:CCP}

The basic idea underlying the CCP paradigm is that computation
progresses via monotonic accumulation of information in a global
store. The information is produced (in form of constraints) by the
concurrent and asynchronous activity of several agents which can
\emph{add} a constraint $\sf c$ to the store by performing the basic
action $\sf tell(c)$.  Dually, agents can also \emph{check} whether a
constraint $\sf c$ is entailed by the store by using an $\sf ask(c)$
action. This allows the synchronization of different agents.

Concurrent constraint languages are defined parametrically \wrt the
notion of \emph{constraint system}, which is usually formalized in an
abstract way following the guidelines of Scott's treatment of
information systems (see \cite{SR90}).  Here, we consider a more
concrete notion of constraint which is based on first-order logic and
which coincides with the one used for constraint logic programming
(e.g. see \cite{JM94}). This will allow us to define the
transformation operations in a more comprehensible way, while
retaining a sufficient expressive power.
We could equally well define the transformations in terms of the
abstract notion of constraint system given in \cite{SR90}\footnote{To
  this aim, essentially we should replace equations of the form $X=Y$
  for diagonal elements $d_{XY}$.}.

Thus, assume given a signature $\Sigma$ defining a set of
function and predicate symbols and associating an arity with
each symbol. A \emph{constraint} $c$ is a first-order
$\Sigma$-formula built by using symbols of $\Sigma$, variables
from a given (countable) set $V$ and the logical connectives and
quantifiers $(\wedge,\vee,\neg,\exists$) in the usual way. The
interpretation for the symbols in $\Sigma$ is provided by a
$\Sigma$-structure ${\cal D}$ consisting of a set $D$ and an
assignment of functions and relations on $D$ to the symbols in
$\Sigma$ which respect the arities. So, ${\cal D}$ defines the
computational domain on which constraints are interpreted.
Usually, in order to model parameter passing, $\Sigma$ is
assumed to contain the binary predicate symbol $=$ which is
interpreted as the identity in ${\cal D}$. We will follow this
assumption, which allows us to avoid the use of most general
unifiers (indeed, for many computation domains ${\cal D}$ the
most general unifier of two terms does not exist).

The formula $\sf {\cal D}\models c$ states that $\sf c$ is valid
in the interpretation provided by ${\cal D}$, i.e.\ that it is true
for every valuation of the free variables of $\sf c$.
The empty conjunction of primitive constraints will be identified
with $\sf true$.  We also denote by $\sf \Var(e)$ the set of free variables
occurring in the expression $\sf e$.


In the sequel, constraints will be considered up to equivalence
in the domain ${\cal D}$, i.e. we write $\sf c_1 = c_2$ in case
$\sf {\cal D}\models c_1 \leftrightarrow c_2$.  Terms will be denoted
by $\sf t,s, \ldots$, variables with $\sf X, Y, Z, \ldots$,
further, as a notational convention, $\sf \ol{t}$ and $\sf
\ol{X}$ denote a tuple of terms and a tuple of distinct
variables, respectively.  $\sf \exists_{-\ol{X}}\ c$ stands for
the existential closure of $\sf c$ except for the variables in
$\sf \ol{X}$ which remain unquantified.
We also assume that the reader is acquainted with the notion of
substitution and of most general unifier (see \cite{Llo87}). We
denote by $\sf e\sigma$ the result application of a substitution
$\sigma$ to an expression $\sf e$.  Given a substitution $\sigma$,
the \emph{domain} of $\sigma$, \Dom($\sigma$), is the finite set
of variables $\sf \{X\mid X\sigma \neq X\}$, the \emph{range} of
$\sigma$ is defined as $\sf \Ran(\sigma) = \bigcup_{X\in
\Dom(\sigma)} \Var(X\sigma)$.

The notation and the semantics of programs and agents is virtually the
same one of \cite{SR90}. In particular, the $\sf \parallel$ operator
allows one to express parallel composition of two agents and it is
usually described in terms of interleaving, while non-determinism
arises by introducing a (global) choice operator $\sf \sum_{i=1}^n
ask(c_i)\rightarrow A_i$: the agent $\sf \sum_{i=1}^n
ask(c_i)\rightarrow A_i$ nondeterministically selects one $\sf
ask(c_i)$ which is enabled in the current store, and then behaves like
$\sf A_i$. Thus, the syntax of CCP \emph{declarations} and
\emph{agents} is given by the following grammar:
\[
\begin{array}{lll}
\sf \hbox{\sl Declarations} &
\sf D::= & \sf \epsilon\;|\; p(\tuple{t})\la A \;|\; D,D
\\
\sf \hbox{\sl Agents} &
\sf A ::= &\sf stop \;|\; tell(c) \;|\;
\sum_{i=1}^n ask(c_i)\rightarrow A_i \;| A\parallel A \;|\; p(\tuple{t})
\\
\sf \hbox{\sl Processes} &
\sf Proc ::= &\sf D.A
\end{array}
\]
where $\sf c$ and $ \sf c_i$'s are constraints.
Note that here we allow terms both as formal and actual parameters.

Usually this is not the case, since the procedure call $\sf
p(\tuple{t})$ can be equivalently written as $\sf
p(\tuple{X})\parallel tell(\tuple{X} = \tuple{t})$, while the
declaration $\sf p(\tuple{t})\leftarrow A$ is equivalent to $\sf
p(\tuple{X})\leftarrow A \parallel tell (\tuple{X} = \tuple{t})$.  We
make this assumption only because this simplifies the writing of
programs in the examples.

Due to the presence of an explicit choice operator, as usual we assume
(without loss of generality) that each predicate symbol is defined by
exactly one declaration.  A \emph{program} is a set of declarations.
In the following examples we assume that the operator $\sf \sum$ binds
tighter  than $\sf \parallel$ (so, $\sf ask(a) \rightarrow A \Par ask(b)
\rightarrow B + ask(d) \rightarrow C$ means $\sf (ask(c) \rightarrow A
) \Par (ask(b) \rightarrow B + ask(d) \rightarrow C))$.  In case some
ambiguity arises we will use brackets to indicate the scope of the
operators.

An important aspect for which we slightly depart from the usual
formalization of CCP regards the notion of \emph{locality}. In
\cite{SR90} locality is obtained by using the operator $\sf \exists$,
and the behavior of the agent $\sf {\exists_X}\, A$ is defined like
the one of $\sf A$, with the variable $\sf X$ considered as
\emph{local} to it.  Here we do not use such an explicit operator:
analogously to the standard CLP setting, locality is introduced
implicitly by assuming that if a process is defined by $\sf
p(\tuple{t})\la A$ and a variable $\sf Y$ occurs in $\sf A$ but not in
$\sf \tuple{t}$, then $\sf Y$ has to be considered local to $\sf A$.

The operational model of CCP is described by a transition system
$\sf T= (Conf, \rightarrow )$ where configurations (in) {\sf Conf}
are pairs consisting of a process and a constraint (representing the
common \emph{store}), while the transition relation
$\sf \rightarrow\ \subseteq Conf \times Conf$ is described by
the (least relation satisfying the) rules {\bf R1-R4} of Table \ref{t1}
which should be self-explanatory.  Here and in the following we assume
given a set $\sf D$ of declarations and we denote by $\sf defn_D(p)$
the set of variants\footnote{A variant of a declaration $\sf d$ is
obtained by replacing the tuple $\sf \tuple{X}$ of all the variables
appearing in $\sf d$ for another tuple $\sf \tuple{Y}$.} of the (unique)
declaration in $\sf D$ for the predicate symbol $\sf p$. Due to the
presence of terms as arguments to predicates symbols, differently from
the standard setting in rule {\bf R4} parameter passing is performed
by a tell action. We also assume the presence of a renaming mechanism
that takes care of using fresh variables each time a declaration is
considered\footnote{For the sake of simplicity we do not describe this
renaming mechanism in the transition system.  The interested reader
can find in \cite{SR90,SRP91} various formal approaches to this problem.}.




We denote by $\rightarrow^*$ the reflexive-transitive closure of the
relation $\rightarrow$ defined by the transition system, and we denote
by $\sf Stop$ any agent which contains only $\sf stop$ and $\parallel$
constructs.  A finite derivation (or computation) containing only
satisfiable constraints is called
\emph{successful} if it is of the form $\sf \langle
D.A,c\rangle\rightarrow^*\langle D.Stop,d \rangle\not \rightarrow$
while it is called \emph{deadlocked} if it is of the form $\sf \langle
D.A,c\rangle\rightarrow^*\langle D.B,d \rangle\not \rightarrow$ with
$\sf B$ different from $\sf Stop$ (i.e., $\sf B$ contains at least one
suspended agent). A derivation producing eventually false is called failed.
Note that we consider here the so called ``eventual
tell'' CCP, i.e. when adding constraints to the store (via tell
operations) there is no consistency check. Our results could
be adapted to the CCP language with consistency check (``atomic tell''
CCP)
by minor modifications of the transformation operations.


\begin{table}
\begin{center}
\begin{tabular}{|lllll|}  \hline
&&&&\\[-3mm]
&\mbox{\bf R1}&
$\sf {\langle D.tell(c),d\rangle\rightarrow\langle D.stop,c \wedge
d \rangle}$&&
\\
&&&&
\\
&\mbox{\bf R2}&
$\sf {\langle D.\sum_{i=1}^n ask(c_i)\rightarrow
A_i,d\rangle\rightarrow\langle D.A_j,d\rangle}$
&\  if $\sf j\in [1,n]\;{\it and} \; {\cal D}\models d\rightarrow c_j$&\\
&&&&
\\
&\mbox{\bf R3}&
$\sf \frac
{\displaystyle \langle D.A,c\rangle
\rightarrow
\langle D.A',  c' \rangle}
{\displaystyle
\begin{array}{l}
\sf \langle D.(A \Par B),c\rangle\rightarrow
\sf \langle D.(A'\Par B),  c'\rangle
\\
\sf \langle D.(B\Par A),c\rangle\rightarrow
\sf \langle D.(B\Par A'),   c'\rangle
\end{array}}$
&&\\&&&
&\\
&\mbox{\bf R4}&
 $\sf \langle D.p(\tuple{t}),c\rangle\rightarrow\langle
D.A\Par tell(\tuple{t}=\tuple{s}) , c\rangle$&
\ if $\sf p(\tuple{s})\la A \in {\sf defn}_D(p)$&\\
\hline
\end{tabular}
\caption{The (standard) transition system.}\label{t1}
\end{center}
\end{table}

\section{The Transformation}
\label{sec:transformation} In order to illustrate the application
of our method we will adopt a working example. We consider an
auction problem in which two bidders participate:
\textsf{bidder\_a} and \textsf{bidder\_b}; each bidder takes as
input the list of the bids of the other one and produces as
output the list of his own bids. When one of the two bidders
wants to quit the auction, it produces in its own output stream
the token \textsf{quit}. This protocol is implemented by the
following program \textsf{AUCTION}. Here and in the following
examples we do not make any assumption on the specific constraint
domain being used, apart from the fact that it should allow us to
use lists of elements. This is the case for most existing general
purpose constraint languages, which usually incorporate also some
arithmetic domain (see \cite{JM94}).

\begin{programss}
  auction(LeftBids,RightBids)$\,\leftarrow\,$bidder\_a([0$|$RightBids],LeftBids)$\,\|\,$bidder\_b(LeftBids,RightBids)\\

\\
bidder\_a(HisList, MyList) \la \\
\> \HS  ask($\exists_{\sf HisBid, HisList'}$ HisList = [HisBid$|$HisList']
\A HisBid = quit) \ra stop\\
\> +  ask($\exists_{\sf HisBid, HisList'}$ HisList = [HisBid$|$HisList']
\A HisBid $\neq$ quit) \ra \\
\> \>     (tell(HisList = [HisBid$|$HisList']) \Par\\
\> \>     make\_new\_bid\_a(HisBid,MyBid) \Par\\
\> \>         \HS  ask(MyBid = quit) \ra (tell(MyList = [MyBid$|$MyList'])
\Par
                 broadcast(``a quits''))\\
\> \>         +  ask(MyBid $\neq$ quit) \ra 
                    (tell(MyList = [MyBid$|$MyList']) \Par \\
\> \> \>            tell(MyBid $\neq$ quit) \Par\\
\> \> \>            bidder\_a(HisList',MyList')))\\[2mm]
\textrm{plus an analogous definition for \textsf{bidder\_b}}.
\end{programss}

\NI Here, the agent \textsf{make\_new\_bid\_a(HisBid,MyBid)} is in
charge of producing a new offer in presence of the competitor's offer
\textsf{HisBid}; the agent will produce \textsf{MyBid = quit} if it
evaluates that \textsf{HisBid} is too high to be topped, and decides
to leave the auction. This agent could be further specified by using
arithmetic constraints. In order to avoid deadlock, \textsf{auction}
initializes the auction by inserting a fictitious zero bid in the
input of bidder a. Notice that in the above program the agent {\sf
  tell(HisList = [HisBid$|$HisList'])} is needed to bind the local
variables ({\sf HisBid, HisList'}) to the global one ({\sf HisList}):
In fact, as a result of the operational semantics, such a binding
is not performed by the \textsf{ask} agent.  On the contrary the agent
{\sf tell(MyBid $\neq$ quit)} is redundant: We have introduced it in
order to slightly simplify the following transformations (the
transformations remain possible also without such a tell). The
introduction of redundant {\sf tell}'s is a transformation operation
which will be formally defined in Subsection~\ref{subsec:atsemp}.

\subsection{Introduction of a new definition}

The introduction of a new definition is virtually always the first
step of a transformation sequence. Since the new definition is going
to be the main target of the transformation operation, this step will
actually determine the very direction of the subsequent
transformation, and thus the degree of its effectiveness.

Determining which definitions should be introduced is a very difficult
task which falls into the area of \emph{strategies}. To give a simple
example, if we wanted to apply \emph{partial evaluation} to our
program \wrt a given agent $\sf A$ (i.e.\ if we wanted to specialize
our program so that it would execute the partially instantiated agent
$\sf A$ in a more efficient way), then a good starting point would
most likely be the introduction of the definition $\sf p(\ol X) \la
A$, where $\sf \ol X$ is an appropriate tuple of variables and $\sf p$
is a new predicate symbol. A different strategy would probably
determine the introduction of a different new definition. For a survey
of the other possibilities we refer to \cite{PP94}.

In this paper we are not concerned with the strategies, but only with
the basic transformation operations and their correctness: we aim at
defining a transformation system which is general enough so to be
applied in combination with different strategies.

In order to simplify the terminology and the technicalities, we assume
that these new declarations are added once for all to the original
program before starting the transformation itself.  Note that this is
clearly not restrictive. As a notational convention we call $\sf D_0$
the program obtained after the introduction of new definitions.  In
the case of program {\sf AUCTION}, we assume that the following new
declarations are added to the original program.
\begin{programss}
  \footnotesize \sf
  auction\_left(LastBid) \la tell(LastBid $\neq$ quit)  \Par
bidder\_a([LastBid$|$Bs],As) \Par bidder\_b(As,Bs).\\
  \footnotesize \sf auction\_right(LastBid) \la tell(LastBid $\neq$
  quit) \Par bidder\_a(Bs,As) \Par bidder\_b([LastBid$|$As],Bs).
\end{programss}
The agent \textsf{auction\_left(LastBid)} engages an auction starting
from the bid \textsf{LastBid} (which cannot be \textsf{quit}) and
expecting the bidder ``a'' to be the next one in the bid. The agent
\textsf{auction\_right(LastBid)} is symmetric.

\subsection{Unfolding}
The first transformation we consider is the \emph{unfolding}.  This
operation consists essentially in the replacement of a procedure call
by its definition. The syntax of CCP agents allows us to define it in
a very simple way by using the notion of context. A \emph{context},
denoted by $\sf C[\;]$, is simply an agent with a ``hole'', where the
hole can contain any expression of type agent. So, for example, $\sf
[\ ] \parallel A$ and $\sf ask(c) \rightarrow A + ask(b)\rightarrow [\
]$ are contexts, while $\sf ask(a) \rightarrow A + [\ ]$ is not.  $\sf
C[A]$ denotes the agent obtained by replacing the hole in $\sf C[\;]$
for the agent $\sf A$, in the obvious way.

\begin{definition}[\thetheorem\ (Unfolding)]
\label{def:unfolding}
Consider a set of declarations $\sf D$ containing
\begin{eqnarray*}
\sf  d: && \sf H \la  C[p(\ol{t})]\\
\sf u: && \sf p(\ol{s})\la  B
\end{eqnarray*}
Then \emph{unfolding} $\sf p(\ol{t})$ in $\sf d$ consists in
replacing $\sf d$ by
\begin{eqnarray*}
\sf d':&&  \sf H \la  C[B \Par tell(\ol{t}=\ol{s})]
\end{eqnarray*}
\noindent in $\sf D$. Here $\sf d$ is the \emph{unfolded}
definition and $\sf u$ is the unfolding one; \textsf{d} and
\textsf{u} are assumed to be renamed so that they do not share
variables.
\end{definition}


After an unfolding we often need to simplify some of the newly
introduced \textsf{tell}'s in order to ``clean up'' the resulting
declarations.  This is accomplished via a tell elimination.  Recall
that a most general unifier $\sigma$ of the terms $\sf t$ and $\sf s$
is called \emph{relevant} if $(\Dom(\sigma) \U \Ran(\sigma)) \subseteq
\Var({\sf t,s})$.

\begin{definition}[\thetheorem\ (Tell Elimination and Tell
  Introduction)]\label{def:tellelin} The declaration
\begin{eqnarray*}
  \sf d: && \sf H \la C[tell(\ol{s}=\ol{t}) \Par B]
\end{eqnarray*}
can be transformed via a \emph{tell elimination} into
\begin{eqnarray*}
\sf d':&&  \sf H \la  C[B\sigma]
\end{eqnarray*}
where $\sigma$ is a relevant most general unifier of $\sf \ol{s}$ and $\sf \ol{t}$, provided that the variables in the domain of $\sigma$ do not occur
neither in $\sf C[\ ]$ nor in $\sf H$. This operation is applicable
either when the computational domain ${\cal D}$ admits a most general
unifier, or
when   $\sf \ol{s}$ and $\sf \ol{t}$ are sequences of distinct variables, in
which case $\sigma$ is simply a renaming.  On the other hand, the
declaration
\begin{eqnarray*}
  \sf d: && \sf H \la C[B\sigma]
\end{eqnarray*}
can be transformed via a \emph{tell introduction} into
\begin{eqnarray*}
\sf d':&&  \sf H \la  C[tell(\ol{X}=\ol{X}\sigma) \Par B]
\end{eqnarray*}
provided that $\sigma$ is a substitution such that $\sf \ol{X}=
\Dom(\sigma)$ and $\Dom(\sigma) \I (\Var({\sf C[\ ],H}) \cup
\Ran(\sigma)) = \emptyset$.
\end{definition}

Notice that, in particular, we can always exchange $\sf C[\sf
tell(true)\parallel A]$ with $\sf C[A]$ and vice-versa. The presence
of $Ran(\sigma)$ in the above condition is needed to ensure that
$\sigma$ is idempotent: in fact, using substitutions $\sigma$ of the
form $X/f(X)$ would not be correct in general. In practice, the
constraints on the domain of $\sigma$ can be weakened by appropriately
renaming some local variables; this is also shown in the upcoming
example. In fact, if all the occurrences of a local variable in $\sf
C[\ ]$ are in choice branches different from the one the ``hole'' lies
in, then we can safely rename apart each one of these occurrences.


In our \textsf{AUCTION} example, we start working on the definition of
\textsf{auction\_right}, and we unfold the agent
\textsf{bidder\_b([LastBid$|$As], Bs)} and then we perform the
subsequent tell eliminations (we eliminate the \textsf{tell}s
introduced by the unfolding).  The result of these operations is the
following program.

\begin{programss}
auction\_right(LastBid) \la tell(LastBid $\neq$ quit) \Par \\
\> bidder\_a(Bs, As) \Par \\
\> \HS  ask($\exists_{\sf HisBid, HisList'}$ [LastBid$|$As] =
[HisBid$|$HisList'] \A HisBid = quit) \ra stop\\
\> +  ask($\exists_{\sf HisBid, HisList'}$ [LastBid$|$As] =
[HisBid$|$HisList'] \A HisBid $\neq$ quit) \ra \\
\> \>      tell([LastBid$|$As] = [HisBid$|$HisList']) \Par\\
\> \>      make\_new\_bid\_b(HisBid,MyBid) \Par\\
\> \>      \HS  ask(MyBid = quit) \ra tell(Bs = [MyBid$|$Bs']) \Par 
                    broadcast(``b quits'')\\
\> \>      +  ask(MyBid $\neq$ quit) \ra 
                   tell(Bs = [MyBid$|$Bs']) \Par \\
\> \> \>           tell(MyBid $\neq$ quit) \Par\\
\> \> \>           bidder\_b(HisList',Bs')
\end{programss}

\subsection{Backward Instantiation}

The new operation of \emph{backward instantiation}, is somehow similar
to the one of unfolding.  We immediately begin with its definition.
\begin{definition}[\thetheorem\ (Backward instantiation)]
  Let $\sf D$ be a set of definitions and
\[\begin{array}{rclcl}
\sf d: && \sf H \la C[p(\ol t)]\\
\sf b: && \sf p(\ol s) \la tell(c)\Par  B\\
\end{array}
\]
be two definitions of $\sf D$.
The \emph{backward instantiation} of $\sf p(\ol t)$ in $\sf d$
consists in replacing $\sf d$ by $\sf d'$, which is either
\[\begin{array}{rclcl}
  \sf d': && \sf H \la C[p(\ol t) \Par tell(c)\Par tell(\ol
  t = \ol s)]\\
\end{array}
\]
or
\[\begin{array}{rclcl}
  \sf d': && \sf H \la C[p(\ol t)\Par tell(\ol t = \ol s)]\\
\end{array}
\]
(it is assumed here that $\sf d$ and $\sf b$ are renamed so
that they have no variables in common).
\II

\NI More generally, the operation can also be applied when $\sf
b$ is not of the form $\sf p(\ol s) \la tell(c)\Par B$ by
considering $\sf c$ to be $\sf true$.
\end{definition}
Intuitively, this operation can be regarded as a
``half-unfolding'' for the following reason: performing an
unfolding is equivalent to applying a derivation step to the
atomic agent under consideration, here we do not quite do it, yet
we carry out (part of) the two first phases that the derivation
step requires.

In the Section \ref{sec:example} we will show an application of this
operation (Example \ref{exa:monitor}).

\subsection{Ask and Tell Simplification}\label{subsec:atsemp}

A new important operation is the one which allows us to modify the
$\sf ask$ guards and the $\sf tell$'s occurring in a program. Let us
call \emph{produced constraint} of $\sf C[\ ]$ the conjunction of all
the constraints appearing in $\sf ask$ and $\sf tell$ actions which
can be evaluated before $[\ ]$ is reached (in the context $\sf C[\ 
]$).  Now, if $\sf a$ is the produced constraint of $\sf C[\ ]$ and
$\sf {\cal D}\models a \rightarrow c$, then clearly we can simplify an
agent of the form $\sf C[ask(c)\rightarrow A + ask(d)\rightarrow B]$
to $\sf C[ask(true)\rightarrow A + ask(d)\rightarrow B]$\footnote{Note
  that in general the further simplification to $\sf C[A +
  ask(d)\rightarrow B]$ is not correct, although we can transform $\sf
  C[ask(true) \rightarrow A]$ into $\sf C[A]$.}. Moreover, under the
previous hypothesis, we can clearly transform $\sf C[tell(c)\parallel
A]$ to $\sf C[A]$ and, conversely, $\sf C[A]$ to $\sf
C[tell(c)\parallel A]$ (as previously mentioned, this latter
transformation consisting in the introduction of a redundant tell
might be needed to prepare a program for the folding operation).

In general, if $\sf a$ is the  produced constraint of $\sf C[\
]$ and for some constraint $\sf c'$ we have that $\sf {\cal D}\models
\exists_{-\ol Z}\ (a \And c) \lra (a \And c')$ (where $\sf \ol Z =
\Var(C,A)$), then we can replace $\sf c$ with $\sf c'$ in
$\sf C[ask(c)\rightarrow A]$ and in
$\sf C[tell(c)]$. In particular,
if we have that $\sf a \And c$ is unsatisfiable, then $\sf c$ can
immediately be replaced with $\sf false$ (the unsatisfiable
constraint). In order to formalize this intuitive idea, we start with
the following definition.





\begin{definition} \label{def:pc}
  Given an agent $\sf A$, the \emph{produced constraint} of $\sf A$
  is denoted by $\sf pca (A)$ and is defined by structural induction
  as follows:
\II

$\begin{array}[b]{lll} \sf pca(tell(c)) &=& \sf c
  \\
  \sf pca(A \Par B) &=& \sf pca(A) \wedge pca(B)
  \\
  \sf pca(A) &=& \sf true \ \ \begin{array}[t]{l}\mbox{\rm for any agent $\sf A$ which is
    neither of the form $\sf tell(c)$}
    \\
    \mbox{nor a parallel composition.}
    \end{array}
\end{array}$
\II

\noindent
By extending the definition we use for agents to contexts, given a
context $\sf C[\ ]$ the \emph{produced constraint} of $\sf C[\ ]$ is
denoted by $\sf pc(C[\ ])$ and is inductively defined as follows:
\II

$\begin{array}[b]{lll}
\sf pc([\ ]) &=& \sf true
\\
\sf pc(C'[\ ]\Par B)  &=& \sf pc(C'[\ ]) \wedge pca(B)
\\
\sf pc(\sum_{i=1}^n ask(c_i) \rightarrow A_i)  &=&
\sf c_j\wedge pc(C'[\ ]) \
\mbox{ where }\ {\sf j \in [1,n]} \ \mbox{and}\ {\sf  A_j = C'[\ ]}
\end{array}$\mbox{}
\end{definition}

The following definition allows us to determine when two constraints
are \emph{equivalent} within a given context $\sf C[\ ]$.

\begin{definition}
  Let $\sf c$, $\sf c'$ be constraints, $\sf C[\ ]$ be a context, and $\sf
\ol Z$ be a set of
  variables.  We say that $\sf c$ is \emph{equivalent to} $\sf c'$
  \emph{within} \textsf{C[\ ]} and \emph{w.r.t. the variables in}
  $\sf \ol Z$ iff ${\cal D} \models \sf \exists_{- \ol Z}\
  (pc(C[\ ]) \And c)$ $\sf \lra
  \exists_{- \ol Z}\ (pc(C[\ ]) \And c')$
\end{definition}

This definition is employed in the following operation, which allows
us to simplify the constraints in the ask and tell guards.

\begin{definition}[\thetheorem\ (Ask and Tell Simplification)]
\label{def:guard_simplification}
Let $\sf D$ be a set of declarations.
\begin{enumerate}
\item Let $\sf d:\ H \la C[\sum_{i=1}^n ask(c_i) \rightarrow A_i]$ be
  a declaration of $\sf D$. Suppose that $\sf c'_1,\ldots,c'_n$ are
  constraints such that for $j\in [1,n]$, $\sf c'_j$ is equivalent to
  $\sf c_j$ within $\sf C[\ ]$ and w.r.t.\ the variables in
  $\Var(\sf{C,H,A_j})$.

Then we can replace $\sf d$ with $\sf d':\ H \la C[\sum_{i=1}^n
ask(c'_i) \rightarrow A_i]$ in $\sf D$. We call this an \emph{ask
simplification}
operation.

\item Let $\sf d: \ H \la C[tell(c)]$ be a declaration of $\sf D$.
  Suppose that the constraint $\sf c'$ is equivalent to $\sf c$ within
  $\sf C[\ ]$ and w.r.t.\ the variables in $\sf \Var(C,H)$.

  Then we can replace $\sf d$ with $\sf d':\ H \la C[tell(c')]$ in $\sf
D$. We
  call this a \emph{tell simplification} operation.
\end{enumerate}
\end{definition}
In our \textsf{AUCTION} example, we can consider the  produced
constraint of \textsf{tell(LastBid $\neq$ quit)}, and modify the
subsequent \textsf{ask} constructs as follows:
\begin{programss}
auction\_right(LastBid) \la tell(LastBid $\neq$ quit) \Par \\
\> bidder\_a(Bs, As) \Par \\
\> \HS  ask($\exists_{\sf HisBid, HisList'}$ [LastBid$|$As] =
[HisBid$|$HisList'] \A  LastBid $\neq$ quit \A HisBid = quit)
$\rightarrow$ \\
\> \> stop\\
\> \HS +  \\
\> \HS ask($\exists_{\sf HisBid, HisList'}$ [LastBid$|$As] =
[HisBid$|$HisList']) \ra \\
\> \>      tell([LastBid$|$As] = [HisBid$|$HisList']) \Par $\ldots$
\end{programss}
Via the same operation, we can immediately simplify this to.
\begin{programss}
auction\_right(LastBid) \la tell(LastBid $\neq$ quit) \Par bidder\_a(Bs,
As) \Par \\
\> \HS  ask(false) \ra stop\\
\> +  ask(true) \ra 
      tell([LastBid$|$As] = [HisBid$|$HisList']) \Par $\ldots$
\end{programss}

\subsection{Branch Elimination and Conservative Ask Elimination}

In the above program we have a guard \textsf{ask(false)} which of
course will never be satisfied.  The first important application of
the guard simplification operation regards then the elimination of
unreachable branches.

\begin{definition}[\thetheorem\ (Branch Elimination)]
\label{def:branch_elimination}
Let $\sf D$ be a set of declarations and let
\II

$\sf d:\ H \la C[\sum_{i=1}^n ask(c_i) \rightarrow A_i]$
\II

\NI be a declaration of $\sf D$. Assume that $n>1$ and that for
some $j\in [1,n]$, we have that $\sf c_j = false$, then we can
replace $\sf d$ with \II

$\sf d':\ H \la C[ (\sum_{i=1}^{j-1} ask(c_i) \rightarrow A_i) \
+ \ (\sum_{i=j+1}^n ask(c_i) \rightarrow A_i) ]$.
\end{definition}
The condition that $n>1$ means that we cannot eliminate all the
branches of a choice and it is needed to ensure the correctness of
the system (otherwise one could transform a deadlock into a success:
For example, the agent $\sf tell(c) \parallel ask(false) \rightarrow
stop$ when evaluated in the empty store produces the constraint $\sf
c$ and deadlocks, while the agent $\sf tell(c)$ produces $\sf c$ and
succeeds).

By applying this operation to the above piece of example, we can
eliminate \textsf{ask(false) \ra stop}, thus obtaining
\begin{programss}
auction\_right(LastBid) \la tell(LastBid $\neq$ quit) \Par \\
\> bidder\_a(Bs, As) \Par \\
\>   ask(true) \ra 
      tell([LastBid$|$As] = [HisBid$|$HisList']) \Par\\
\> \> $\ldots$
\end{programss}
Now we do not see any reason for not eliminating the guard
\textsf{ask(true)} altogether.  This can indeed be done via the
following operation.

\begin{definition}[\thetheorem\ (Conservative Ask Elimination)]
\label{def:cons_ask_eval}
  Consider the declaration \II

  $\sf d:\ H \la C[ask(true) \rightarrow B] $ \II

  \NI We can transform $\sf d$ into the
  declaration \II

  $\sf d':\ H \la C[B] $.
\end{definition}
This operation, although trivial, is subject of debate. In fact,
Sahlin \citeyear{Sah95} defines a similar operation, with the crucial
distinction that the choice might still have more than one branch, in
other words, in the system of \cite{Sah95} one is allowed to simplify
the agent $\sf C[ask(true) \ra A + ask(b) \ra B]$ to the agent $\sf
C[A]$, even if $\sf b$ is satisfiable. Ultimately, one is allowed to
replace the agent $\sf C[ask(true) \ra A + ask(true) \ra B]$ either
with $\sf C[A]$ or with $\sf C[B]$, indifferently. Such an operation
is clearly more widely applicable than the one we have presented but
is bound to be \emph{incomplete}, i.e.\ to lead to the loss of
potentially successful branches.  Nevertheless, Sahlin argues that an
ask elimination such as the one defined above is potentially too
restrictive for a number of useful optimization. We agree with the
statement only partially, nevertheless, the system we propose could
easily be equipped also with an ask elimination as the one proposed by
Sahlin (which of course, if employed, would lead to weaker correctness
results).

In our example program, the application of these branch elimination
and conservative ask elimination leads to the following:
\begin{programss}
auction\_right(LastBid) \la tell(LastBid $\neq$ quit) \Par \\
\> bidder\_a(Bs, As) \Par \\
\> tell([LastBid$|$As] = [HisBid$|$HisList']) \Par\\
\> make\_new\_bid\_b(HisBid,MyBid) \Par\\
\> \HS ask(MyBid = quit) \ra (tell(Bs = [quit$|$Bs']) \Par 
          broadcast(``b quits''))\\
\> +  ask(MyBid $\neq$ quit) \ra 
          (tell(Bs = [MyBid$|$Bs']) \Par \\
\> \>     tell(MyBid $\neq$ quit) \Par\\
\> \>     bidder\_b(HisList',Bs'))
\end{programss}
Via a tell elimination of \textsf{tell([LastBid$|$As] =
  [HisBid$|$HisList'])}, this simplifies to:
\begin{programss}
auction\_right(LastBid) \la tell(LastBid $\neq$ quit) \Par \\
\> bidder\_a(Bs, As) \Par \\
\> make\_new\_bid\_b(LastBid,MyBid) \Par\\
\> \HS  ask(MyBid = quit) \ra (tell(Bs = [quit$|$Bs']) \Par 
           broadcast(``b quits''))\\
\> +  ask(MyBid $\neq$ quit) \ra 
          (tell(Bs = [MyBid$|$Bs']) \Par \\
\> \>     tell(MyBid $\neq$ quit) \Par\\
\> \>     bidder\_b(As,Bs'))
\end{programss}

\subsection{Distribution}
A crucial operation in our transformation system is the
\emph{distribution}, which consists of bringing an agent inside a
choice as follows: from the agent $\sf A \Par \sum_i ask(c_i)
\rightarrow B_i$, we want to obtain the agent $\sf \sum_i ask(c_i)
\rightarrow (A\Par B_i)$. This operation
requires delicate
applicability conditions, as it can easily introduce deadlocks:
consider for instance the following contrived program $\sf D$.

\begin{programss}
p(Y) \la\ q(X) \Par
ask(X $>=$ 0) \ra tell(Y=0)\\
q(0) \la\ stop
\end{programss}
In this program, the process $\sf D.p(Y)$ originates the derivation
$\langle \sf D.p(Y), true \rangle \rightarrow^* \langle \sf D.Stop,
Y=0 \rangle$. Now, if we blindly apply the distribution operation
to the first definition we would change $\sf D$ into:
\begin{programss}
p(Y) \la\ 
ask(X $>=$ 0) \ra (q(X) \Par  tell(Y=0))
\end{programss}
and now we have that $\langle \sf D.p(Y), true \rangle$ generates only
deadlocking derivations.  This situation is avoided by demanding that
the agent being distributed will not be able to produce any output,
unless it is completely determined which branches of the choices might
be entered.

To define the applicability conditions for the distribution
operation we then need the notion of
\emph{productive configuration}. Here and in the following we say that a
derivation $\sf \langle D.A, c\rangle \rightarrow^* \langle D.A', c'
\rangle$ is \emph{maximal} if $\sf \langle D.A', c' \rangle
\not\rightarrow$.

\begin{definition}[\thetheorem\ (Productive)]
\label{def:productive} Given a process $\sf D.A$ and a
satisfiable constraint $\sf c$, we say that $\sf \langle D.A,
c\rangle$ is \emph{productive} iff either it has no (finite)
maximal derivations or there exists a derivation $\sf \langle
D.A, c\rangle \rightarrow^* \langle D.A', c' \rangle$ such that
$\sf {\cal D}\models \neg (\exists_{-\ol{Z}}c \rightarrow
\exists_{-\ol{Z}}c')$, where $\sf \ol{Z} = \Var(A)$.
\end{definition}

So, a configuration is productive if its evaluation can (strictly) augment
the information
contained in the global store. For technical reasons which
will be clear after the next definition, we call productive also those
configurations which have no finite maximal derivations.

We can now provide the definition of the distribution operation.

\begin{definition}[\thetheorem\ (Distribution)]
\label{def:distribution}
  Let $\sf D$ be a set of declarations and let
  \II

  $\sf d:\ H \la C[A \Par \sum_{i=1}^n ask(c_i) \rightarrow B_i]$ \II

  \NI
  be a declaration in $\sf D$, where
  $\sf e = pc(C[\ ])$.
  The \emph{distribution} of $\sf A$ in $\sf d$ yields
  as result the definition\II

  $\sf d':\   H \la  C[ \sum_{i=1}^n ask(c_i) \rightarrow (A \Par B_i)]$
\II

\NI provided that for every constraint $\sf c$ such that $\sf \Var(c)
\cap \Var(d) \subseteq \Var(H,C)$, if
$\sf  \langle D.A, c \wedge e \rangle$ is productive
then both the following conditions hold:
\begin{enumerate}\parentalphi
\item There exists at least one $i\in [1,n]$ such that $\sf {\cal
    D}\models (c \wedge e ) \rightarrow c_i$,

\item for each $i\in [1,n]$, either $\sf {\cal D}\models (c \wedge e )
  \rightarrow c_i$ or $\sf {\cal D}\models (c \wedge e ) \rightarrow
  \neg c_i$.
\end{enumerate}
\end{definition}
Intuitively, the constraint $\sf c$ models the possible ways of
``calling'' 
$\sf A \Par \sum_{i=1}^n ask(c_i) \rightarrow B_i.$
Condition (b) basically requires that if the store $\sf c$ is
such that $\sf A$ might produce some output (that is, the
configuration $\sf  \langle D.A, c \wedge e \rangle$ is
productive), then for each branch of the choice it is already
determined whether we can follow it or not.  This guarantees
that the constraints possibly added to the store by the
evaluation of $\sf A$ cannot influence the choice.  Moreover,
condition (a) guarantees that we do not apply the operation to a
case such as $\sf tell(X=a) \Par ask(false) \ra stop$, which
would clearly be wrong. If $\sf \langle D.A, c \wedge e \rangle$
is not productive then we do not impose any condition, since the
evaluation of $\sf \langle D.A, c \wedge e \rangle$ cannot
affect the choice. As previously mentioned, we call productive
also those configurations which have no finite maximal
derivations, that is, those configurations which originate
non-terminating computations only (possibly with no output). In
fact, also in this case we need conditions (a) and (b), since
otherwise bringing $\sf A$ inside the choice might transform a
looping program into a deadlocking one.

The above applicability conditions are a strict improvement on the
ones we presented in \cite{EGM98}, in which we used the concept of
\emph{required variable}. We now report this definition, both for
simplifying the explanation for some examples and for comparing the
above definition of distribution with the one in \cite{EGM98}.

\begin{definition}[\thetheorem\ (Required Variable)]
\label{def:required} We say that the process $\sf D.A$
\emph{requires} the variable $\sf X$ iff, for each satisfiable
constraint $\sf c$ such that $\sf {\cal D}\models \exists_X
c\leftrightarrow c$, $\sf \langle D.A, c\rangle$ is not
productive.
\end{definition}


In other words, the agent $\sf A$ requires the variable $\sf X$
if, in the moment that the global store does not contain any
information on $\sf X$, $\sf A$ cannot produce any information
which affects the variables occurring in \textsf{A} and has at
least one finite maximal derivation. Even though the above notion
is not decidable in general, it is easy to find wide-applicable
(decidable) sufficient conditions guaranteeing that a certain
variable is required. For example it is immediate to see that, in
our program, {\sf bidder\_a(Bs, As)} requires $\sf Bs$: in fact
the derivation starting in {\sf
  bidder\_a(Bs, As)} suspends (without having provided any output)
after one step and resumes only when more information for the variable
$\sf Bs$ has been produced.

The following remark clarifies how the concept of required variable
might be used for ensuring the applicability of the distributive
operation. Its proof is straightforward.

\begin{remark}
\label{rem:distribution}
Referring to Definition \ref{def:distribution}.  If $\sf A$ requires a
variables which does not occur in $\sf H, C[\ ]$, then the
distribution operation is applicable.
\end{remark}
\begin{proof} In this case, there exists no constraint $\sf c$
such that $\sf \Var(c) \cap \Var(d) \subseteq \Var(H,C)$ and
$\sf \langle D.A, c \wedge e \rangle$ is productive. \end{proof}


In our example, since the agent \textsf{bidder\_a(Bs, As)} requires
the variable \textsf{Bs}, which occurs only inside the \textsf{ask}
guards, we can safely apply the distributive operation.  The result is
the following program.

\begin{programss}
auction\_right(LastBid) \la tell(LastBid $\neq$ quit) \Par 
make\_new\_bid\_b(LastBid,MyBid) \Par\\
\> \HS  ask(MyBid = quit) \ra  tell(Bs = [quit$|$Bs']) \Par 
          broadcast(``b quits'') \Par
          bidder\_a(Bs, As) \\
\> +  ask(MyBid $\neq$ quit) \ra 
          (tell(Bs = [MyBid$|$Bs']) \Par \\
\> \>     tell(MyBid $\neq$ quit) \Par\\
\> \>     bidder\_a(Bs, As) \Par \\
\> \>     bidder\_b(As, Bs'))
\end{programss}

In this program we can now eliminate the construct \textsf{tell(Bs =
  [MyBid$|$Bs'])}: In fact, even though the variable $\sf Bs$ here
occurs also elsewhere in the definition, we can assume it to be
renamed since it occurs only on choice-branches different than the one
on which the considered agent lies. Thus we obtain:

\begin{programss}
auction\_right(LastBid) \la tell(LastBid $\neq$ quit) \Par 
make\_new\_bid\_b(LastBid,MyBid) \Par\\
\> \HS  ask(MyBid = quit) \ra  tell(Bs = [quit$|$Bs']) \Par 
          broadcast(``b quits'') \Par
          bidder\_a(Bs, As) \\
\> +  ask(MyBid $\neq$ quit) \ra  (tell(MyBid $\neq$ quit) \Par\\
\> \>     bidder\_a([MyBid$|$Bs'], As) \Par \\
\> \>     bidder\_b(As, Bs'))
\end{programss}
Before we introduce the fold operation, let us clean up the program a
bit further:
we can now first apply a tell elimination to \textsf{tell(Bs =
  [quit$|$Bs'])}, and then properly transform (by unfolding, and
simplifying the result) the agent \textsf{bidder\_a([quit$|$Bs'], As)}
in the first ask branch. We easily obtain:
\begin{programss}
auction\_right(LastBid) \la tell(LastBid $\neq$ quit) \Par 
make\_new\_bid\_b(LastBid,MyBid) \Par\\
\> \HS  ask(MyBid = quit) \ra  broadcast(``b quits'')\Par 
stop\\
\> +  ask(MyBid $\neq$ quit) \ra 
          (tell(MyBid $\neq$ quit) \Par\\
\> \>     bidder\_a([MyBid$|$Bs'], As) \Par \\
\> \>     bidder\_b(As, Bs'))
\end{programss}
The just introduced \textsf{stop} agent can safely be removed
(see Proposition \ref{pro:toglistop}) and we  are left with:
\begin{programss}
auction\_right(LastBid) \la tell(LastBid $\neq$ quit) \Par 
make\_new\_bid\_b(LastBid,MyBid) \Par\\
\> \HS  ask(MyBid = quit) \ra  broadcast(``b quits'')\\
\> +  ask(MyBid $\neq$ quit) \ra 
          (tell(MyBid $\neq$ quit) \Par\\
\> \>     bidder\_a([MyBid$|$Bs'], As) \Par \\
\> \>     bidder\_b(As, Bs'))
\end{programss}

\subsection{Folding}\label{subs:folding}
The folding operation has a special role in the suite of the
transformation operations. This is due to the fact that it
allows us to introduce recursion in a definition, often making it
independent from the definitions it depended on. As previously
mentioned, the applicability conditions that we use here for the
folding operation do not depend on the transformation history,
nevertheless, we require that the declarations used to fold an
agent appear in the initial program.  Thus, before defining the
fold operation, we need the following.

\begin{definition}
  A \emph{transformation sequence} is a sequence of programs $\sf D_0,
  \ldots, D_n$, in which $\sf D_0$ is an \emph{initial program} and
  each $\sf D_{i+1}$ is obtained from $\sf D_i$ via one of the
  following transformation operations: unfolding, backward
  instantiation, tell elimination, tell introduction, ask and tell
  simplification, branch elimination, conservative ask elimination,
  distribution and folding.
\end{definition}

Recall that we assume that the new declarations introduced by using the
definition introduction operation are added once for all to the original
program $\sf D_0$ before starting the transformation itself.
We also need the notion of \emph{guarding context}.
Intuitively, a context $\sf C[\ ]$ is \emph{guarding} if
the ``hole'' appears in the scope of an $\sf ask$ guard.


\begin{definition}[\thetheorem\ (Guarding Context)]
  We call $\sf C[\ ]$ is a \emph{guarding context} iff
\II

$ \sf C[\ ] \ =\ \sf C'[\sf \sum_{i=1}^n ask(c_i)\rightarrow A_i]
\sf  \mbox{\ \ \ \rm and }  A_j = C''[\ ]\mbox{\ \ \rm for some }
j\in [1,n]. $
\end{definition}
So, for example, $\sf ask(c)\rightarrow (A \parallel [\ ])$ is a guarding
context, while $\sf (ask(c)\rightarrow A) \parallel [\ ]$ is not.
We can finally give the definition of folding:

\begin{definition}[\thetheorem\ (Folding)]
\label{def:folding}
Let $\sf D_0, \ldots, D_i$, $\sf i \ge 0$, be a transformation
sequence. Consider two definitions.
\[\begin{array}{rclcl}
\sf d: && \sf H \la C[A] &&\sf  \in  D_i\\
\sf f: && \sf B \la A && \sf \in D_0
\end{array}
\]
If $\sf C[\ ]$ is a \emph{guarding context}, $\sf B$ contains only distinct variables
as arguments and $\Var(\textsf{A}) \I
\Var(\textsf{C,H}) \subseteq \Var(\textsf{B})$ then \emph{folding} $\sf
A$ in $\sf d$ consists of replacing $\sf d$ by
\[\begin{array}{rclcl}
\sf d': && \sf H \la C[B] &&\sf  \in  D_{i+1}\\
\end{array}
\]
(it is assumed here that $\sf d$ and $\sf f$ are suitably
renamed so that the variables they have in common are only the
ones occurring in $\sf A$).
\end{definition}
In many situations this operation is actually applicable also in
absence of a guarding context as discussed below.

\begin{remark}

\label{rem:propagation} We can apply the fold operation also in
case $\sf C[\ ]$ is not guarding context (referring to the
notation of the previous definition), provided that the
definition $\sf H \la C[A]$ was not modified nor used during the
transformation. In fact, in this case we can simply assume that
the original definition of $\sf H \la C[A]$ contained a dummy
ask guard as in
$$\sf \sf H \la ask(true)\rightarrow C[A] $$
 that the
folding operation is applied to this definition, and that the
guard $\sf ask(true)$ will eventually be removed by an ask
elimination operation.

 Actually, in many cases this reasoning
can be applied also to definitions that \emph{are} used during
the transformation.  This kind of folding is called
\emph{propagation folding} (as opposed to the \emph{recursive}
folding): it is not employed to introduce recursion, but to
\emph{propagate} to other contexts the efficiency that was
hopefully gained by the transformation. Usually, transformation
systems provide a special condition for the propagation folding
operation. For instance, in \cite{TS84}, a distinction is made
between \emph{new} and \emph{old} predicates.  Here we decided
not to do so. This allows us to have a definition of folding
operation which is particularly simple.
\end{remark}

We refer to the end of Example \ref{exa:monitor} for an example
 of application of folding without guarding context.

The reach of the folding operation is best shown via our example.
We can now fold \textsf{auction\_left(MyBid)} in the above
definition, and obtain:
\begin{programss}
auction\_right(LastBid) \la tell(LastBid $\neq$ quit) \Par 
make\_new\_bid\_b(LastBid,MyBid) \Par\\
\> \HS  ask(MyBid = quit) \ra
       broadcast(``b quits'')\\
\> +  ask(MyBid $\neq$ quit) \ra  auction\_left(MyBid)
\end{programss}
Now, by performing an identical optimization on
\textsf{auction\_left}, we can also obtain:
\begin{programss}
auction\_left(LastBid) \la tell(LastBid $\neq$ quit) \Par 
make\_new\_bid\_a(LastBid,MyBid) \Par\\
\> \HS ask(MyBid = quit) \ra
broadcast(``a quits'')\\
\> + ask(MyBid $\neq$ quit) \ra auction\_right(MyBid)
\end{programss}

This part of the transformation shows in a striking way one of the
main benefits of the folding operation: the saving of synchronization
points. Notice that in the initial program the two bidders had to
``wait'' for each other. In principle they were working in parallel,
but in practice they were always acting sequentially, since one had
always to wait for the bid of the competitor. The transformation
allowed us to discover this sequentiality and to obtain an equivalent
program in which the sequentiality is exploited to eliminate all
suspension points, which are known to be one of the major overhead
sources.  Furthermore, the transformation allows a drastic saving of
computational \emph{space}. In fact, in the initial definition the
parallel composition of the two bidders leads to the construction of
two lists containing all the bids done so far.  After the
transformation we have a definition which does not build the list any
longer, and which, by exploiting a straightforward optimization can
employ only \emph{constant} space.

Concerning the syntax of the operation, in our setting the folding
operation reduces to a mere replacement. To people familiar with this
operation, this might seem restrictive: one might wish to apply the
folding also in the case that the definition to be folded contains an
\emph{instance} of \textsf{A}, i.e.\ when \textsf{d} has the form $\sf
H \la C[A\sigma]$ (in this case the folding operation is applicable
only if $\sigma$ satisfies specific conditions described in
\cite{TS84} for logic programs and in \cite{EG96-tcs} for CLP). This
extended operation would actually correspond to the (most) usual
definition of folding as in \cite{TS84,EG96-tcs,BG98}. In our system
such an extended operation is formally not needed, as it can be
obtained by combining together the folding operation with the tell
introduction.

In fact, assume that we would like to fold the definition
\[\begin{array}{rclcl}
\sf d: && \sf H \la C[A\sigma] &&\sf  \in  D_i
\end{array}
\]
by using the definition
\[\begin{array}{rclcl}
\sf f: && \sf B \la A &&\sf  \in  D_0
\end{array}
\]
In the first place, via a tell introduction, we can modify definition
\textsf{d} as follows
\[\begin{array}{rclcl}
\sf d^*: && \sf H \la C[A \Par tell(\ol X = \ol X\sigma)],\\
\end{array}
\]
Clearly, we assume here that $\sf \ol X$ and $\sigma$ fulfill the
applicability conditions given in Definition \ref{def:tellelin}.
Then, via a normal folding operation we obtain
\[\begin{array}{rclcl}
\sf d^{**}: && \sf H \la C[B \Par tell(\ol X = \ol X\sigma)]
\end{array}
\]
(provided that the applicability conditions for the folding are
satisfied) which is equivalent to the definition
\[\begin{array}{rclcl}
\sf d': && \sf H \la C[B\sigma].\\
\end{array}
\]
obtained in the case of the folding operation as defined in
\cite{TS84,EG96-tcs,BG98}. Actually, in case the constraint domain
admit most general unifiers, the definition $\sf d'$ can be obtained from
$\sf \sf d^{**}$ by using a tell elimination operation
(also in this case we assume that the
applicability conditions for the tell elimination
are satisfied).

For the sake of simplicity, we do not give the explicit definition of
this (derived) extended folding operation and of its applicability
conditions. Therefore, the occurrences of this operation in the last
example of Section \ref{sec:example} have to be considered as a
shorthands for the sequence of operations described above.  \II

\section{Correctness}
\label{sec:correctness}
Any transformation system must be useful (i.e.\ allow useful
transformations and optimization) and -- most importantly --
\emph{correct}, i.e., it must guarantee that the resulting program is
in some sense equivalent to the one we have started with.

Having at hand the transition system in Table \ref{t1}, we
provide now the intended semantics to be preserved by the
transformation system by defining a suitable notion of
``observables''.  We start with the following definition which
takes into account terminating and failed computations only. In
the next Section we will consider also non-terminating
computations. Here and in the sequel we say that a constraint
$\sf c$ is \emph{satisfiable} iff ${\cal D}\sf \models \exists\
c$.

\begin{definition}[\thetheorem\ (Observables)]
\label{def:semantics}
  Let $\sf D.A$ be a CCP process. We define
\II
\nopagebreak

$\begin{array}[b]{llll}
\sf {\cal O}(D.A)= &
\sf \{ \langle c,\exists_{-\Var(A,c)} d,ss\rangle  & \mid &
\sf c \  \hbox{\rm and } d \ \hbox{\rm are satisfiable, and there exists}\\
&&&\sf \hbox{\rm a derivation } \langle D.A, c\rangle
\rightarrow^*\langle D.Stop ,d\rangle
\}
\\
&\cup
\\
&\sf \{ \langle c,\exists_{-\Var(A,c)} d,dd\rangle  &\mid&
\sf c \  \hbox{\rm and } d \ \hbox{\rm are satisfiable, and there exists}\\
&&&\sf \hbox{\rm a derivation }
\langle D.A, c\rangle\rightarrow^*\langle D.B,d\rangle\not\rightarrow,
B\neq Stop
\}
\\
&\cup
\\
&\sf \{ \langle c,false,ff\rangle  &\mid&
\sf c \   \ \hbox{\rm is satisfiable, and there exists}\\
&&&\sf \hbox{\rm a derivation } \langle D.A,
c\rangle\rightarrow^*\langle D.B,false\rangle \}.
\end{array}
$ \mbox{}
\end{definition}

Thus what we observe are the results of terminating computations (if
consistent), abstracting from the values of the local variables in the
results, and distinguishing the successful computations from the
deadlocked ones (by using the termination modes $\sf ss$ and $\sf dd$,
respectively). We also observe failed computations, i.e. those
computations which produce an inconsistent store.

Having defined a formal semantics for our paradigm, we can now define more
precisely the notion of \emph{correctness} for the transformation system:
we say that a transformation sequence $\sf D_0, \ldots, D_n$ is
\emph{partially correct} iff, for each agent $\sf A$, we have that
\[\sf {\cal O}(D_0.A) \supseteq {\cal O}(D_n.A)\]
holds, that is, nothing is added
to the semantics of the initial program.
Dually, we say that $\sf D_0, \ldots, D_n$ is
\emph{complete} iff, for each agent $\sf A$,  we have that
\[\sf {\cal O}(D_0.A) \subseteq {\cal O}(D_n.A)\]
holds, that is, no
semantic information is lost during the transformation.
Finally a transformation sequence is called
\emph{totally correct} iff it is both partially correct and complete.

In the following we prove that the our transformation system is
\emph{totally correct}. As previously mentioned, for the sake of
readability some proofs are only sketched and their full versions
can be found in the Appendix.

The proof of this result is originally inspired by the one of
Tamaki and Sato for pure logic programs \cite{TS84} and has
retained some of its notation, in particular we also use the
notions of weight and of split derivation. Of course the
similarities do not go any further, as demonstrated by the fact
that in our transformation system the applicability conditions of
folding operation do not depend on the transformation history
(while allowing the introduction of recursion), and that the
folding definitions are allowed to be recursive (the distinction
between $P_{new}$ and $P_{old}$ of \cite{TS84} is now
superfluous).

We start with the following proposition allows us to eliminate $\sf
stop$ agents in programs. 

\begin{proposition}
\label{pro:toglistop}
For any agent $\sf A$ and set of declarations $\sf D$,
${\cal O}({\sf D.A\parallel stop}) = {\cal O}({\sf D.A})$.
\end{proposition}
\begin{proof} The proof follows immediately from the definition of
observables by noting that, according to rules {\bf R1}-{\bf
R4}, the agent $\sf stop$ has no transition and $\sf \langle D.
A\parallel stop,c \rangle \rightarrow^* \langle D. B\parallel
stop,d \rangle $ iff $\sf \langle D. A,c \rangle \rightarrow^*
\langle D. B,d \rangle$, where obviously $\sf B\parallel stop$
is equal to $\sf Stop$ iff $\sf B= Stop$ (recall that $\sf Stop$
is the generic agent containing only $\parallel$ and $\sf stop$).
\end{proof}
The following notion of mode will be useful to shorten the notation.

\begin{definition}
\label{def:mode}
  Let $\sf D_0,\ldots,D_n$ be a transformation sequence, $\sf A$ be an
  agent and $\sf d$ be a constraint.  We define the \emph{mode} $\sf
  m(A,d)$ of the agent $\sf A$ w.r.t. the constraint $\sf d$ as
  follows
\II

$      \begin{array}[b]{rcl}
      \sf m(A,d) &=&
      \left\{ \begin{array}{ll}
      \sf ss & \sf \mbox{\rm  if $\sf d$ is satisfiable and \ \ }
      {\sf A = Stop}\\
      \sf dd & \sf \mbox{\rm  if $\sf d$ is satisfiable, \ \ }
      \langle D_0.A,d\rangle\not\rightarrow
      \mbox{\rm \ \ and \ \ } {\sf A \neq Stop}\\
      \sf ff & \sf \mbox{\rm  if $\sf d$ is not satisfiable }
      \end{array}
      \right.
      \end{array}
 $\\
\mbox{}
\end{definition}

Note that the notion of mode does not depend on the set of
declarations $\sf D_i$ we are considering, that is, in the above
definition we could equivalently use $\sf D_i$ rather than $\sf D_0$.
This is the content of the following.

\begin{proposition}
  Let $\sf D_0,\ldots,D_n$ be a transformation sequence, $\sf A$ be an
  agent and $\sf d$ be a constraint.  Then $\sf \langle
  D_0.A,d\rangle\not\rightarrow$ iff $\sf \langle
  D_i.A,d\rangle\not\rightarrow$, for any $i\in[1,n]$.
\end{proposition}
\begin{proof} Immediate by observing that a procedure call can be evaluated
in $\sf D_0$ iff it can be evaluated in $\sf D_i$, for any
$i\in[1,n]$.  \end{proof}

In what follows, we are going to refer to a fixed
\emph{transformation sequence} $\sf D_0, \ldots, D_n$. We start with
the following result, concerning partial correctness.

\begin{proposition}[\thetheorem\ (Partial Correctness)]
\label{pro:partial}
If, for each agent $\sf A$, $\sf {\cal O}(D_0.A) = {\cal O}(D_i.A)$ holds
then, for each agent $\sf A$,
$\sf {\cal O}(D_i.A) \supseteq {\cal O}(D_{i+1}.A)$.
\end{proposition}
\begin{proof} (Sketch). We show that given an agent $\sf A$ and a
satisfiable constraint $\sf c_{I}$, if there exists a derivation
$\sf \xi = \langle D_{i+1}.A,c_{I}\rangle \rightarrow^* \langle
D_{i+1}.B,c_{F}\rangle$, with $\sf m(B,c_{F}) \in \{ss, dd,
ff\}$, then there exists also a derivation $\sf \xi' = \langle
D_{i}.A,c_{I}\rangle \rightarrow^* \langle
D_{i}.B',c_{F}'\rangle$ with $\sf \exists_{-\Var(A,c_{I})}c_{F}'
= \exists_{-\Var(A,c_{I})}c_{F}$ and ${\sf m(B',c'_{F})}={\sf
m(B,c_{F})}$.  By Definition~\ref{def:semantics}, this will
imply the thesis.  The proof is by induction on the length $l$
of the derivation.  \II

$(l=0)$. In this case $\sf \xi = \langle D_{i+1}.A,c_{I}\rangle$.
By the definition $\sf \langle D_{i}.A,c_{I}\rangle$ is also a
derivation of length $0$ and then the thesis holds.  \II

$(l>0)$. If the first step of derivation $\xi$ does not use rule
${\bf R4}$, then the proof follows from the inductive hypothesis.


Now, assume that the first step of derivation $\sf \xi$ uses rule
${\bf R4}$ and let $\sf d' \in D_{i+1}$ be the declaration used in the
first step of $\sf \xi$. If $\sf d'$ was not modified in the
transformation step from $\sf D_i$ to $\sf D_{i+1}$ (that is, $\sf d'
\in D_{i}$), then the result follows from the inductive hypothesis. We
assume then that $\sf d' \not\in D_{i}$, $\sf d'$ is then the result
of the transformation operation applied to obtain $\sf D_{i+1}$. The
proof proceeds by distinguishing various cases according to the
operation itself. Here we consider only the operations of unfolding,
tell elimination, tell introduction and folding. The other cases are
deferred to the Appendix.\II

\NI {\bf Unfolding:} If $\sf d'$ is the result of an unfolding
operation then proof is immediate.  \II

\NI {\bf Tell elimination and introduction:} If $\sf d'$ is the
result of a tell elimination or of a tell introduction the thesis
follows from a straightforward analysis of the possible
derivations which use \textsf{d} or \textsf{d'}. First, observe
that for any derivation which uses a declaration $\sf H \la
C[tell(\ol{s}=\ol{t}) \Par B]$, we can construct another
derivation such that the agent $\sf tell(\ol{s}=\ol{t})$ is
evaluated before $\sf B$. Moreover for any constraint $\sf c$
such that $\sf \exists_{dom(\sigma)} c = \exists_{dom(\sigma)} c
\sigma$, (where $\sigma$ is a relevant most general unifier of
$\sf \ol{s}$ and $\sf \ol{t}$), there exists a derivation step $\sf \langle
D_{i}.B_{1}\sigma,c \sigma \rangle \rightarrow \langle
D_{i}.B_{2}\sigma,c'\rangle$ if and only if there exists a
derivation step $\sf \langle D_{i}.B_{1},c \wedge (\ol{s}=\ol{t})
\rangle \rightarrow \langle D_{i}.B_{2},c''\rangle$, where, for
some constraint \textsf{e}, $\sf c'=e \sigma $, $\sf c''= e
\wedge (\ol{s}=\ol{t})$ and therefore $\sf c'
=\exists_{dom(\sigma)} c''$. Finally, since by definition
$\sigma$ is idempotent and the variables in the domain of
$\sigma$ do not occur neither in $\sf C[\ ]$ nor in $\sf H$, for
any constraint $\sf e$  we have that $\sf
\exists_{-\Var(A,c_{I})}e \sigma= \exists_{-\Var(A,c_{I})} (e
\wedge (\ol{s}=\ol{t}))$. \II

\NI {\bf Folding:} If $\sf d'$ is the result of a folding then let

- $\sf d: \ q(\ol r) \la C[H]$ be the folded declaration ($\sf
\in D_i$),

- $\sf f: \ p(\ol X) \la H$ be the folding declaration ($\in \sf
D_{0}$),

- $\sf d': \ q(\ol r) \la C[p(\ol X)]$ be the result of the
folding operation ($\in \sf D_{i+1}$)
\\
where, by hypothesis, $\sf \Var(d) \cap \Var(\ol X)\subseteq
\Var(H)$ and $\sf \Var(H) \cap (\Var(\ol r)\cup \Var(C))\subseteq
\Var(\ol X)$. In this case $\sf \xi = \langle D_{i+1}.C_{I}[q
(\ol{v})],c_{I}\rangle \rightarrow \langle D_{i+1}.C_{I}[C[p(\ol
X)] \Par tell(\ol v = \ol r)],c_{I}\rangle \rightarrow^* \langle
D_{i+1}.B,c_{F}\rangle$ and we can assume, without loss of
generality, that $\sf \Var (C_{I}[q (\ol{v})],c_{I}) \cap
\Var(H)=\emptyset$.

By the inductive hypothesis, there exists a derivation
\[
\sf \chi = \langle D_{i}.C_{I}[C[p(\ol X)] \Par tell(\ol v = \ol
r)],c_{I}\rangle \rightarrow^* \langle D_{i}.B'',c_{F}''\rangle,
\]
with $\sf \exists_{-\Var(C_{I}[C[p(\ol X)] \Par tell(\ol v = \ol
r)],c_{I})} c_{F}'' = \exists_{-\Var(C_{I}[C[p(\ol X)] \Par
tell(\ol v = \ol r)],c_{I})} c_{F}$ and
\begin{equation}
        {\sf m(B'',c''_{F})= m(B,c_{F})}.
        \label{eq:25nov199}
\end{equation}
Since $\sf \Var(C_{I}[q (\ol{v})],c_{I}) \subseteq
\Var(C_{I}[C[p(\ol X)] \Par tell(\ol v = \ol r)],c_{I})$, we have
that
\begin{equation}
        \sf \exists_{-\Var(C_{I}[q (\ol{v})],c_{I})} c_{F}'' =
    \exists_{-\Var(C_{I}[q (\ol{v})],c_{I})} c_{F}.
        \label{eq:25nov299}
\end{equation}
Since by hypothesis for any agent $\sf A'$, $\sf {\cal O}(D_0.A')
= {\cal O}(D_i.A')$, there exists a derivation
\[\sf \xi_{0}=\langle D_{0}.C_{I}[C[p(\ol X)] \Par
tell(\ol v = \ol r)],c_{I}\rangle \rightarrow^* \langle
D_{0}.B_{0}, c_{0}\rangle
\]
such that $\sf \exists_{-\Var(C_{I}[C[p(\ol X)] \Par tell(\ol v =
\ol r)],c_{I})}c_{0} = \exists_{-\Var(C_{I}[C[p(\ol X)] \Par
tell(\ol v = \ol r)],c_{I})} c_{F}''$ and ${\sf m( B_{0},
c_{0})}$ = ${\sf m(B'',c''_{F})}$.

By (\ref{eq:25nov199}), (\ref{eq:25nov299}) and since $\sf
\Var(C_{I}[q (\ol{v})],c_{I}) \subseteq \Var(C_{I}[C[p(\ol X)]
\Par tell(\ol v = \ol r)],c_{I})$, we have that
\begin{equation}
        {\sf \exists_{-\Var(C_{I}[q (\ol{v})],c_{I})} c_0 =
    \exists_{-\Var(C_{I}[q (\ol{v})],c_{I})} c_{F}}
    \mbox{ and } {\sf m( B_{0}, c_{0})} = {\sf m(B,c_{F})}.
        \label{eq:25nov1099}
\end{equation}
Let $\sf f': \ p(\ol X') \la H'$ be an appropriate renaming of $\sf
f$, which renames only the variables in $\sf \ol X$, such that $\sf
\Var(d) \cap \Var(f')= \emptyset$ (note that this is possible, since
$\sf \Var(H) \cap (\Var(\ol r)\cup \Var(C))\subseteq \Var(\ol X)$). By
hypothesis, $\sf \Var (C_{I}[q (\ol{v})],c_{I}) \cap
\Var(H)=\emptyset$. Then, without loss of generality we can assume
that $\sf \Var(\xi_{0}) \cap \Var (f') \neq \emptyset$ if and only if
the procedure call $ \sf p(\ol X)$ is evaluated, in which case
declaration $\sf f'$ is used.
\\
Thus there exists a derivation
\[\sf \langle D_{0}.
C_{I}[ C [H'\Par tell(\ol X = \ol X') ] \Par tell(\ol v = \ol
r)],c_{I}\rangle \rightarrow^* \langle D_{0}. B'_{0},
c_{0}\rangle,
\]
where ${\sf m(B'_{0}, c_{0})} = {\sf m(B_{0}, c_{0})}$. By
(\ref{eq:25nov1099}) we have
\begin{equation}
    {\sf m(B'_{0}, c_{0})} = {\sf m(B,c_{F})}.
    \label{eq:1ott2}
\end{equation}
We now show that we can substitute $\sf H$ for $\sf H' \Par tell(\ol X
= \ol X') $ in the previous derivation. Since $\sf f': \ p(\ol X') \la
H'$ is a renaming of $\sf f: \ p(\ol X) \la H$, the equality $\sf \ol X
 = \ol X'$ is conjunction of equations involving only distinct variables.
Then, by replacing the variables $\sf \ol X$ with $\sf \ol X'$ and vice versa
in the previous derivation we obtain the derivation $\sf \chi_{0}= \sf
\langle D_{0}. C_{I}[C [H \Par tell(\ol X' = \ol X) ] \Par tell(\ol v
= \ol r)],c_{I}\rangle \rightarrow^* \langle D_{0}.  B''_{0},
c_{0}'\rangle$ where $\sf \exists_{-\Var(C_{I}[C[H \Par tell(\ol X' =
  \ol X)] \Par tell(\ol v = \ol r)],c_{I})} c_{0}' =
\exists_{-\Var(C_{I}[C [H \Par tell (\ol X' = \ol X) ] \Par tell(\ol v
  = \ol r)],c_{I})} c_{0}$ and ${\sf m(B''_{0}, c'_{0})} = {\sf
  m(B'_{0}, c_{0})}$.

>From (\ref{eq:1ott2}) it follows that
\begin{equation}
        {\sf m(B''_{0}, c'_{0})} = {\sf m(B,c_{F}). }
        \label{eq:1ott3}
\end{equation}
Then, from (\ref{eq:25nov1099}) and since $\sf \Var(C_{I}[q(\ol
v)],c_{I}) \subseteq \Var(C_{I}[C [H \Par tell (\ol X' = \ol X) ]
\Par tell(\ol v = \ol r)],c_{I})$ we obtain
\begin{equation}
        \sf \exists_{-\Var(C_{I}[q(\ol v)],c_{I})} c_{0}' =
       \exists_{-\Var(C_{I}[q(\ol v)],c_{I})}c_{F}.
        \label{eq:28genn10}
\end{equation}
Moreover, we can drop the constraint $\sf tell(\ol X' = \ol X)$,
since the declarations used in the derivation are renamed apart
and, by construction, $\sf \Var(C_{I}[C [H] \Par tell(\ol r = \ol
v)], c_{I}) \cap \Var(\ol X') =\emptyset$.
Therefore there exists a derivation $\sf \langle D_{0}. C_{I}[C
[H] \Par tell(\ol v = \ol r)],c_{I}\rangle \rightarrow^*  \langle
D_{0}.\bar B_{0},\bar c_{0}\rangle$ which performs exactly the
same steps of $\sf \chi_{0}$, (possibly) except for the
evaluation of $\sf tell(\ol X' = \ol X)$, and such that $\sf
\exists_{-\Var(C_{I}[C [H] \Par tell(\ol v = \ol r)],c_{I})} \bar
c_{0} = \exists_{-\Var(C_{I}[C [H] \Par tell(\ol v = \ol
r)],c_{I})} c_{0}'$ and ${\sf m(\bar B_{0},\bar c_{0})}= {\sf
m(B''_{0}, c'_{0})}$. From (\ref{eq:1ott3}),  (\ref{eq:28genn10})
and since $\sf \Var(C_{I}[q(\ol v)],c_{I}) \subseteq \Var(C_{I}[C
[H] \Par tell(\ol v = \ol r)],c_{I})$, it follows that
\begin{equation}
    {\sf m(\bar B_{0},\bar c_{0})= m (B,c_{F})}
    \mbox{ and }
    {\sf \exists_{-\Var(C_{I}[q(\ol v)],c_{I})}\bar c_{0} =
    \exists_{-\Var(C_{I}[q(\ol v)],c_{I})}c_{F}}.
    \label{eq:1ott4}
\end{equation}
Since $\sf {\cal O}(D_0.A') = {\cal O}(D_i.A')$ holds by
hypothesis for any agent $\sf A'$, there exists a derivation
\[
\sf \langle D_{i}. C_{I}[C [H] \Par tell(\ol v = \ol
r)],c_{I}\rangle \rightarrow^*  \langle D_{i}.B',c_{F}'\rangle
\]
where
$$ \sf \exists_{-\Var(C_{I}[C [H] \Par tell(\ol v = \ol
r)],c_{I})} c_{F}'= \exists_{-\Var(C_{I}[C [H] \Par tell(\ol v =
\ol r)],c_{I})} \bar c_{0}$$ and ${\sf m(B',c'_{F})} = {\sf
m(\bar B_{0},\bar c_{0})}$.

>From (\ref{eq:1ott4}) and since $\sf \Var(C_{I}[q(\ol v)],c_{I})
\subseteq \Var(C_{I}[C [H] \Par tell(\ol v = \ol r)],c_{I})$, we
obtain
\begin{equation}
     {\sf m(B',c'_{F})} = {\sf m (B,c_{F})}
     \mbox{ and }
     {\sf \exists_{-\Var(C_{I}[q(\ol v)],c_{I})}c_{F}'=
     \exists_{-\Var(C_{I}[q(\ol v)],c_{I})}c_{F}}.
     \label{eq:1ott5}
\end{equation}
Finally, since $\sf d: \ q(\ol r) \la C[H] \in \sf D_i$, there
exists a derivation
\[
\sf \xi' = \langle D_{i}.C_{I}[q ( \ol v)],c_{I}\rangle
\rightarrow \langle D_{i}.C_{I}[C [H] \Par tell(\ol v = \ol
r)],c_{I}\rangle \rightarrow^* \langle D_{i}.B',c_{F}'\rangle
\]
and then the thesis follows from (\ref{eq:1ott5}).
\end{proof}

In order to prove total correctness we need the following.

\begin{definition}[\thetheorem\ (Weight)]
\label{def:weight}
Let $\xi$ be a derivation.  We denote by $wh(\xi)$ the number of
derivation steps in $\xi$ which use rule ${\bf R2}$. Given an agent $\sf A$
and a pair of satisfiable constraints $\sf c$, $\sf d$, we then
define the \emph{success weight} $\sf w_{ss}(A,c,d)$ of the agent $\sf A$
w.r.t. the constraints $\sf c$ and $\sf d$ as follows
\II

${\sf w_{ss}(A,c,d)} = min\{ n \mid \begin{array}[t]{lll} n = wh(\xi)
\hbox{ and }
\xi \hbox{ is a derivation }
\\
\sf \langle D_0.A,c\rangle \rightarrow^* \langle D_0.Stop,d'\rangle
\not\rightarrow \\
\hbox{with } \sf \exists_{-\Var(A,c)}d' = \exists_{-\Var(A,c)}d & \}
\end{array}
$
\II

\noindent
Analogously, we define the \emph{deadlock weight} $\sf w_{dd}(A,c,d)$
of the agent $\sf A$ w.r.t. the constraints $\sf c$ and $\sf d$
\II

${\sf w_{dd}(A,c,d)} = min\{ n \mid \begin{array}[t]{lll} n = wh(\xi)
\hbox{ and }
\xi \hbox{ is a derivation }
\\
\sf \langle D_0.A,c\rangle \rightarrow^* \langle D_0.B,d'\rangle
\not\rightarrow \\
\hbox{ with } {\sf B\neq Stop} \hbox{ and }
\sf \exists_{-\Var(A,c)}d' = \exists_{-\Var(A,c)}d & \}
\end{array}
$
\II

\noindent
and the \emph{failure weight} $\sf w_{ff}(A,c,d')$ of the
agent $\sf A$ w.r.t. the constraints $\sf c$ and $\sf d'$
\II

${\sf  w_{ff}(A,c,d')} = min\{ n \mid  \begin{array}[t]{lll} n =
wh(\xi) \hbox{ and } \xi \hbox{ is a derivation } {\sf \langle
D_0.A,c\rangle \rightarrow^* \langle D_0.B,d'\rangle}\\
 \hbox{
with } {\sf d' = false}    & \} \end{array}$ \mbox{}
\end{definition}
Notice that ${\sf w_{ss}(A,c,d')}$ is undefined in case there is
no successful derivation corresponding to the given constraints
(and analogously for $\sf w_{dd}$ and $\sf w_{ff}$). Also, both
${\sf w_{ss}(A,c,false)}$ and ${\sf w_{dd}(A,c,false)}$ are
undefined, as the success and deadlock weight consider only non
failed derivations (i.e. derivations which do not produce the
constraint $\sf false$).

As previously mentioned, this notion of weight is rather
different from the one in \cite{TS84}, since the latter is based
on the number of nodes in a proof tree for an atom, by taking
into account the fact that the predicate symbol appearing in that
atom is ``new'' or ``old''.

In the total correctness proof we also make use of the concept of
\emph{split  derivations}. Intuitively, these are derivations which
can be split into two parts: the first one, up to the first
\textsf{ask} evaluation, is performed in the program $\sf D_i$ while
the second one is carried out in $\sf D_0$.

\begin{definition}[\thetheorem\ (Split derivation)]
\label{def:splitder} Let $\sf D_0,\ldots,D_i$ be a transformation sequence.
We call a derivation in $\sf D_i\cup D_0$ a \emph{successful split
derivation} if it has the form
\II

$\sf \langle D_i.A_1,c_1\rangle \rightarrow^* \langle
D_i.A_m,c_m\rangle \rightarrow \langle D_0.A_{m+1},c_{m+1}\rangle
\rightarrow^* \langle D_0.Stop,c_n\rangle\not\rightarrow $
\II

\NI where $\sf c_n$ is a satisfiable constraint,
$m\in[1,n]$\footnote{If $m=n$ we can write indifferently
  $\sf \langle D_i.Stop,c_n\rangle$ or $\sf \langle
  D_0.Stop,c_n\rangle$ to denote the last configuration of the
  derivation.} and the following conditions hold:
\begin{enumerate}
\parentalphi
\item the first $m-1$ derivation steps do not use rule ${\bf R2}$;
\item the $m$-th derivation step $\sf \langle D_i.A_m,c_m\rangle
\rightarrow \langle D_0.A_{m+1},c_{m+1}\rangle$ uses rule ${\bf R2}$;
\item $\sf w_{ss}(A_1,c_1,c_n) >  w_{ss}(A_{m+1},c_{m+1},c_n)$.
\end{enumerate}
A \emph{deadlocked split derivation} is defined analogously,
by replacing $\sf w_{ss}$ for $\sf w_{dd}$ and  {\sf Stop} for a generic
agent $\sf B\neq Stop$
in the last configuration of the derivation above.
Finally a \emph{failed split derivation} is defined by replacing
$\sf w_{ss}$ for $\sf w_{ff}$ and {\sf Stop} for a generic agent
(which is not necessarily terminated) and by assuming that $\sf
c_{n}=false$ in the last configuration of the derivation above.
\end{definition}

In the following we call split derivations both successful, deadlocked
and failed split derivations.  The previous definition is inspired by
the definition of \emph{descent clause} of \cite{KK88}; however, here
we use a different notion of weight and rather different conditions on
them.  We need one final concept.
\begin{definition}
  \label{def:progcompl}
  We call the program $\sf D_i$ \emph{weight complete} iff, for any
  agent $\sf A$, for any satisfiable constraint $\sf c$  and
  for any constraint $\sf d$, the
  following hold: if there exists a derivation
\II

$\sf \langle D_0.A,c\rangle
\rightarrow^* \langle D_0.B,d\rangle$
\II

\NI
such that $\sf  m(B,d)\in \{ss,\ dd, \ ff\}$ then there
exists a split derivation in $\sf D_i \cup D_0$
\II

$\sf
\langle D_i.A,c\rangle \rightarrow^* \langle D_0.B',d' \rangle
$
\II

\NI where $\sf \exists_{-\Var(A,c)}d' = \exists_{-\Var(A,c)}d$
and $\sf  m(B',d') = \sf m(B,d)$.
\end{definition}

So $\sf D_i$ is weight complete if we can reconstruct the semantics of
$\sf D_0$ by using only (successful, deadlocked and failed) split
derivations in $\sf D_i \cup D_0$. We now show that if $\sf D_i$ is
weight complete then no observables are lost during the transformation
(i.e.,  the transformation is complete). This is the
content of the following.

\begin{proposition}
\label{pro:total1}
If $\sf D_i$ is weight complete then, for any agent $\sf A$, $\sf
{\cal O}(D_0.A) \subseteq {\cal O}(D_{i}.A)$.
\end{proposition}
\begin{proof}
We consider only the case of successful derivations, since
the case of deadlocked (failed) derivations can be proved analogously
by considering the notions of deadlock (failure) weight and deadlocked
(failed) split derivation.  Assume that
there exists a (finite, successful) derivation $\sf \langle D_0.A,c\rangle
\rightarrow^* \langle D_0.Stop,d\rangle$.  We show, by induction on the
success weight of $\sf (A,c,d)$, that there exists a derivation $\sf
\langle
D_i.A,c\rangle \rightarrow^* \langle D_i.Stop,d'\rangle $, where $\sf
\exists_{-\Var(A,c)}d' = \exists_{-\Var(A,c)}d$. \II

\NI \emph{Base Case}. If $\sf w_{ss}(A,c,d) =0$ then,
since $\sf D_i$ is weight complete, from Definition~\ref{def:splitder}
and Definition~\ref{def:progcompl} it follows that
there exists a (successful) split derivation in $\sf D_i \cup D_0$
of the form
$\sf \langle D_i.A,c\rangle \rightarrow^* \langle D_i.Stop,d'\rangle $
where $\sf \exists_{-\Var(A,c)}d' = \exists_{-\Var(A,c)} d$,
rule ${\bf R2}$ is not used and therefore each derivation
step is done in $\sf D_i$.
\II

\NI \emph{Inductive Case}. Assume that $\sf w_{ss}(A,c,d) =n$.
Since $\sf D_i$ is weight complete there exists a (successful)
split derivation in $\sf D_i \cup D_0$
\[
\sf \xi:\ \langle D_i.A,c\rangle \rightarrow^* \langle D_0.Stop,d'\rangle,
\]
where $\sf \exists_{-\Var(A,c)}d' = \exists_{-\Var(A,c)}d$.
If rule ${\bf R2}$ is not used in $\xi$ then the proof is the same
as in the previous case. Otherwise $\xi$ has the form
\[
\sf \langle D_i.A,c\rangle \rightarrow^* \langle D_i.A_m,c_m\rangle
\rightarrow \langle D_0.A_{m+1},c_{m+1}\rangle
 \rightarrow^* \langle D_0.Stop,d' \rangle
\]
where $\sf w_{ss}(A,c,d')> w_{ss}(A_{m+1},c_{m+1},d')$.
Let $\xi'$ be the derivation
\[
\xi': \sf
\langle D_i.A,c\rangle \rightarrow^* \langle D_i.A_m,c_m\rangle
\rightarrow \langle D_i.A_{m+1},c_{m+1}\rangle.
\]
By the inductive hypothesis, there exists a derivation
\[
\xi'':\ \sf \langle D_i.A_{m+1},c_{m+1}\rangle
 \rightarrow^* \langle D_i.Stop,d'' \rangle
\]
where $\sf \exists_{-\Var(A_{m+1},c_{m+1})}d'' =
\exists_{-\Var(A_{m+1},c_{m+1})}d'$.  Without loss of generality, we
can assume that $\sf \Var(\xi' )\cap \Var(\xi'' ) =
\Var(A_{m+1},c_{m+1})$ and hence there exists a derivation
\[
\sf
\langle D_i.A,c\rangle \rightarrow^* \langle D_i.Stop,d'' \rangle.
\]
Finally, by our hypothesis on the variables and by construction,
\[
\begin{array}{ll}
\sf \exists_{-\Var(A,c)}d'' & =
\\
\sf \exists_{-\Var(A,c)}(c_{m+1} \wedge
\exists_{-\Var(A_{m+1},c_{m+1})}d'')
& =
\\
\sf \exists_{-\Var(A,c)}(c_{m+1} \wedge \exists_{-\Var(A_{m+1},c_{m+1})}d')
& =
\\
\sf \exists_{-\Var(A,c)}d' & =
\\
\sf \exists_{-\Var(A,c)}d
\end{array}
\]
which concludes the proof.  \end{proof}

Before proving the total correctness result we need some
technical lemmata. Here and in the following we use the notation
$\sf w_t$ (with $\sf t \in \{ss,dd,ff\}$) as a shorthand for
indicating the success weight $\sf w_{ss}$, the deadlock weight
$\sf w_{dd}$ and the failure weight $\sf w_{ff}$.

\begin{lemma}
    \label{lem:pesoclausola}
    Let $ \sf q (\ol{r})\la H \in D_{0}$, $\sf t \in \{ss,dd,ff\}$ and
    let $\sf C[\ ]$ be a context. For any satisfiable constraint $\sf
    c$ and for any constraint $\sf c'$, such that $\sf
    \Var(C[q(\ol{t})],c) \cap \Var(\ol{r}) =\emptyset$ and $\sf w_t(
    C[q (\ol{t}) ],c,c')$ is defined, there exists a constraint $\sf
    d'$ such that $\sf w_t( C[q (\ol{r}) \Par
    tell(\ol{t}=\ol{r})],c,d')\leq w_t( C[q (\ol{t}) ],c,c')$ and $\sf
    \exists_{-\Var( C[q (\ol{t}) ],c)}d' = \exists_{-\Var( C[q
      (\ol{t}) ],c)}c'$.
\end{lemma}
\begin{proof}
Immediate.  \end{proof}

\begin{lemma}
    \label{lem:pesiind0}
    Let $\sf q (\ol{r})\la H \in D_{0}$ and $\sf t \in \{ss,dd,ff\}$.
    For any context $\sf C_{I}[\ ]$, any satisfiable
    constraint $\sf c$ and for any constraint $\sf c'$, the following
holds.
    \begin{enumerate}
        \item\label{pt:pesiind01}
        If $\sf \Var(H) \cap \Var(C_{I},c) \subseteq \Var(\ol r)$
        and $\sf w_t( C_{I}[q (\ol{r}) ],c,c')$ is defined, then there
        exists a constraint $\sf d'$, such that
        $\sf \Var(d') \subseteq \Var(C_{I}[H],c)$,
        $\sf w_t( C_{I}[H],c,d') \leq  w_t( C_{I}[q (\ol{r}) ],c,c')$
        and $\sf \exists_{-\Var( C_{I}[q (\ol{r}) ],c)}d' =
        \exists_{-\Var( C_{I}[q (\ol{r}) ],c)}c'$.

        \item\label{pt:pesiind02}
        If $\sf \Var(H) \cap \Var(C_{I},c) \subseteq \Var(\ol r)$,
        $\sf \Var(c') \cap \Var(\ol r) \subseteq \Var(C_{I}[H],c)$
        and
        \\$\sf w_t( C_{I}[H],c,c')$ is defined, then there exists a
        constraint $\sf d'$, such that
        \\
        $\sf w_t( C_{I}[q (\ol{r}) ],c,d') \leq  w_t( C_{I}[H],c,c')$
        and $\sf \exists_{-\Var( C_{I}[q (\ol{r}) ],c)}d' =
        \exists_{-\Var( C_{I}[q (\ol{r}) ],c)}c'$.
    \end{enumerate}
\end{lemma}
\begin{proof}
Immediate. \end{proof}

The following Lemma is crucial in the proof of completeness.

\begin{lemma}
    \label{lem:pesiindi}
    Let $0 \leq i \leq n$, $\sf t \in \{ss,dd,ff\}$,
    $ \sf cl:\  q (\ol{r})\la H \in D_{i}$,
    and let $ \sf cl':\  q (\ol{r})\la H' $ be the corresponding
    declaration in $\sf D_{i+1}$ (in the case $i<n$).
    For any context $\sf C_{I}[\ ]$ and any satisfiable
    constraint $\sf c$ and for any constraint $\sf c'$ the following holds:
    \begin{enumerate}
       \item\label{pt:pesiindi1}
       If $\sf \Var(H) \cap \Var(C_{I},c) \subseteq \Var(\ol r)$
       and $\sf w_t( C_{I}[q (\ol{r}) ],c,c')$ is defined,
       then there exists a constraint $\sf d'$, such that
       $\sf \Var(d') \subseteq \Var(C_{I}[H],c)$,
       $\sf w_t( C_{I}[H],c,d') \leq
       w_t( C_{I}[q (\ol{r}) ],c,c')$
       and $\sf \exists_{-\Var( C_{I}[q (\ol{r}) ],c)}d' =
       \exists_{-\Var( C_{I}[q (\ol{r}) ],c)}c'$;
       \item\label{pt:pesiindi2}
       If $\sf \Var(H,H') \cap \Var(C_{I},c) \subseteq \Var(\ol r)$,
       $\sf \Var(c') \cap \Var(\ol r)\subseteq \Var(C_{I}[H],c)$ and
       $\sf w_t( C_{I}[H],c,c')$ is defined, then there exists a
       constraint $\sf d'$, such that $\sf \Var(d') \subseteq
       \Var(C_{I}[H'],c)$,
       $\sf w_t( C_{I}[H'],c,d') \leq  w_t( C_{I}[H],c,c')$
       and
       \\
       $\sf \exists_{-\Var( C_{I}[q (\ol{r}) ],c)}d' =
       \exists_{-\Var( C_{I}[q (\ol{r}) ],c)}c'$.
    \end{enumerate}
\end{lemma}
\begin{proof} (Sketch).
Observe that, for $i=0$, the proof of 1 follows from the first
part of Lemma~\ref{lem:pesiind0}. We prove here that, for each $i
\geq 0$,
\begin{description}
        \item[{\bf a)}] If 1 holds for $i$ then 2 holds for $i$;

        \item[{\bf b)}] If 1 and 2 hold for $i$ then 1 holds for $i+1$.
\end{description}
The proof of the Lemma then follows from straightforward inductive
argument.

{\bf a)}. If $\sf cl$ was not affected by the transformation step
from $\sf D_i$ to $\sf D_{i+1}$ then the result is obvious by
choosing $ \sf d'= \exists_{-\Var( C_{I}[H],c)} c'$. Assume then
that $\sf cl$ is affected when transforming $\sf D_i$ to $\sf
D_{i+1}$. We have various cases according to the operation used
to perform the transformation. Here we show only the proofs for
the unfolding and the folding operations, the other cases being
deferred to  the Appendix.

\II

\NI {\bf Unfolding:} Assume $\sf cl' \in D_{i+1}$ was obtained from
$\sf D_i$ by unfolding.  In this case, the situation is the following:

- $\sf cl:\ q (\ol{r}) \la  C[p(\ol{t})] \in D_{i}$

- $\sf u:\ p(\ol{s})\la  B  \in D_{i}$

- $\sf cl':\ q (\ol{r}) \la C[B \Par tell(\ol{t}=\ol{s})] \in
D_{i+1}$
\\
where \textsf{cl} and \textsf{u} are assumed to be renamed so
that they do not share variables. Let $\sf n =
w_t(C_{I}[C[p(\ol{t})]],c,c')$. By the definition of
transformation sequence, there exists a declaration $\sf
p(\ol{s})\la B_{0} \in D_{0}$. Moreover, by the hypothesis on the
variables, $\sf \Var(C[p(\ol{t})],C[B \Par tell(\ol{t}=\ol{s})])
\cap \Var(C_{I},c) \subseteq \Var(\ol r)$ and then $\sf
\Var(C_{I}[C[p(\ol{t})]],c) \cap \Var(\ol s)=\emptyset$.
Therefore, by Lemma~\ref{lem:pesoclausola}, there exists a
constraint $\sf d_{1}$, such that
\begin{equation}
        \sf w_t( C_{I}[C[p(\ol{s})\Par
        tell(\ol{t}=\ol{s})]],c,d_{1})
        \leq w_t( C_{I}[C[p(\ol{t})]],c,c')=n
        \label{eq:14ott1}
\end{equation}
and
\begin{equation}
        \sf \exists_{-\Var( C_{I}[C[p(\ol{t})]],c)}d_{1} =
        \exists_{-\Var( C_{I}[C[p(\ol{t})]],c)}c'.
        \label{eq:14ott2}
\end{equation}
By the hypothesis on the variables and since \textsf{u} is
renamed apart from \textsf{cl}, $\sf \Var(B) \cap
\Var(C_{I},C,\ol{t},c) = \emptyset$ and therefore $\sf \Var(B)
\cap \Var(C_{I}[C[\ ]\Par tell(\ol{t}=\ol{s})],c) \subseteq
\Var(\ol s)$. Then, by Point 1, there exists a constraint $\sf
d'$, such that
\begin{eqnarray*}
\sf \Var(d') &\sf  \subseteq &
\sf \Var(C_{I}[C[B\Par tell(\ol{t}=\ol{s})]],c)
\\
\sf w_t(C_{I}[C[B\Par tell(\ol{t}=\ol{s})]],c,d') & \leq &
\sf w_t(C_{I}[C[p(\ol{s})\Par tell(\ol{t}=\ol{s})]],c,d_{1}) \\
\sf \exists_{-\Var( C_{I}[C[p(\ol{s})\Par
tell(\ol{t}=\ol{s})]],c)}d' & = &
\sf \exists_{-\Var(C_{I}[C[p(\ol{s})\Par
tell(\ol{t}=\ol{s})]],c)}d_{1}.
\end{eqnarray*}
By (\ref{eq:14ott1}), $\sf w_t( C_{I}[C[B\Par
tell(\ol{t}=\ol{s})]],c,d')\leq n$.  Furthermore, by hypothesis and
construction,
$$\sf \Var(c',d')
\cap \Var(\ol r) \subseteq \Var(C_{I}[C[p(\ol{t})]],c)$$ and,
without loss of generality, we can assume that $$\sf \Var(d_{1})
\cap \Var(\ol r) \subseteq \Var(C_{I}[C[p(\ol{t})]],c).$$
Then, by (\ref{eq:14ott2}) and since $\sf \Var(
C_{I}[C[p(\ol{t})]],c) \subseteq \Var( C_{I}[C[p(\ol{s})\Par
tell(\ol{t}=\ol{s})]],c)$, we have that $\sf \exists_{-\Var(
C_{I}[q(\ol{r})],c)}d' = \exists_{-\Var( C_{I}[q(\ol{r})],c)}c'$
and this completes the proof.

 \II
\NI{\bf Folding:} Let

- $\sf cl: \ q(\ol r) \la C[B]$ be the folded declaration ($\in
\sf D_i$),

- $\sf f: \ p(\ol X) \la B$ be the folding declaration ($\in \sf
D_{0}$),

- $\sf cl': \ q(\ol r) \la C[p(\ol X)]$ be the result of the
folding operation $(\sf \in D_{i+1})$,
\\
where, by hypothesis, $\sf \Var(cl) \cap \Var(\ol X) \subseteq
\Var(B)$, $\sf \Var(B) \cap \Var(\ol r,C) \subseteq \Var(\ol X)$,
$\sf \Var(C[B],C[p(\ol X)]) \cap \Var(C_{I},c) \subseteq \Var(\ol
r)$, $\sf \Var(c') \cap \Var(\ol r) \subseteq
\Var(C_{I}[C[B]],c)$ and there exists $n$ such that $\sf w_t(
C_{I}[C[B]],c,c')=n$.  Then,
\begin{equation}
        \sf \Var(B)  \cap \Var(C_{I}[C[\ ]],c)  \subseteq
        \Var(B)  \cap \Var(\ol r,C) \subseteq \Var(\ol X)
        \label{eq:18nov3}
\end{equation}
and
\begin{equation}
        \sf \Var(c') \cap \Var(\ol r) \subseteq \Var(C_{I}[C[B]],c)
        \cap \Var(\ol r) \subseteq
        \Var(C_{I}[C[p(\ol X)]],c)
        \label{eq:14ott10}
\end{equation}
hold. Moreover, we can assume without loss of generality that
$\sf \Var(c') \cap \Var(\ol X) \subseteq \Var(C_{I}[C[B]],c)$.\\
Since $\sf f \in \sf D_{0}$, from (\ref{eq:18nov3}) and Point 2
of Lemma~\ref{lem:pesiind0} it follows that there exists a
constraint $\sf d'$  such that $\sf w_t( C_{I}[C [p(\ol
X)]],c,d') \leq w_t( C_{I}[C[B]],c,c')$ and
\begin{equation}
        \sf \exists_{-\Var( C_{I}[C [p(\ol X)]],c)}d' =
        \exists_{-\Var( C_{I}[C [p(\ol X)]],c)}c'.
        \label{eq:14ott11}
\end{equation}
We can assume, without loss of generality, that $\sf \Var(d')
\subseteq \Var(C_{I}[C[p(\ol X)]],c)$. Then by using
(\ref{eq:14ott10}) and (\ref{eq:14ott11}) we obtain that $\sf
\exists_{-\Var( C_{I}[q(\ol r)],c)} d' = \exists_{-\Var(
C_{I}[q(\ol r)],c)}c'$
which concludes the proof of {\bf a)}.\\

{\bf b)}. Assume that the parts 1 and 2 of this Lemma hold for $i
\geq 0$.
We prove that 1 holds for $i+1>0$.\\
Let $ \sf cl:\  q (\ol{r})\la H \in D_{i+1}$, and let $ \sf
\bar{cl}:\  q (\ol{r})\la \bar{H} $ be the corresponding
declaration in $\sf D_{i}$. Moreover let $\sf C_{I}[\ ]$ be a
context, $\sf c$ a satisfiable constraint and let $\sf c'$ be a
constraint, such that $\sf \Var(H) \cap \Var(C_{I},c) \subseteq
\Var(\ol r)$ and  $\sf w_t( C_{I}[q (\ol{r}) ],c,c')$ is defined.
Without loss of generality, we can assume that $\sf \Var(\bar{H})
\cap \Var(C_{I},c) \subseteq \Var(\ol r)$. Then, since by
inductive hypothesis, part 1 holds for $i$, there exists a
constraint $\sf d_{1}$ such that $\sf \Var(d_{1}) \subseteq
\Var(C_{I}[\bar{H}],c)$,
\begin{equation}
        {\sf w_t( C_{I}[\bar{H}],c,d_{1}) \leq  w_t( C_{I}[q (\ol{r})
],c,c')}
        \mbox{ and }
        {\sf \exists_{-\Var( C_{I}[q (\ol{r}) ],c)}d_{1} =
       \exists_{-\Var( C_{I}[q (\ol{r}) ],c)}c'}.
        \label{eq:42nov99}
\end{equation}
Since by inductive hypothesis part 2 holds for $i$, there exists a
constraint $\sf d'$, such that $\sf \Var(d') \subseteq
\Var(C_{I}[H],c)$, $\sf w_t( C_{I}[H],c,d') \leq w_t(
C_{I}[\bar{H}],c,d_{1})$ and $\sf \exists_{-\Var( C_{I}[q (\ol{r})
  ],c)}d' =
\exists_{-\Var( C_{I}[q (\ol{r}) ],c)}d_{1}$.
By (\ref{eq:42nov99}), $\sf w_t( C_{I}[H],c,d') \leq w_t( C_{I}[q
(\ol{r}) ],c,c')$ and $\sf \exists_{-\Var( C_{I}[q (\ol{r}) ],c)}d' =
\exists_{-\Var( C_{I}[q (\ol{r}) ],c)}c'$ and then the thesis
follows.\end{proof}

 We finally obtain our first
main theorem.

\begin{theorem}[\thetheorem\ (Total Correctness)]
\label{thm:correctness} Let $\sf D_0, \ldots, D_n$ be a
transformation sequence. Then, for any agent $\sf A$, $\sf {\cal
O}(D_0.A) = {\cal O}(D_n.A)$.
\end{theorem}
\begin{proof} (Sketch). The proof proceeds by showing
simultaneously, by induction on $i$, that for $i \in [0,n]$:
\begin{enumerate}
    \item for any agent $\sf A$, $\sf {\cal O}(D_0.A) = {\cal O}(D_i.A)$;
    \item $\sf D_i$ is weight complete.
\end{enumerate}

\NI \emph{Base case}. We just need to prove that $\sf D_0$ is
weight complete.  Assume that there exists a derivation $\sf
\langle D_0.A,c_{I}\rangle \rightarrow^* \langle
D_0.B,c_{F}\rangle$, where $\sf c_{I}$ is a satisfiable
constraint and ${\sf m(B,c_{F})} \in {\sf \{ss,dd,ff\}}$. Then
there exists a derivation $\sf \xi:\ \langle D_0.A,c_{I}\rangle
\rightarrow^* \langle D_0.B',c'_{F}\rangle $, such that ${\sf
m(B',c'_{F})} = {\sf m(B,c_{F})}$, whose weight is minimal and
where $\sf \exists_{-\Var(A,c_{I})} c'_{F} =
\exists_{-\Var(A,c_{I})} c_{F}$. It follows from
Definition~\ref{def:splitder} that $\xi$ is a split derivation.
\II

\NI \emph{Induction step}.

\NI By the inductive hypothesis for any agent $\sf A$, $\sf {\cal
  O}(D_0.A) = {\cal O}(D_{i-1}.A)$ and $\sf D_{i-1}$ is weight
complete. From propositions~\ref{pro:partial} and \ref{pro:total1} it
follows that if $\sf D_{i}$ is weight complete then for any agent $\sf
A$, $\sf {\cal O}(D_0.A) = {\cal O}(D_{i}.A)$. So, in order to prove
parts 1 and 2, we only have to show that $\sf D_{i}$ is weight
complete.

Assume then that there exists a derivation $\sf \langle
D_0.A,c_{I}\rangle \rightarrow^* \langle D_0.B,c_{F}\rangle$ such
that $\sf c_{I}$ is a satisfiable constraint and ${\sf
m(B,c_{F})} \in {\sf \{ss,dd,ff\}}$. From the inductive
hypothesis it follows that there exists a split derivation
\[\sf \chi=
\langle D_{i-1}.A,c_{I}\rangle \rightarrow^* \langle
D_{i-1}.A_m,c_m\rangle \rightarrow \langle
D_{0}.A_{m+1},c_{m+1}\rangle \rightarrow^* \langle
D_0.B'',c''_{F} \rangle
\]
where
\begin{equation}
        {\sf \exists_{-\Var(A,c_{I})}c''_{F} =
        \exists_{-\Var(A,c_{I})}c_{F}}
        \mbox{ and }
        {\sf m(B'',c''_{F})} = {\sf m(B,c_{F})}.
        \label{eq:7ott6}
\end{equation}
Let $\sf d \in D_{i-1} \backslash D_{i}$ be the modified clause
in the transformation step from $\sf D_{i-1}$ to $\sf D_{i}$.

If in the first $m$ steps of $\chi$ there is no procedure call
which uses $\sf d$ then clearly there exists a split derivation
$\xi$ in $\sf D_{i} \U D_{0}$,

$$\sf \xi=\langle D_{i}.A,c_{I}\rangle \rightarrow^* \langle
D_{i}.A_m,c_m\rangle \rightarrow \langle D_{0}.A_{m+1},c_{m+1}\rangle
\rightarrow^* \langle D_0.B'',c''_{F} \rangle$$
which performs the
same steps of $\chi$ and then the thesis holds.

Otherwise, assume without loss of generality that ${\bf R4}$ is
the rule used in the first step of derivation $\chi$ and that
$\sf d$ is the clause employed in the first step of $\chi$. We
also assume that the declaration $\sf d$ is used only once in
$\chi$, since the extension to the general case is immediate.

We have to distinguish various cases according to what happens to
the clause $\sf d$ when moving from  $\sf D_{i-1}$ to $\sf D_{i}$.
As before, we consider here only the unfolding and the folding
cases, the others being deferred to the Appendix.
 \II

\NI {\bf Unfolding:} Assume that $\sf d$ is unfolded and let $\sf
d'$ be the corresponding declaration in $\sf D_i$. The situation
is the following:

- $\sf d: \  \sf q (\ol{r})\la C[p(\ol{t})] \in D_{i-1}$,

- $\sf u: \   \sf p (\ol{s}) \la  H \in D_{i-1}$, and

- $\sf d': \  \sf q (\ol{r})\la C[H\Par tell(\ol t = \ol s)]
\in  D_{i}$,\\
where $\sf d$ and $\sf u$ are assumed to be renamed apart.  By the
definition of split derivation, $\chi$ has the form
\[\begin{array}{ll}
      \sf \langle D_{i-1}.C_I[q(\ol v)],c_I\rangle \rightarrow
      \langle D_{i-1}.C_I[C[p(\ol t)]\Par tell(\ol{v}=\ol{r})],
      c_I\rangle \rightarrow^*
      \langle D_{i-1}.A_m,c_m\rangle \rightarrow \\
      \sf \langle D_0.A_{m+1},c_{m+1}\rangle \rightarrow^*
      \langle D_0.B'',c''_F\rangle.
\end{array}\]
Without loss of generality, we can assume that $\sf \Var(\chi)\cap
\Var(u) \neq \emptyset$ if and only if $\sf p(\ol t)$ is
evaluated in the first $m$ steps of $\chi$, in which case $\sf u$
is used for evaluating it.  We have to distinguish two cases.

\emph{1)}\ \ There exists $k<m$ such that the $k$-th derivation
step of $\chi$ is the procedure call $\sf p(\ol t)$.  In this
case $\chi$ has the form
\[\begin{array}{ll}
      \sf \langle D_{i-1}.C_I[q(\ol v)],c_I\rangle \rightarrow
      \langle D_{i-1}.C_I[C[p(\ol t)]\Par tell(\ol{v}=\ol{r})],
      c_I\rangle \rightarrow^*
      \langle D_{i-1}.C_k[p(\ol t)], c_k\rangle\rightarrow \\
      \sf \langle D_{i-1}.C_k[H\Par tell(\ol t = \ol s)],
      c_k\rangle
      \rightarrow^* \langle D_{i-1}.A_m,c_m\rangle \rightarrow
      \langle D_0.A_{m+1},c_{m+1}\rangle \rightarrow^*
      \langle D_0.B'',c''_F\rangle.
\end{array}\]
Then there exists a corresponding derivation in $\sf D_{i} \U
D_{0}$
\[\begin{array}{ll}
      \sf \xi=& \sf
      \langle D_{i}.C_I[q(\ol v)],c_I\rangle \rightarrow
      \langle D_{i}.C_I[C[H\Par tell(\ol t = \ol s)]\Par
      tell(\ol{v}=\ol{r})], c_I\rangle \rightarrow^*\\
       &\sf \langle D_{i}.C_k[H\Par tell(\ol t = \ol s)], c_k\rangle
      \rightarrow^*
      \sf \langle D_{i}.A_m,c_m\rangle \rightarrow
      \langle D_0.A_{m+1},c_{m+1}\rangle \rightarrow^*
      \langle D_0.B'',c''_F\rangle,
\end{array}\]
which performs exactly the same steps of $\chi$ except for a
procedure call to $\sf p(\ol t)$. In this case the proof follows
by observing that, since by the inductive hypothesis  $\chi$ is a
split derivation, the same holds for $\xi$.

\emph{2)}\ \ There is no procedure call to $\sf p(\ol t)$ in the
first $m$ steps. Therefore $\chi$ has the form
\[\begin{array}{ll}
      \sf \langle D_{i-1}.C_I[q(\ol v)],c_I\rangle \rightarrow
      \langle D_{i-1}.C_I[C[p(\ol t)]\Par tell(\ol{v}=\ol{r})],
      c_I\rangle  \rightarrow^*
      \langle D_{i-1}.C_m[p(\ol t)], c_m\rangle\rightarrow\\
      \sf \langle D_{0}.C_{m+1}[p(\ol t)], c_m\rangle\rightarrow^*
      \langle D_0.B'',c''_F\rangle.
\end{array}\]
Then, by the definition of $\sf D_{i}$, there exists a derivation
\[\begin{array}{ll}
     \sf \xi_{0}= & \sf \langle D_{i}.C_I[q(\ol v)],c_I\rangle
     \rightarrow
     \langle D_{i}.C_I[C[H\Par tell(\ol t = \ol s)]\Par
     tell(\ol{v}=\ol{r})], c_I\rangle \rightarrow^* \\
     &\sf\langle D_{i}.C_m[H\Par tell(\ol t = \ol s)],
     c_m\rangle\rightarrow
     \langle D_{0}.C_{m+1}[H\Par tell(\ol t = \ol s)], c_m\rangle.
\end{array}\]

Observe that from the derivation $\sf \langle D_{0}.C_{m+1}[p(\ol
t)], c_m\rangle\rightarrow^* \langle D_0.B'',c''_F\rangle$ and
(\ref{eq:7ott6}) it follows that
\begin{equation}
        {\sf w_t(C_{m+1}[p(\ol t)],c_{m},c''_F)}
        \mbox{ is defined, where }
        {\sf t} ={\sf m(B,c_{F})}.
        \label{eq:23nov199}
\end{equation}
The hypothesis on the variables implies that $\sf \Var
(C_{m+1}[p(\ol t)], c_m) \cap \Var(u) =\emptyset$. Then, by the
definition of transformation sequence and since $\sf u \in
D_{i-1}$, there exists a declaration $\sf p (\ol{s}) \la H_{0}
\in D_{0}$.  By Lemma~\ref{lem:pesoclausola} and part 1 of
Lemma~\ref{lem:pesiindi} it follows that there exists a
constraint $\sf d_{F}$ such that
\begin{equation}
    \sf w_t(C_{m+1}[H\Par tell(\ol t = \ol s)],c_{m},d_F) \leq
        w_t(C_{m+1}[p(\ol t)],c_{m},c''_F)
    \label{eq:7ott3}
\end{equation}
and
\begin{equation}
    \sf\exists_{-\Var(C_{m+1}[p(\ol t)],c_{m})} d_F =
    \exists_{-\Var(C_{m+1}[p(\ol t)], c_{m})} c''_F .
        \label{eq:7ott1}
\end{equation}
Therefore, by the definition of $\sf w_t$, by (\ref{eq:7ott3}) and
since  $\sf w_t(C_{m+1}[p(\ol t)],c_{m},c''_F)$ is defined, there
exists a derivation
\[
\sf \xi_{1} = \langle D_0.C_{m+1}[H\Par tell(\ol t = \ol
s)],c_{m}\rangle \rightarrow^* \langle D_0.B',c'_F\rangle,
\]
where $\sf \exists_{-\Var(C_{m+1}[H\Par tell(\ol t = \ol s)],
c_{m})} c'_{F} = \exists_{-\Var(C_{m+1}[H\Par tell(\ol t = \ol
s)], c_{m})} d_{F}$ and, by (\ref{eq:23nov199}),
\begin{equation}
        {\sf  m(B',c'_F)} ={\sf m(B,c_{F})}.
        \label{eq:23nov299}
\end{equation}
By (\ref{eq:7ott1})
\begin{equation}
      {\sf \exists_{-\Var(C_{m+1}, c_{m})} c'_{F} =
      \exists_{-\Var(C_{m+1}, c_{m})} c''_{F}}
      \label{eq:7ott7}
\end{equation}
holds and, by definition of weight, we obtain
\begin{equation}
     \sf w_t(C_{m+1}[H\Par tell(\ol t = \ol s)], c_{m},c'_{F})
     =  w_{t}(C_{m+1}[H\Par tell(\ol t = \ol s)], c_{m},  d_{F}).
     \label{eq:7ott2}
\end{equation}
Moreover, we can assume without loss of generality that $\sf
\Var(\xi_{0}) \cap \Var(\xi_{1})= \Var(C_{m+1}[H\Par tell(\ol t =
\ol s)],c_{m})$. Then, by the definition of procedure call
\begin{equation}
    \sf \Var(C_I[q(\ol v)],c_I) \cap (\Var(c'_{F})
    \cup \Var(c''_{F}))
    \subseteq \Var(C_{m+1},c_{m})
    \label{eq:7ott8}
\end{equation}
and there exists a derivation
\[\begin{array}{ll}
     \sf \xi=&\sf \langle D_{i}.C_I[q(\ol v)],c_I\rangle
     \rightarrow \langle D_{i}.C_I[C[H\Par tell(\ol t = \ol s)]
     \Par tell(\ol{v}=\ol{r})],
     c_I\rangle \rightarrow^*\\
     &\sf \langle D_{i}.C_m[H\Par tell(\ol t = \ol s)],
     c_m\rangle\rightarrow
     \langle D_{0}.C_{m+1}[H\Par tell(\ol t = \ol s)], c_m\rangle
     \rightarrow^* \langle D_0.B',c'_F\rangle
\end{array}\]
such that the first $m-1$ derivation steps do not use rule ${\bf
R2}$ and the $m$-th derivation step uses the rule ${\bf R2}$.
Now, we have the following equalities
\[\begin{array}{lll}
     \sf \exists_{-\Var(C_I[q(\ol v)],c_I)} c'_{F} & = &
     \mbox{(by (\ref{eq:7ott8}) and by construction)}  \\
     \sf \exists_{-\Var(C_I[q(\ol v)],c_I)} (c_{m} \wedge
     \exists_{-\Var(C_{m+1},c_{m})}c'_{F}) & = &
     \mbox{(by (\ref{eq:7ott7}))}  \\
     \sf \exists_{-\Var(C_I[q(\ol v)],c_I)} (c_{m} \wedge
     \exists_{-\Var(C_{m+1},c_{m})}c''_{F}) & = &
     \mbox{(by (\ref{eq:7ott8}) and by construction)}  \\
     \sf \exists_{-\Var(C_I[q(\ol v)],c_I)} c''_{F} & = &
     \mbox{(by the first statement in (\ref{eq:7ott6}))}\\
     \sf \exists_{-\Var(C_I[q(\ol v)],c_I)} c_{F}.
\end{array}
\]
By the definition of weight, $\sf w_t(C_I[q(\ol
v)],c_I,c'_F)=w_t(C_I[q(\ol v)],c_I,c''_F)$, by (\ref{eq:7ott2})
and (\ref{eq:7ott3}), $\sf w_t(C_{m+1}[H\Par tell(\ol t = \ol
s)],c_{m},c'_F) \leq w_t(C_{m+1}[p(\ol t)],c_{m},c''_F)$ and $\sf
w_t(C_{m+1}[p(\ol t)],c_{m},c''_F) < w_t(C_I[q(\ol
v)],c_I,c''_F)$, since $\chi$ is a split derivation. Therefore
$\sf w_t(C_{m+1}[H\Par tell(\ol t = \ol s)],c_{m},c'_F) <
w_t(C_I[q(\ol v)],c_I,c'_F)$ and then, by definition, $\xi$ is a
split derivation in $\sf D_{i} \cup D_{0}$. This, together with
(\ref{eq:23nov299}), implies the thesis. \II

\noindent
{\bf Folding:} Assume that $\sf d$ is folded and let

- $\sf d: \ q(\ol r) \la C[H]$ be the folded declaration ($\in \sf
D_{i-1}$),

- $\sf f: \ p(\ol X) \la H$ be the folding declaration ($\in \sf
D_{0}$),

- $\sf d': \ q(\ol r) \la C[p(\ol X)]$ be the result of the
folding operation $ (\in \sf D_{i})$,
\\
where, by definition of folding, $\sf \Var(d) \cap \Var(\ol X)
\subseteq \Var(H)$ and $\sf \Var(H) \cap (\Var(\ol r)\cup
\Var(C))\subseteq \Var(\ol X)$. Since $\sf C[\ ]$ is a guarding
context, the agent $\sf H$ in $\sf C[H]$ appears in the scope of
an $\sf ask$ guard. By definition of split derivation $\chi$ has
the form
\[\begin{array}{ll}
\sf \langle D_{i-1}.C_I[q(\ol v)],c_I\rangle \rightarrow \langle
D_{i-1}.C_I[C[H]\Par tell(\ol{v}=\ol{r})],c_I\rangle
\rightarrow^* \langle
D_{i-1}.C_{m}[H],c_m\rangle \rightarrow \\
\sf \langle  D_0.C_{m+1}[H],c_{m}\rangle \rightarrow^* \langle
D_0.B'',c''_F\rangle ,
\end{array}\]
where $\sf C_{m}[\ ]$ is a guarding context. Without loss of
generality we can assume that $\sf \Var (\chi)\cap \Var(\ol X)\subseteq \Var(H)$. Then, from the definition of $\sf D_{i}$ it
follows that there exists a derivation
\[
\begin{array}[t]{ll}
\sf \xi_{0} = & \sf \langle D_{i}.C_I[q(\ol v)],c_I\rangle
\rightarrow \langle D_{i}.C_I[C[p(\ol X)]\Par
tell(\ol{v}=\ol{r})],c_I\rangle \rightarrow^*
\\
& \sf \langle \sf D_i.C_{m}[p(\ol X)],c_m\rangle \rightarrow
\langle D_0.C_{m+1}[p(\ol X)],c_{m}\rangle,
\end{array}
\]
which performs exactly the first $m$ steps as $\chi$. Since $\sf
\langle D_0.C_{m+1}[H],c_{m}\rangle \rightarrow^* \langle
D_0.B'',c''_F\rangle$, the definition of weight implies that $\sf
w_{t}(C_{m+1}[H],c_{m},c''_F)$ is defined, where $\sf
t=m(B'',c''_F)$. Then, by (\ref{eq:7ott6}), we have that
\begin{equation}
        {\sf t= m(B, c_{F})}.
        \label{eq:29ott5}
\end{equation}
The definitions of derivation and folding  imply that $\sf \Var(H
) \cap \Var(C_{m+1},c_{m}) \subseteq \sf \Var(H ) \cap
(\Var(C,\ol r)) \subseteq \Var(\ol{X})$ holds.  Moreover, from
the assumptions on the variables, we obtain that $\sf
\Var(c''_{F} ) \cap \Var(\ol{X})\subseteq \Var(H)$. Thus, from
part 2 of Lemma~\ref{lem:pesiind0} it follows that there exists a
constraint $\sf d'$ such that
\begin{eqnarray}
        {\sf w_{t}(C_{m+1}[p(\ol{X})],c_{m},d')
        \leq w_{t}(C_{m+1}[H],c_{m},c''_F)}
        \mbox{ and }
        \label{eq:29ott2}
\\
        {\sf \exists _{-\Var(C_{m+1}[p(\ol{X})],c_{m})}d' =
        \exists _{-\Var(C_{m+1}[p(\ol{X})],c_{m})}c''_{F}.}
        \label{eq:29ott2bis}
      \end{eqnarray}
>From the definition of weight and the fact that $\sf
w_{t}(C_{m+1}[H],c_{m},c''_F)$ is defined it follows that there
exists a derivation $\sf \xi_{1}= \langle
D_0.C_{m+1}[p(\ol{X})],c_{m}\rangle \rightarrow^* \langle
D_0.B',c'_F\rangle$, where $\sf m(B',c'_F)=t$ and $\sf \exists
_{-\Var(C_{m+1}[p(\ol{X})],c_{m})}c'_{F}= \exists
_{-\Var(C_{m+1}[p(\ol{X})],c_{m})}d'$. Then, by the definition of
weight, $\sf w_{t}(C_{m+1}[p(\ol{X})],c_{m},c'_{F})=
w_{t}(C_{m+1}[p(\ol{X})],c_{m},d')$ and therefore, by
(\ref{eq:29ott2}) and (\ref{eq:29ott2bis}),
\begin{eqnarray}
        {\sf \exists _{-\Var(C_{m+1}[p(\ol{X})],c_{m})}c'_{F}=
        \exists _{-\Var(C_{m+1}[p(\ol{X})],c_{m})}c''_{F}}
        \mbox{ and }
        \label{eq:29ott1}
\\
        {\sf w_{t}(C_{m+1}[p(\ol{X})],c_{m},c'_{F}) \leq
        w_{t}(C_{m+1}[H],c_{m},c''_F)}
        \label{eq:29ott1bis}
\end{eqnarray}
hold. Moreover, from (\ref{eq:29ott5}) we obtain
\begin{equation}
            {\sf  m(B',c'_F)} = {\sf m(B, c_{F})}.
        \label{eq:29ott51}
\end{equation}
Without loss of generality, we can now assume that
$$\sf
\Var(\xi_{0}) \cap \Var(\xi_{1}) =
\Var(C_{m+1}[p(\ol{X})],c_{m})$$.
 Then, by (\ref{eq:29ott1}),
(\ref{eq:29ott1bis}) and (\ref{eq:7ott6}) it follows that
\begin{eqnarray}
         &  & \sf
         \exists _{-\Var(C_I[q(\ol v)],c_I)}c'_{F}=
        \exists _{-\Var(C_I[q(\ol v)],c_I)}(c_{m} \wedge
        \exists _{-\Var(C_{m+1}[p(\ol{X})],c_{m})} c'_{F})=
        \nonumber \\
         &  & \sf \exists _{-\Var(C_I[q(\ol v)],c_I)}(c_{m} \wedge
        \exists _{-\Var(C_{m+1}[p(\ol{X})],c_{m})} c''_{F})=
        \nonumber \\
         &  & \sf
        \exists _{-\Var(C_I[q(\ol v)],c_I)}c''_{F} =
        \exists _{-\Var(C_I[q(\ol v)],c_I)}c_{F}.
        \label{eq:29ott3}
\end{eqnarray}
>From the definition of weight $\sf w_{t}(C_I[q(\ol
v)],c_I,c'_{F})= w_{t}(C_I[q(\ol v)],c_I,c''_{F})$ and since
$\chi$ is a split derivation we obtain $\sf w_{t}(C_I[q(\ol
v)],c_I,c''_{F}) > w_{t}(C_{m+1}[H],c_{m},c''_F)$. Then, from
(\ref{eq:29ott3}) it follows that
\begin{equation}
        \sf \sf w_{t}(C_I[q(\ol v)],c_I,c'_{F})>
        w_{t}(C_{m+1}[p(\ol{X})],c_{m},c'_{F})
        \label{eq:29ott4}
\end{equation}
and therefore, by construction,
\begin{eqnarray*}
        \xi & = &
        \sf \langle D_{i}.C_I[q(\ol v)],c_I\rangle \rightarrow
        \langle D_{i}.C_I[C[p(\ol X)]\Par
        tell(\ol{v}=\ol{r})],c_I\rangle \rightarrow^*
        \langle  D_i.C_{m}[p(\ol X)],c_m\rangle \rightarrow  \\
        &  & \sf \langle D_0.C_{m+1}[p(\ol X)],c_{m}\rangle
        \rightarrow^* \langle D_0.B',c'_F\rangle
\end{eqnarray*}
is a derivation in $\sf D_{i} \cup D_{0}$ such that: (a) rule
${\bf R2}$ is not used in the first $m-1$ steps; (b) rule ${\bf
R2}$ is used in the $m$-th step. The thesis then follows from
(\ref{eq:29ott3}), (\ref{eq:29ott51}) and (\ref{eq:29ott4}) thus
concluding the proof.
 \end{proof}

 It is important to notice that -- given the definition
of observables we are adopting (Definition~\ref{def:semantics}) --
the initial program $\sf D_0$ and the final one $\sf D_n$ have
exactly the same successful derivations, the same deadlocked
derivations and the same failed derivations. The first feature
(regarding successful derivations) is to some extent the one we
expect and require from a transformation, because it corresponds
to the intuition that $\sf D_n$ ``produces the same results'' as
$\sf D_0$.  Nevertheless, also the second feature (preservation
of deadlock derivations) has an important role. Firstly, it
ensures that the transformation does not introduce deadlock
points, which is of crucial importance when we are using the
transformation for optimizing a program. Secondly, as exemplified
in the Section \ref{sec:example}, this feature allows us to use
the transformation as a tool for proving deadlock freeness (i.e.,
absence of deadlock). In fact, if, after the transformation we
can prove or see that the process $\sf D_n.A$ does never
deadlock, then we are also sure that $\sf D_0.A$ does not
deadlock either.

\section{Correctness for non-terminating computations}
\label{sec:intermediate}

The correctness results obtained so far consider terminating
(successful and deadlocked) and failed computations only.  This is
satisfactory for many applications of concurrent constraint
programming which have a ``transformational'' behaviour, i.e.  which
are supposed to produce a (finite) output for a given (finite) input. 
In this respect,
it is worth noting that the two main semantic models of CCP consider
essentially the same notion of observables we used. In fact, the model
based on linear sequences defined in
\cite{BP91} characterizes (in a fully abstract way) the results of
terminating computations, together with a termination mode indicating
success, deadlock or failure\footnote{There are irrelevant differences
  between the observables considered in \cite{BP91} and the ones we
  used, due to the treatment of failure and to the existential
  quantification on local variables.}.  Such a model has been proved
(\cite{BP92}) to be isomorphic to the semantics based on (bounded)
closure operators introduced in \cite{SRP91}, provided that the
termination mode and the consistency checks are eliminated.

So, our correctness results are adequate in the sense that they ensure
that the standard semantics of CCP is preserved.  On the other hand,
as in the case of any other concurrent programming paradigm, CCP
programs may have a ``reactive'' nature: rather than producing a final
result they produce a (possibly non-terminating) sequence of
intermediate results in response to some external stimuli. For these
programs the notion of observables employed in Theorem~\ref{thm:correctness}
and the related results are not adequate, since
they exclude non-terminating computations.

When considering non-terminating computations one is interested
in observing (possibly in terms of traces) the intermediate
results, that is the constraints produced also by non-maximal
derivations, rather than the final limit of the computation (note
however that in CCP such a notion of limit makes sense, as the
store grows monotonically). Therefore, in the remainder of this
section we first discuss the correctness of our system w.r.t.\
this new class of observables. Then, we show a modification of
our transformation system and we present a stronger correctness
result, which guarantees that (traces of) intermediate results
are preserved.

\subsection{Partial preservation of intermediate results}

It is easy to see that the system we have proposed does not preserve
the intermediate results of computations. More precisely,
let us define these observables as follows:

$$\begin{array}[b]{llllll}
\sf {\cal O}_i(D.A) & =  & \sf \{ \langle c,\exists_{-\Var(A,c)}
d,pp\rangle  & \mid &
\sf c \  \hbox{\rm and } d \ \hbox{\rm are satisfiable, and there exists}\\
&&&&\sf \hbox{\rm a derivation }
\langle D.A, c\rangle \rightarrow^*\langle D.B ,d\rangle & \}
\end{array}
$$
(the symbol \textsf{pp} indicates here that we consider results
obtained from ``partial'', that is, possibly not maximal,
derivations).  Now, it is easy to see that the operations of ask and
tell simplification are neither partially nor totally correct w.r.t.
the semantics $\sf {\cal O}_i(D.A)$. In fact, the ask simplification
allows one to transform the agent
\[
\sf A:\ tell(c) \parallel ask(true)\rightarrow tell(d)
\]
into the agent
\[
\sf A':\ tell(c) \parallel ask(c)\rightarrow tell(d).
\]
While the agent $\sf A$, when evaluated in the empty store,
produces the intermediate result $\sf d$, this is not the case for the
agent $\sf A'$ (we assume that $\sf c\wedge d \neq d$).  Analogously,
assuming that $\sf {\cal D} \models d\rightarrow c$ and $\sf {\cal D}
\models d\rightarrow c'$, the tell simplification allows one to
transform
\[
\sf B:\ tell(c) \parallel tell(d)
\]
into the agent
\[
\sf B':\ tell(c') \parallel tell(d)
\]
and the agents $\sf B$ and $\sf B'$ have different intermediate
results. Other operations which are not correct w.r.t.\ the above
semantics are the distribution and the tell elimination and
introduction.
\medskip

Nevertheless, the system we have defined does preserve already a
form of intermediate results. This is shown by the following
theorem.

\begin{theorem}[\thetheorem\ (Total Correctness 2)]
\label{thm:correctness2}
Let $\sf D_0, \ldots, D_n$ be a transformation sequence, and $\sf A$
be an agent.\begin{itemize}
  \item  If there exists a derivation $\sf \langle
  D_0.A,c\rangle \rightarrow^* \langle D_0.B,d\rangle$ then there
  exists a derivation $\sf \langle D_n.A,c\rangle \rightarrow^*
  \langle D_n.B',d'\rangle$ such that $\sf {\cal D}\models
  \exists_{-Var(A,c)}d'\rightarrow \exists_{-Var(A,c)}d$.
\item Conversely, if there exists a derivation $\sf \langle
  D_n.A,c\rangle \rightarrow^* \langle D_n.B,d \rangle$ then there
  exists a derivation $\sf \langle D_0.A,c\rangle \rightarrow^*
  \langle D_0.B',d'\rangle$ with $\sf {\cal D}\models
  \exists_{-Var(A,c)}d'\rightarrow \exists_{-Var(A,c)}d$.
\end{itemize}
\end{theorem}
\begin{proof}  The proof of this result is essentially the same as
that one of the total correctness Theorem \ref{thm:correctness}
provided that in such a proof, as well as in the proofs of the
related preliminary results, we perform the following changes:
\begin{enumerate}
\item Rather than considering terminating derivations, we consider any
(possibly non-maximal) finite derivation.
\item Whenever in a proof we write that, given a derivation $\xi$, a
  derivation $\xi'$ is constructed which performs the same steps $\xi$
  does, possibly in a different order, we now write that a derivation
  $\xi''$ is constructed which performs the same step of $\xi$
  (possibly in a different order) plus some other additional steps.
  Since the store grows monotonically in CCP derivations, clearly if a
  constraint $\sf c$ is the result of the derivation $\xi$, then a
  constraint $\sf c''$ is the result of $\xi''$ such that $\sf {\cal
    D}\models c''\rightarrow c$ holds. For example, for case 2 in the
  proof of Proposition \ref{pro:partial}, when considering a
  (non-maximal) derivation $\xi$ which uses the declaration $\sf
  H\leftarrow C[tell(\ol{s} = \ol{t})]\parallel B]$ we can always
  construct a derivation $\xi''$ which performs all the steps of $\xi$
  (possibly plus others) and such that the $\sf tell(\ol{s} = \ol{t})$
  agent is evaluated before $\sf B$.  Differently from the previous
  proof, now we are not ensured that the result of $\xi$ is the same
  as that one of $\xi''$, since $\xi$ is non-maximal (thus, $\xi$
  could also avoid the evaluation of $\sf tell(\ol{s} = \ol{t})$).
  However, we are ensured that the result of $\xi''$ is stronger (i.e.
  implies) that one of $\xi$.
\end{enumerate}
\end{proof}

This result ensures that the original and the transformed program have
the same intermediate results up to logical implication: If the
evaluation of an agent in the original program produces a constraint
$\sf d$, then a constraint stronger than $\sf d$ is produced in the
transformed program and vice versa. The vice versa is important, as it
ensures that the transformed program will never produce something that
could not be produced by the original program, up to implication.
Clearly, this result is relevant in presence of non-terminating
computations (which were not covered by Theorem
\ref{thm:correctness}).

In order to maintain a consistent notation throughout the paper, the
above result can be reformulated in terms of the following class of
observables
$$\begin{array}[b]{llllll}
\sf {\cal O}_{ic}(D.A) & =  &
\sf \{ \langle c,\exists_{-\Var(A,c)} d,pp\rangle  & \mid &
\sf c \ \ \hbox{\rm is satisfiable, there exists a derivation }
\\
&&&&
\sf \langle D.A, c\rangle \rightarrow^*\langle D.B ,d'\rangle\\
&&&&\hbox{and }\sf {\cal D}\models \exists_{-\Var(A,c)} d'\rightarrow
\exists_{-\Var(A,c)} d  & \}
\end{array}
$$
where the subscript \textsf{ic} stands for \emph{implication closure}
(of intermediate results). We then have 
following Corollary whose proof is immediate.

\begin{corollary}
  Let $\sf D_0, \ldots, D_n$ be a transformation sequence. Then, for
  any agent $\sf A$, $\sf {\cal O}_{\sf ic}({\sf D_0.A}) = {\cal
O}_{ic}({\sf
    D_n.A})$.\\ \mbox{}
\end{corollary}

This result guarantees a degree of correctness which should be
sufficient for many reactive programs employing non-terminating
computations. In fact, when transforming a program, probably one
should not expect to be able to preserve exactly each intermediate
result the original program was producing.

Nevertheless, it is of interest to check if it is possible to modify
the system in order to obtain stronger correctness results. We do this
in the following section.

\subsection{Full preservation of intermediate results}

In this section we introduce a few restrictions on our transformation
system and we prove that they guarantee the preservation of the whole
sequence of intermediate results of a program.

As previously mentioned, the only operations not preserving the
intermediate results are the ask and tell simplification, the
distribution and the tell elimination and introduction. As it possibly
appears from the example above, the problem using the ask and tell
simplification lies in the fact that one can modify the arguments of
ask and tell agents by taking into account (via the ``produced
constraint'') also the constraints introduced by tell actions
appearing in the parallel context (see Definitions \ref{def:pc} and
\ref{def:guard_simplification}).  This clearly can affect the
intermediate results of the computations, since no order is imposed on
the evaluation of parallel agents. This reasoning applies to the
distribution operation as well.

We have then to modify the ask and tell
simplification and the distribution by considering a weaker notion of
``produced constraint'', which includes only those constraints which
have \emph{certainly} been produced before reaching the ask or tell
agent we are simplifying. Such a notion is defined as follows.

\begin{definition}
\label{def:wpc} Given a context $\sf C[\ ]$
the \emph{weakest produced constraint} $\sf wpc(C[\ ])$ of $\sf C[\ ]$
is inductively defined as follows: \II

$\begin{array}[b]{ll}
\sf wpc([\ ]) = true
\\
\sf wpc(C'[\ ]\Par B) = wpc(C'[\ ])
\\
\sf wpc(\sum_{i=1}^n ask(c_i) \rightarrow A_i) =
c_j\wedge wpc(C'[\ ]) &
\mbox{where}\ {\sf j \in [1,n]} \ \mbox{and} \ {\sf  A_j = C'[\ ]}.\\
\end{array}
$ \mbox{ }
\end{definition}

For example, the weakest produced constraint of $\sf [\ ]
\parallel tell(c)$ is \textsf{true}, while the weakest produced constraint of
$\sf tell(c) \parallel
ask(d) \ra (ask(e) \ra [\ ])$ is $\sf d \A e$.  We can then define the
weak equivalence of two constraints within a given context $\sf C[\ ]$
as follows.

\begin{definition}
  Let $\sf c$, $\sf c'$ be constraints,
 $\sf C[\ ]$ be a context, and $\sf \ol Z$ be a set of
  variables.  We say that $\sf c$ is \emph{weakly equivalent to} $\sf c'$
  \emph{within} \textsf{C[\ ]} and \emph{w.r.t. the variables in}
  $\sf \ol Z$ iff ${\cal D} \models \sf \exists_{- \ol Z}\
  (wpc (C[\ ]) \And c)$ $\sf \lra
  \exists_{- \ol Z}\ (wpc (C[\ ]) \And c')$.
\end{definition}

Using this definition we can modify the operations of ask and tell
simplification and of distribution by simply replacing the context
equivalence used in Definition \ref{def:guard_simplification} with the
above notion of weak context equivalence.  For the sake of clarity we
state below the resulting definitions.

\begin{definition}[\thetheorem\ (Restricted Ask and Tell Simplification)]
\label{def:guard_simplificationrestricted}
Let $\sf D$ be a set of declarations.
\begin{enumerate}
\item Let $\sf d:\ H \la C[\sum_{i=1}^n ask(c_i) \rightarrow A_i]$
be a declaration of $\sf D$.
Suppose that $\sf c'_1,\ldots,c'_n$ are constraints such that
for $j\in [1,n]$, $\sf c'_j$ is weakly equivalent to $\sf c_j$ within $\sf
C[\ ]$ and w.r.t.\ the variables in $\Var(\sf{C,H,A_j})$.
Then we can replace $\sf d$ with $\sf d':\ H \la C[\sum_{i=1}^n
ask(c'_i) \rightarrow A_i]$ in $\sf D$. We call this a
\emph{restricted ask simplification} operation.

\item Let $\sf d: \ H \la C[tell(c)]$ be a declaration of $\sf D$.
  Suppose that the constraint $\sf c'$ is weakly equivalent to $\sf c$
within
  $\sf C[\ ]$ and w.r.t.\ the variables in $\sf \Var(C,H)$.
  Then we can replace $\sf d$ with $\sf d':\ H \la C[tell(c')]$ in
  $\sf D$. We call this a \emph{restricted tell simplification}
  operation.
\end{enumerate}
\end{definition}

%
\begin{definition}[\thetheorem\ (Restricted Distribution)]
\label{def:distributionrestricted}
  Let $\sf D$ be a set of declarations and let
  \II

  $\sf d:\ H \la C[A \Par \sum_{i=1}^n ask(c_i) \rightarrow B_i]$ \II

  \NI be a declaration in $\sf D$. Let also $\sf e = wpc(C[\ ])$.  The
  \emph{restricted distribution} of $\sf A$ in $\sf d$ yields the
  definition\II

  $\sf d':\   H \la  C[ \sum_{i=1}^n ask(c_i) \rightarrow (A \Par B_i)]$
\II

\NI provided that for every constraint $\sf c$ such that $\sf \Var(c)
\cap \Var(d) \subseteq \Var(H,C)$, if
$\sf  \langle D.A, c \wedge e \rangle$ is productive
then both the following conditions hold:
\begin{enumerate}\parentalphi
\item There exists at least one $i\in [1,n]$ such that $\sf {\cal
    D}\models (c \wedge e ) \rightarrow c_i$,

\item for each $i\in [1,n]$, either $\sf {\cal D}\models (c \wedge e )
  \rightarrow c_i$ or $\sf {\cal D}\models (c \wedge e ) \rightarrow
  \neg c_i$.
\end{enumerate}
\end{definition}

Remark \ref{rem:distribution} is also sufficient for guaranteeing that
the restricted distribution operation is applicable. Thus
we have the following.

\begin{remark}
\label{rem:distributionrestricted} Referring to Definition
\ref{def:distributionrestricted}.  If $\sf A$ requires a
variables which does not occur in $\sf H, C[\ ]$, then the
restricted distribution operation is applicable.
\end{remark}

Also the tell elimination and the tell introduction operations do not
preserve the intermediate results of computations.  This is not due to
the presence of the \emph{produced constraint}, but rather to the very
nature of the operation which can eliminate or introduce constraints
which, via the local variables, can
(temporarily)
affect also the values of global variables. For example, the
declaration
\begin{eqnarray*}
  \sf d: && \sf p(Y) \la tell(Z = a) \parallel tell(Y = f(Z))
\end{eqnarray*}
can be transformed via a \emph{tell elimination} into
\begin{eqnarray*}
\sf d':&& \sf p(Y) \la tell(Y = f(a))
\end{eqnarray*}
The evaluation of $\sf p(Y)$ in the empty store and using $\sf d$
produces the (intermediate) result $\sf Y = f(Z)$, while this is not
the case if one uses the declaration $\sf d'$.  We can solve this
problem by simply requiring that if we eliminate a tell by applying
the resulting substitution to the parallel context $\sf B$, then $\sf
B$ does not contain any variable appearing the head or in the
outer context. Thus we have the following.


\begin{definition}[\thetheorem\ (Restricted Tell Elimination and Tell
  Introduction)]\label{def:tellelinrestricted} The declaration
\begin{eqnarray*}
  \sf d: && \sf H \la C[tell(\ol{s}=\ol{t}) \Par B]
\end{eqnarray*}
can be transformed via a \emph{restricted tell elimination} into
\begin{eqnarray*}
\sf d':&&  \sf H \la  C[B\sigma]
\end{eqnarray*}
where $\sigma$ is a relevant most general unifier of $\sf \ol{s}$ and 
$\sf \ol{t}$, provided that the variables in the domain of $\sigma$ do not occur
neither in $\sf C[\ ]$ nor in $\sf H$, and that $\sf \Var(B) \I
\Var(H,C) = \emptyset$. Again, this operation is applicable either when
the computational domain admits a most general unifier, or when
$\sf \ol{s}$ and $\sf \ol{t}$ are sequence of distinct variables, in which
case $\sigma$ is simply a renaming. On the other hand, the declaration
\begin{eqnarray*}
  \sf d: && \sf H \la C[B\sigma]
\end{eqnarray*}
can be transformed via a \emph{restricted tell introduction} into
\begin{eqnarray*}
\sf d':&&  \sf H \la  C[tell(\ol{X}=\ol{X}\sigma) \Par B]
\end{eqnarray*}
provided that $\sigma$ is a substitution such that $\sf \ol{X}=
\Dom(\sigma)$ and $\Dom(\sigma) \I (\Var({\sf C[\ ],H}) \cup
\Ran(\sigma)) = \emptyset$, and that $\sf \Var(B) \I \Var(H,C) =
\emptyset$.
\end{definition}

At this point it is worth recalling that the tell elimination is often
used for making variable bindings explicit after an unfolding
operation: In fact we start from a definition of the form $\sf d:\ H
\la C[p(\ol{t})]$ and by unfolding $\sf p(\ol{t})$ we end with $\sf
d':\ H \la C[B \Par tell(\ol{t}=\ol{s})]$ (provided that $\sf p$ is
defined by $\sf u:\ p(\ol{s})\la B$). Then we want to eliminate $\sf
tell(\ol{t}=\ol{s})$ from $\sf d'$ in order to perform the ``parameter
passing''. Since $\sf d$ and $\sf u$ are always renamed apart, clearly
the additional condition of the restricted tell elimination ($\sf
\Var(B) \I \Var(H,C) = \emptyset$) is always satisfied here.  So, in
general, this operation is applicable every time that $\sf \ol t$ is
an instance of $\sf \ol s$.

We can finally define the restricted transformation system as follows.

\begin{definition}
  A \emph{restricted transformation sequence} is a sequence of
  programs $\sf D_0, \ldots, D_n$ in which $\sf D_0$ is a initial
  program and each $\sf D_{i+1}$ is obtained from $\sf D_i$ via one of
  the following operations: unfolding, backward instantiation,
  restricted tell elimination, restricted tell introduction,
  restricted ask and tell simplification, branch elimination,
  conservative ask elimination, restricted distribution and folding.
\end{definition}

Clearly, the restricted transformation operations are applicable in fewer
situations than
their non-restricted counterparts, yet they are useful in many cases.
Example \ref{exa:read-write} shows a case of an unfold-fold
transformation sequence using only restricted operations and the other
examples contain several occurrences of them. We now prove that the
restricted system is correct w.r.t.
the trace semantics of CCP. Here and in the following we denote by $\sf
c_1;c_2; \ldots ;c_n$ a sequence of constraints, also called trace.

\begin{definition}[\thetheorem\ (Traces)]
\label{def:semanticstraces}
  Let $\sf D.A$ be a CCP process. We define $\sf {\cal O}_t(D.A)=$
\II

$\begin{array}[b]{llll} \sf \{ \langle c_1;c_2; \ldots
;c_n,ss\rangle  & \mid &
\sf \hbox{\rm there exists a derivation}\\
&& \sf \langle D.A, d_1\rangle \rightarrow \langle D.A_2,
d_2\rangle
\rightarrow \ldots \rightarrow \langle D.Stop ,d_n\rangle \\
&& \sf d_i\ \hbox{\rm is satisfiable for each } i\in [1,n],
\\
&& \sf c_1 = d_1 \hbox { and } \sf c_j = \exists_{-\Var(A,c_1)}
d_j \hbox{ for each } j\in[2,n] \}
\\
\cup
\\
\sf \{ \langle c_1;c_2; \ldots; c_n,dd\rangle  & \mid &
\sf \hbox{\rm there exists a derivation }\\
&& \sf \langle D.A, d_1\rangle \rightarrow \langle D.A_2,
d_2\rangle
\rightarrow \ldots \rightarrow \langle D.A_n ,d_n\rangle \not\rightarrow \\
&& \sf A_n\neq Stop,\ d_i\ \hbox{\rm is satisfiable for each }
i\in [1,n],
\\
&& \sf c_1 = d_1 \hbox { and } \sf c_j = \exists_{-\Var(A,c_1)}
d_j \hbox{ for each } j\in[2,n] \}
\\
\cup
\\
\sf \{ \langle c_1;c_2; \ldots; c_n,pp\rangle  & \mid &
\sf \hbox{\rm there exists a derivation }\\
&& \sf \langle D.A, d_1\rangle \rightarrow \langle D.A_2,
d_2\rangle
\rightarrow \ldots \rightarrow \langle D.A_n ,d_n\rangle \\
&& \sf d_i\ \hbox{\rm is satisfiable for each } i\in [1,n],
\\
&& \sf c_1 = d_1 \hbox { and } \sf c_j = \exists_{-\Var(A,c_1)}
d_j \hbox{ for each } j\in[2,n] \}
\\
\cup
\\
\sf \{ \langle c_1;c_2; \ldots; c_n,ff\rangle  & \mid &
\sf \hbox{\rm there exists a derivation }\\
&& \sf \langle D.A, d_1\rangle \rightarrow \langle D.A_2,
d_2\rangle
\rightarrow \ldots \rightarrow \langle D.A_n ,d_n\rangle \not\rightarrow \\
&&
\sf d_i\ \hbox{\rm is satisfiable for each } i\in
[1,n-1], \ d_n = false
\\
&& \sf c_1 = d_1 \hbox { and } c_j = \exists_{-\Var(A,c_1)} d_j
\hbox{ for each } j\in[2,n]. \}
\end{array}
$ \mbox{}
\end{definition}
Thus what we observe are the finite traces consisting of the
constraints produced by any (possibly non-terminating)
derivation.  As before, we abstract from the values for the
local variables in the results, and we make distinction between
the successful traces (termination mode $\sf ss$), the
deadlocked ones ($\sf dd$), the partial (i.e. possibly non
maximal) traces ($\sf pp$) and the failed ones ($\sf ff$).  Note
that, due to the monotonic computational model of CCP which does
not allow us to retract information from the global store, the
traces we observe are monotonically increasing.  That is, given
a trace $\sf c_1;c_2; \ldots; c_n$ appearing in the
observables, we have that $\sf {\cal D} \models c_i\rightarrow
c_j$ for each $i,j\in [1,n]$ such that $i\geq j$. Before giving
the correctness result, we need one last definition.

\begin{definition}
  We say that a trace $\sf c_1;c_2; \ldots; c_n$ is \emph{simulated
    by} a trace
    \\$\sf d_1;d_2;\ldots;d_m$, notation $\sf
  c_1;c_2;\ldots;c_n \preceq d_1;d_2;\ldots;d_m$, iff there exists
  \\
  $\{ j_1,\ldots j_n\} \subseteq \{1,2,\ldots,m\}$ such that
\begin{enumerate}
\item $\sf c_i = d_{j_i}$ for each $i \in [1,n]$;
\item $\sf j_1 = 1, j_n = m$ and $j_i \leq j_k$ iff $i < k$.
\end{enumerate}
\end{definition}

So, a trace $s$ is simulated by a trace $s'$ iff they have the same
first and last element and, all components appearing in $s$ appear, in
the same order, in $s'$.

We can now state our strongest correctness result. Its proof,
contained in the Appendix, follows the
guidelines of that one of Theorem \ref{thm:correctness}. In fact, the
definitions of mode, weight, split derivation and weight complete
program can readily be extended to consider traces and weakest
produced constraints, rather than input/output pairs and produced
constraints. Then it is easy to extend all the technical lemmata
needed for Theorem \ref{thm:correctness} in order to obtain the
preliminary results needed in the proof of the following.

\begin{theorem}[\thetheorem\ (Strong Total Correctness)]
\label{thm:correctnessstrong}
Let $\sf D_0, \ldots, D_n$ be a restricted transformation sequence, and
$\sf
A$ be an agent.
\begin{itemize}
\item If $\sf \langle s,x\rangle \in \sf {\cal O}_t(D_0.A)$
  (with $\sf x \in \{ss,dd,pp,ff\}$)
then there exists $\sf \langle s',x\rangle \in \sf {\cal O}_t(D_n.A)$
such that $\sf s\preceq s'$.
\item  Conversely, if $\sf \langle
s,x\rangle \in \sf {\cal O}_t(D_n.A)$ then there exists $\sf
\langle s',x\rangle \in \sf {\cal O}_t(D_0.A)$ such that $\sf
s\preceq s'$.
\end{itemize}
\end{theorem}

As it results from the definition of $\preceq$, we do not have exactly
the equality of traces since in some traces we might introduce some
intermediate steps. However, notice that these additional steps do not
introduce new values, rather they can be seen as different
``approximation'' to obtain a given constraint, since we consider here
monotonically increasing traces. This can best be
explained by means of an example. Consider the following one-line
program $\sf D_0$: \textsf{p(Y) \la tell(X = f(a,W)) \Par tell(X =
  f(Z,b)) \Par tell(X = Y)}. Its trace semantics $\sf {\cal
  O}_t(D_0.p(Y))$ contains $\sf \langle t,\ ss \rangle$, where $\sf t$
is the trace $\sf (true;\ true;\ true;\ Y=f(a,b))$. If we apply here a
restricted tell evaluation to $\sf tell(X = Y)$ we obtain the program
$\sf D_1$: \textsf{p(Y) \la tell(Y = f(a,W)) \Par tell(Y = f(Z,b))}.
Now, $\sf {\cal O}_t(D_1.p(Y))$ does not contain $\sf t$: one cannot
obtain $\sf Y = f(a,b)$ from $\sf true$ in one step. On the other
hand, $\sf {\cal O}_t(D_1.p(Y))$ contains $\sf \langle (true;\
\exists_W\,Y = f(a,W);\ Y = f(a,b)),\ ss\rangle$ and $\sf \langle
(true;\ \exists_Z\,Y = f(Z,b);\ Y = f(a,b)),\ ss\rangle$ and both the
two traces appearing in these pairs simulate $\sf t$.  Notice also
that the intermediate results semantics is now preserved. In fact, the
following is an immediate consequence of Theorem
\ref{thm:correctnessstrong}.

\begin{corollary}
\label{cor:intermediate}
Let $\sf D_0, \ldots, D_n$ be a restricted transformation sequence,
and $\sf A$ be an agent. Then $\sf {\cal O}_i(D_0.A) = {\cal
  O}_i(D_n.A)$.
\end{corollary}


\subsection{Preservation of infinite traces}

It is worth noting that Theorem \ref{thm:correctnessstrong} can be
extended to consider also infinite traces, as we show below.

In the following we indicate by $\sf |s_i|$ the length of a trace
$\sf s_i$ and we say that a configuration $\sf \langle D.A,
c_1\rangle$ produces the trace $\sf c_1;c_2; \ldots; c_n$ iff
there exists a derivation $\sf \langle D.A, d_1\rangle
\rightarrow \langle D.A_2, d_2\rangle \rightarrow \ldots
\rightarrow \langle D.A_n ,d_n\rangle$ such that $\sf c_1 = d_1$
and $\sf c_j = \exists_{-\Var(A,c_1)} d_j$ for each $j\in[2,n]$.
This notion can be extended to consider infinite computations
(and infinite traces) in the obvious way. We also call an infinite
trace $\sf c_1;c_2\ldots$ ``active'' iff, for any $i\geq 1$,
there exists $j>i$ such that $\sf {\cal D} \models \neg
(c_i\rightarrow c_j)$ (on the other hand, the implication $\sf
{\cal D} \models (c_j\rightarrow c_i)$ holds for any $j\geq i$
when considering traces produced by CCP derivations, since they
are monotonically increasing). So, an active trace
is that one produced by a computation which continuously updates 
the store by adding new constraints. Clearly, when considering
infinite computations, one is interested mainly in those
producing active traces, as the others are essentially pure
loops which stop producing new results after a finite number of
steps.

The essential result we use for extending Theorem
\ref{thm:correctnessstrong} to infinite traces is the following:
If a CCP configuration can produce all the finite prefixes of an
infinite trace, then it can produce the infinite trace itself.
The following Lemma contains a slightly stronger version of it.
With a minor abuse of notation, in the following we denote by $;$
also the operator which concatenates traces. Thus, if $\sf s_i$ are
traces and $\sf c_i$ are constraints, for $i\in[1,n]$, then $\sf
c_1;s_1;c_2;s_2\ldots ;s_{n-1};c_n$ denotes the trace obtained by
concatenating the $\sf s_i$ e $\sf c_i$ in the obvious way.

\begin{lemma}\label{lemma:infinite}
Let $\sf D.A$ be a CCP process and $\sf c_0$ be a constraint.
Assume that $\sf\langle D.A, c_0\rangle$ produces the (infinitely
many) finite traces
$$
\begin{array}[t]{l}
\sf c_0
\\
\sf c_0;s_{1,1};c_1
\\
\sf c_0;s_{2,1};c_1;s_{2,2};c_2
\\
\sf c_0;s_{3,1};c_1;s_{3,2};c_2;s_{3,3};c_3
\\
\vdots
\end{array}
$$
where the $\sf c_0,c_1,c_2\ldots $ are different constraints
(i.e. for any $i$,
 $\sf {\cal D} \models \neg (c_i\rightarrow c_{i+1})$)
and the $\sf s_{i,j}$ are (finite) sub-traces such that, for each
$j\geq 1$, the (infinite) set containing the lengths
 $ \{ {\sf |s_{1,j}|}, {\sf |s_{2,j}|},{ \sf |s_{3,j}|}, \ldots \}$
admits a (finite) maximal element. Then $\sf\langle D.A,
c_0\rangle$ produces also the infinite trace $\sf
c_0;s_1;c_1;s_2;c_2;s_3;c_3;\ldots$ where, for each $j\geq 1$,
${\sf s_j }= {\sf s_{i,j}}$ for some $i\geq 1$.
\end{lemma}
\begin{proof} The proof uses the Koenig Lemma and the fact that the
transition system defining the CCP operational semantics is
finitely branching.

Let us denote by $\sf m_j$ the maximal element appearing in the set
$(\{ {\sf |s_{1,j}|},{ \sf |s_{2,j}|}, {\sf |s_{3,j}|}, \ldots \}$,
for each $j\geq 1$, that is, $\sf m_j$ is the maximal length of the
sub-traces $\sf s_{i,j}$ for a fixed $j$ and $i=1,2,\ldots$. We now
construct a tree $T$ representing the (infinitely many) finite traces
$$
\begin{array}[t]{l}
\sf c_0
\\
\sf c_0;s_{1,1};c_1
\\
\sf c_0;s_{2,1};c_1;s_{2,2};c_2
\\
\sf c_0;s_{3,1};c_1;s_{3,2};c_2;s_{3,3};c_3
\\
\vdots
\end{array}
$$
produced by $\sf\langle D.A, c_0\rangle$. The nodes of the tree
$T$ are labeled by configurations of the form $\sf\langle D.B,
c_i\rangle$, for some $i$, and the edges are labeled by the
sub-traces $\sf s_{i,j}$. More precisely, the tree $T$ is defined
inductively as follows:

(Base step). The root (level 0) of $T$  is labeled by $\sf\langle
D.A, c_0\rangle$. For each derivation of the form $\sf\langle
D.A, c_0\rangle\rightarrow^* \sf\langle D.A_{i,1}, c_1\rangle$
which performs at most $\sf m_1 + 1$ transition steps and which
produces the trace $\sf c_0;s_{i,1}$ we add a son $N$ of the root
(at level 1) labeled by $\sf\langle D.A_{i,1}, c_1\rangle$ and
an edge, labeled by $\sf s_{i,1}$, connecting the root and $N$.

(Inductive step). Assume that $T$ has depth $n-1$ and let
$\sf\langle D.A_{i,n-1}, c_{n-1}\rangle$ be a configuration
labeling a node $N$ at level $n-1$. For each derivation of the
form  $\sf\langle D.A_{i,n-1}, c_{n-1}\rangle\rightarrow^*
\sf\langle D.A_{i,n}, c_{n}\rangle$ which performs at most $\sf
m_n + 1$ transition steps we add a son $N'$ of $N$ labeled by
$\sf\langle D.A_{i,n}, c_{n}\rangle$ and we add an edge labeled
by $\sf s_{i,n}$, connecting $N$ and $N'$.

Note that the number of the configurations  $\sf \langle D.A_{i,n},
c_{n}\rangle$ obtained in this way is finite, since we allow at
most $\sf  m_n + 1$ transition steps. Therefore we construct a
finitely branching tree.

On the other hand, such a tree contains infinitely many nodes, as
it contains all the (different) constraints $\sf c_i$ with $i\geq 1$.
Then, from the Koenig Lemma it follows that the tree contains an
infinite branch and this, by construction of the tree, implies
that $\sf\langle D.A, c_0\rangle$ produces the infinite trace $\sf
c_0;s_1;c_1;s_2;c_2 \ldots s_{n}; c_n ; \ldots$ where, for each
$j\geq 1$, ${\sf s_j} = {\sf s_{i,j}}$ for some $i\geq 1$.

\end{proof}

We also need the following Lemma.

\begin{lemma}
\label{lemma:correctnessstrong} Let $\sf D_0, \ldots, D_n$ be a
restricted transformation sequence, and $\sf A$ be an agent. If $\sf
\langle D_0.A,c_0\rangle$ produces the trace $\sf c_0;s_1;c_1;s_2;c_2
\ldots s_{m}; c_m$, where the $\sf c_i$ are different constraints and
the $\sf s_i$ are sub-traces of constraints all equal to $\sf
c_{i-1}$, then $\sf \langle D_n.A,c_0\rangle$ produces the trace $\sf
c_0;s'_1;c_1;s'_2;c_2 \ldots s'_{m}; c_m$ such that, for any $i\in
[1,m]$, there exists $\sf k_i$ such that ${\sf |s'_i|} \leq {\sf |s_i|
  + k_i}$.  Furthermore, the vice versa (obtained by exchanging $\sf
D_0$ with $\sf D_n$ in the previous statement) holds as well.
\end{lemma}
\begin{proof} The first part follows from
Theorem~\ref{thm:correctnessstrong}.  The part concerning the
length is a direct consequence of the definition of the
transformation sequence, since each transformation operation can
at most add or delete a finite number of computation step.

\end{proof}

 We then obtain the following extension of Theorem
\ref{thm:correctnessstrong}. Here we consider the obvious
extension of the relation $\preceq$ to the case of infinite
traces.

\begin{theorem}
\label{thm:correctness2infi} Let $\sf D_0, \ldots, D_n$ be a
restricted transformation sequence and $\sf A$ be an
agent.\begin{itemize}
  \item  If $\sf \langle
D_0.A,c_0 \rangle $ produces the infinite active trace $\sf
s$, then $\sf \langle D_n.A,c_0 \rangle$ produces an infinite
trace $\sf s'$ such that $\sf s\preceq s'$.
\item Conversely, if $\sf \langle
D_n.A,c_0\rangle $ produces the infinite active trace $\sf
s$, then $\sf \langle D_0.A,c_0\rangle $ produces an infinite
trace $\sf s'$ such that $\sf s\preceq s'$.
\end{itemize}
\end{theorem}
\begin{proof} Assume that $\sf \langle D_0.A,c_0 \rangle $ produces
the infinite active trace
$$\sf t: c_0;s_1;c_1;s_2;c_2 ;
s_3;c_3\ldots $$ where, in order to simplify the notation, we
assume that the $\sf c_i$ are different constraints while the
$\sf s_i$ are sequences of constraints all equal to $\sf c_{i-1}$
(so the $\sf s_i$ are sequences of stuttering steps). Clearly, by
definition of produced sequence, $\sf \langle D_0.A,c_0 \rangle $
produces also the (infinitely many) finite prefixes of $\sf t $
$$
\begin{array}[t]{l}
\sf c_0
\\
\sf c_0;s_{1};c_1
\\
\sf c_0;s_{1};c_1;s_{2};c_2
\\
\sf c_0;s_{1};c_1;s_{2};c_2;s_{3};c_3
\\
\vdots
\end{array}
$$
>From Lemma \ref{lemma:correctnessstrong} it follows that $\sf
\langle D_n.A,c_0 \rangle $ produces the traces
$$
\begin{array}[t]{l}
\sf c_0
\\
\sf c_0;s'_{1,1};c_1
\\
\sf c_0;s'_{2,1};c_1;s'_{2,2};c_2
\\
\sf c_0;s'_{3,1};c_1;s'_{3,2};c_2;s'_{3,3};c_3
\\
\vdots
\end{array}
$$
where, for any $j\geq 1$, there exists $\sf k_j$ such that for
any $i\in [1,j]$ we have that ${\sf |s'_{i,j}| }\leq {\sf |s_j| +
k_j}$. Therefore the set $\{{\sf |s_{1,j}|}, {\sf |s_{2,j}|},
{\sf |s_{3,j}|}, \ldots \}$ admits a (finite) maximal element
for each $j$. Lemma~\ref{lemma:infinite} then implies that $\sf
\langle D_n.A,c_0 \rangle $ produces the infinite trace $\sf t':
\sf c_0;s'_1;c_1;s'_2;c_2;s'_3;c_3 \ldots$ and clearly, by
construction, $\sf t\preceq t'$ holds. Analogously for the vice
versa. \end{proof}

\subsubsection{Preservation of Termination}

The results we have presented guarantee the correctness of the
transformation system w.r.t.\ various semantics based on produced
constraints. We should mention however that these results do not
imply that the system preserves non-declarative properties such
as \emph{termination}. In fact, in case of \emph{non-active}
traces (that from a certain point do not generate any new
constraint), the semantics we have considered equate infinite and
finite traces.

A full treatment of infinite computations is beyond the scope of this
paper and is left for future work.

Nevertheless, we claim that the transformation system we have
proposed here cannot introduce non-termination. That is, if the
initial program, for a given configuration, does not produce any
infinite computations then this is the case also for the
transformed program.

We now provide a sketch of a proof of this claim by considering a
specific class of declarations, and by showing the intuitive,
informal, argument that indicates the proof methodology to be used for
the general case.

Let us then assume that declarations does not contain mutually
recursive definitions (note that mutually recursive definitions can
usually be eliminated by means of unfolding).  We also concentrate on
the restricted system, which preserves active traces.  In the
following we say that a configuration $\sf \langle D.A,c\rangle $
\emph{terminates} if it produces only finite computations, while we
say that it does not terminate if it produces also at least one
infinite derivation.

Let $\sf D_0,\ldots,D_n$ be a transformation sequence, and assume that
$\sf \langle D_{n}.A,c\rangle $ has an infinite (non
active)\footnote{In case of active traces, our result on the
  preservation of intermediate results guarantees the preservation of
  termination.} trace.  This implies that there exists a derivation
  $\sf \xi = \langle D_{n}.A,c\rangle \rightarrow \langle
  D_{n}.A_1,c_1\rangle \rightarrow^*\ldots \rightarrow^* \langle
  D_{n}.A_j,c_j\rangle \rightarrow^*\ldots$, where for some $k$, for
  each $i\geq k$, $\sf \exists_{Var(A,c)} c_i= \exists_{Var(A,c)} c_k$
  holds.  Assume also that for each $\sf i \in [0,n-1]$, $\sf \langle
  D_i.A,c\rangle $ terminates.

 It is easy to see that the only operation that might introduce
non-termination is the folding one (all other operations are clearly
``safe'' in this respect). So the situation is the following::

\[\begin{array}{rclcl}
\sf d: && \sf H \la C[A'] &&\sf  \in  D_{n-1}\\
\sf f: && \sf B \la A' && \sf \in D_0\\
\sf d': && \sf H \la C[B] &&\sf  \in  D_{n}\\
\end{array}
\]

This operation can introduce non-termination only when it
introduces recursion, i.e., when the definition of \textsf{B}
depends on the one of \textsf{H}. The typical case is when
\textsf{B} and \textsf{H} have the same predicate and in the
following, for the sake of simplicity, we assume that this is the
case, so we assume that:

\[\begin{array}{rclcl}
\sf d: && \sf p(\ol X) \la C[A'] &&\sf  \in  D_{n-1}\\
\sf f: && \sf p(\ol Y) \la A' && \sf \in D_0\\
\sf d': && \sf p(\ol X) \la C[p(\ol Y)] &&\sf  \in  D_{n}\\
\end{array}
\]

>From the definition of folding we have that $\sf C[\ ]$ is a
\emph{guarding context} and $\sf Var(A')\cap Var(C,\ol X) \subseteq
\ol Y$ ($\sf f$ and $\sf d$ are suitably renamed so that the variables
they have in common are only those occurring in $\sf A'$). Since $\sf
C$ is a guarding context let us assume that $\sf C[\ ] = C'[\sf
\sum_{i=1}^n ask(c'_i)\rightarrow A'_i]$, where $\sf A'_1 = C''[\ ]$
and $\sf C'[\ ]$ and $\sf C''[\ ]$ are non-guarding contexts. If the
infinite computation is due to the folding operation then the
derivation $\xi$ must contain an infinite number of calls of the form
$\sf{p(\ol Y)\sigma_i}$, where, for each $i\geq 1$, $\sigma_i$ is a
renaming and the current the store $\sf d_i$ entails $\sf
c'_1\sigma_i$. Moreover, assume that $\sf A$ is of the form $\sf
C_0[p(\ol v)]$.

Now, by the definition of transformation sequence, the unfolding is
the only operation which can introduce a new ask action, thus the
guard $\sf c'_1$ in the context $\sf C[\ ]$ was certainly introduced
during an unfolding operation of an agent in $\sf A'$ with a recursive
definition (recall that \textsf{d} must be obtained from \textsf{f},
thus, by unfolding $\sf A'$ we must obtain $\sf C[A']$, and that we
are restricting to the case of direct recursion).  Therefore $\sf A'$
must contain an atom $\sf q$, whose definition in $\sf D_0$ is
\[\begin{array}{rclcl}
\sf d: && \sf q(\ol Z) \la D[q(\ol W)] &&\sf  \in  D_{0}
\end{array}
\]
where the weakest produced constraint of $\sf D$ is precisely $\sf
c'_1\rho$, for some appropriate renaming. Notice also that all
\textsf{tell} actions present in \textsf{D} can be skipped (they
are always in parallel with the rest, they don't form a guard).
Because of this, by taking $\sf c$ as initial store, one can show
that there exist an infinite derivation starting from $\sf
\langle D_0.C_0[p(\ol V)],c\rangle$ where, from a certain point
of the derivation $j$, the current store $\sf d$ satisfies $\sf
\exists_{Var(C_0[p(\ol v)],c)} d_j= \exists_{ Var(C_0[p(\ol
v)],c)} c'_1$.

This is in contrast with the hypothesis made on
the original program, thus showing that no new infinite computation
is generated.

\medskip

In the rest of the
paper we are going to provide some extra examples of transformations
and -- in the Appendix -- the technical proofs of the correctness
results.

\section{More Examples}
\label{sec:example}
The following example is inspired by the one in \cite{EGM98}. It shows
that the transformation system can be used to simplify the dynamic
behavior of a program to the point that it can be used to prove
deadlock freeness. All the operations used in it are of the restricted
sort; the transformation preserves thus the semantics of the
intermediate results as well as that of terminating derivations.

Here and in the following we say that a variable $\sf X$ is
\emph{instantiated} to a term $\sf t$ in case the current store
entails $\sf X=t$. Accordingly, we also say that an agent instantiates
a variable $\sf X$ to $\sf t$ in case that the agent adds the
constraint $\sf X=t$ to the store. Finally, we say that $\sf X$ is
instantiated if the store entails $\sf X=t$ for some non variable term
$\sf t$.

\begin{example}
\label{exa:read-write}
Consider the following simple {\sf Collect-Deliver} program, which
uses a buffer of length one:
\begin{programss}
  collect\_deliver \la collect(Xs) \Par deliver(Xs).\\
  \\
  collect(Xs) \la \emph{\% collects tokens and puts them in the queue
    } Xs\\
  \> \HS ask($\exists_{\sf X,Xs'}$ Xs=[X$|$Xs']) \ra
  tell(Xs=[X$|$Xs']) \Par
  get\_token(X) \Par collect(Xs')\\
  \> + ask(Xs=[\ ]) \ra stop.\\
  \\
  deliver([Y$|$Ys]) \la \emph{\% delivers the tokens in the queue} Xs\\
  \> \HS ask(Y=eof) \ra tell(Ys=[\ ])\\
  \> + ask(Y $\neq$ eof) \ra deliver\_token(Y) \Par deliver(Ys).
\end{programss}
The dynamic behavior of this program is not elementary.  {\sf
  collect(Xs)} behaves as follows: (a) it waits until more information
for the variable {\sf Xs} is produced, (b1) if {\sf Xs} is
instantiated to {\sf [X$|$Xs']} (i.e. when the store entails
$\exists_{\sf X, Xs'} \textsf{Xs}=\textsf{[X$|$Xs']}$) then (by using
$\sf get\_token(X)$) it instantiates {\sf X} with the value it
collects, (b2) if {\sf Xs} is instantiated to {\sf [\ ]} it stops.  On
the other hand, the actions {\sf deliver(Xs)} performs are: (a) it
instantiates {\sf Xs} to {\sf [Y$|$Ys]} (this activates {\sf
  collect(Xs)}), then (b) it waits until {\sf Y} is instantiated. Now
there are two possibilities: (c1) if {\sf Y} is the {\sl end of file}
character then it instantiates {\sf Ys} to {\sf [\ ]} (this will also
stop the collector), (c2) otherwise it delivers $\sf Y$ (by using $\sf
deliver\_token(Y)$) and proceeds with the recursive call (which will
further activate {\sf collect}).

Thus, {\sf collect-deliver} actually implements a communication
channel with a buffer of length one, and {\sf Xs} is a bidirectional
communication channel.  Note also that proving that this program is
deadlock-free is not trivial.

We now proceed with the transformation. First we unfold {\sf
  deliver(Xs)} in the body of the first definition. The result, after
cleaning up the definition via a (restricted) tell elimination is.
\begin{programss}
collect\_deliver \la collect([Y$|$Ys]) \Par \\
\> \PS ask(Y=eof) \ra tell(Ys=[\ ])\\
\> + ask(Y $\neq$ eof) \ra deliver\_token(Y) \Par deliver(Ys)).
\end{programss}
Then, we unfold {\sf collect([Y$|$Ys])} in the resulting definition; we obtain
\begin{programss}
collect\_deliver \la \\
\> \> \PS ask($\exists_{\sf X,Xs}$ [Y$|$Ys]=[X$|$Xs]) \ra
tell([Y$|$Ys]=[X$|$Xs]) \Par get\_token(X) \Par collect(Xs)\\
\> \> + ask([Y$|$Ys]=[\ ]) \ra stop)\\
\> \Par \\
\> \> \PS ask(Y=eof) \ra tell(Ys=[\ ])\\
\> \> + ask(Y $\neq$ eof) \ra deliver\_token(Y) \Par deliver(Ys))
\end{programss}
This definition can be simplified: first, by an ask simplification, we
obtain.
\begin{programss}
collect\_deliver \la \\
\> \> \PS ask(true) \ra tell([Y$|$Ys]=[X$|$Xs]) \Par get\_token(X) \Par
collect(Xs)\\
\> \> + ask(false) \ra stop)\\
\> \Par \\
\> \> \PS ask(Y=eof) \ra tell(Ys=[\ ])\\
\> \> + ask(Y $\neq$ eof) \ra deliver\_token(Y) \Par deliver(Ys))
\end{programss}
Then we can eliminate the branch \textsf{ask(false) \ra stop},
eliminate {\sf tell([Y$|$Ys]=[X$|$Xs])} and eliminate the
\textsf{ask(true)}; the result is
\begin{programss}
collect\_deliver \la get\_token(Y) \Par collect(Ys)\Par\\
\> \PS ask(Y=eof) \ra tell(Ys=[\ ])\\
\> + ask(Y $\neq$ eof) \ra deliver\_token(Y) \Par deliver(Ys))
\end{programss}
Now, we apply the restricted distributive operation in order to bring
{\sf collect(Ys)} inside the scope of the \textsf{ask} construct.
Notice that {\sf collect(Ys)} requires \textsf{Ys}.  Remark
\ref{rem:distributionrestricted} allows us to apply the operation.
\begin{programss}
collect\_deliver \la get\_token(Y) \Par \\
\> \PS ask(Y=eof) \ra   collect(Ys) \Par tell(Ys=[\ ])\\
\> + ask(Y $\neq$ eof) \ra deliver\_token(Y) \Par collect(Ys) \Par deliver(Ys))
\end{programss}

We can now fold ${\sf collect(Ys)\parallel deliver(Ys)}$, using the
original definition \textsf{collect\_deliver \la collect(Xs) \Par
  deliver(Xs)}. We obtain.
\begin{programss}
collect\_deliver \la get\_token(Y) \Par \\
\> \PS ask(Y=eof) \ra   collect(Ys) \Par tell(Ys=[\ ])\\
\> + ask(Y $\neq$ eof) \ra deliver\_token(Y) \Par collect\_deliver)
\end{programss}
To clean up the result, we can now eliminate \textsf{tell(Ys=[\ ])},
unfold the obtained \textsf{collect([\ ])} agent, and perform the usual
clean-up operations on the result. Our final program is the simple
\begin{programss}
  collect\_deliver \la get\_token(Y) \Par \\
  \> \PS ask(Y=eof) \ra   stop\\
  \> + ask(Y $\neq$ eof) \ra deliver\_token(Y) \Par collect\_deliver)
\end{programss}

It is important to compare this to the initial program. In particular,
three aspects are worth noticing.

First, that -- as opposed to the initial program -- the
resulting one has a straightforward dynamic behavior. For
instance if we consider the agent {\sf collect\_deliver}, one
can easily see it to be deadlock-free in the latter program
while in the original program this is not at all immediate.
After proving that the transformation does not introduce
\emph{nor eliminate} any deadlocking branch in the semantics of
the program, we are able to state that ``since the resulting
program is deadlock-free then also the initial program is
deadlock-free''. Thus program's transformations can be
profitably used as analysis tool: it is in fact often easier to
prove deadlock freeness for a transformed version of a program
than for the original one.

Secondly, that the resulting program is more efficient than the
initial one: in fact it does not need to use the global store as
heavily as the initial one for passing the parameters between {\sf
  collect} and {\sf deliver}.

Finally, it is straightforward to check that all transformation
operations used here are of the restricted kind, therefore, by the
Strong Total Correctness Theorem \ref{thm:correctnessstrong} this
transformation is correct wrt the sequence of intermediate results.
\end{example}

We show now an application of our methodology with a third example,
containing an \emph{extended folding} operation (see discussion after
Definition~\ref{def:folding}): this is the case when the replaced
agent coincides with an instance of the body of the folding
definition.

\begin{example}
  \label{exa:monitor} We consider a stream protocol problem where two
  input streams are merged into an output stream. An input stream
  consists of lines of messages, and each line has to be passed to the
  output stream without interruption. Input and output streams are
  dynamically constructed by a reader and a monitor process,
  respectively. A reader communicates with the monitor by means of a
  buffer of length one, and is synchronized in such a way that it can
  read a new message only when the buffer is empty (i.e., when the
  previous message has been processed by the monitor). On the other
  hand, the monitor can access a buffer only when it is not empty
  (i.e., when the corresponding reader has put a message into its
  buffer). This protocol is implemented by the following program {\sf
    STREAMER}:

\begin{programss}
streamer \la
  reader(left,Ls) \Par reader(right,Rs) \Par monitor(Ls,Rs,idle)\\[2mm]

reader(Channel,Xs)\la\\
\>   \> ask($\exists_{\sf X,Xs'}$ Xs=[X$|$Xs']) \ra tell(Xs=[X$|$Xs'])
\Par read(Channel,X) \Par  reader(Channel,Xs')\\
\> + \> ask(Xs=[\ ]) \ra  stop.\\[5mm]

monitor([L$|$Ls],[R$|$Rs],State)\la  \\
\> \HS ask(State=idle) \ra  \em \ \ \ \ \% waiting for an input\\
\> \> ( \> ask(char(L)) \ra  monitor([L$|$Ls],[R$|$Rs],left) \\
\> \> + \> ask(char(R)) \ra  monitor([L$|$Ls],[R$|$Rs],right))\\[2mm]
\>+\  ask(State=left) \ra ask(char(L)) \ra write(L) \Par   \em \ \ \ \
\% { processing the left stream}\\
\> \> ( \> ask(L=eof) \ra tell(Ls=[\ ]) \Par onestream([R$|$Rs]) \\
\>\> + \> ask(L=eol) \ra  monitor(Ls,[R$|$Rs],idle)\\
\>\> + \> ask(L $\neq$ eol AND L $\neq$ eof) \ra
monitor(Ls,[R$|$Rs],left))\\[2mm]
\>+\  ask(State=right) \ra ask(char(R)) \ra \ra \ $\ldots$ \ ) \em \ \ \ \
\% { analogously for the right stream}\\[2mm]

onestream([X$|$Xs]) \la \\
\> ask(char(X)) \ra \\
\> \> \PS ask(X=eof) \ra tell(Xs=[\ ]) \\
\> \> + ask(X $\neq$ eof) \ra write(X) \Par onestream(Xs))
\end{programss}

\NI Here, the primitive agent \textsf{read(Channel,X)} is supposed to
read an input token from channel \textsf{Channel} and instantiate
\textsf{X} with the read value; similarly, \textsf{write(X)} writes
the value of \textsf{X} to the (unique) output stream. The primitive
constraint predicate \textsf{char} is \emph{true} if its argument is
either a printable (e.g.\ ASCII) character or if it is equal to
\textsf{eol} or \textsf{eof}, which are constants denoting the
\emph{end of line} and the \emph{end of file} characters,
respectively.  Furthermore, the agent {\sf reader(Channel,Xs)} waits
to process {\sf Channel} until {\sf Xs} is instantiated; {\sf
  monitor(Ls,Rs,State)} takes care of merging {\sf Ls} and {\sf Rs}
and of writing to the output; the agent {\sf onestream(Xs)} takes care
of handling the single stream {\sf Xs} (when one of the streams has
finished).  Finally, the constants {\sf left}, {\sf right} and {\sf
  idle} describe the state of the monitor, i.e., if it is processing a
message from the left stream, from right stream, or if it is in an
idle situation, respectively.

Notice that {\sf reader(Channel,Xs)} suspends until {\sf Xs} is
instantiated and that \textsf{Xs} will eventually be instantiated by
the \textsf{monitor} process.

We can now transform the {\sf STREAMER} program in order to improve
its efficiency. First we add the following new declaration to the
original program.
\begin{programss}
  handle\_two(L,R,State)\la\ reader(left,Ls) \Par reader(right,Rs)
  \Par monitor([L$|$Ls],[R$|$Rs],State)
\end{programss}
Next, we unfold the agent {\sf monitor([L$|$Ls],[R$|$Rs],State)} in
the new declaration and then we perform the subsequent tell
eliminations (these are restricted in virtue of the argument presented
after Definition~\ref{def:tellelinrestricted}).
The result of these operations is
the following program.
\begin{programss}
handle\_two(L,R,State)\la\ reader(left,Ls) \Par  reader(right,Rs) \Par \\
\> \PS ask(State=idle) \ra \em \ \ \ \ \% { waiting for an input}\\
\> \> ( \> ask(char(L)) \ra  monitor([L$|$Ls],[R$|$Rs],left) \\
\> \> + \> ask(char(R)) \ra  monitor([L$|$Ls],[R$|$Rs],right))\\[2mm]
\> + ask(State=left) \ra ask(char(L)) \ra write(L) \Par   \em \ \ \ \ \%
{ processing the left stream}\\
\> \> ( \> ask(L=eof) \ra tell(Ls=[\ ]) \Par    onestream([R$|$Rs]) \\
\>\> + \> ask(L=eol) \ra  monitor(Ls,[R$|$Rs],idle)\\
\>\> + \> ask(L $\neq$ eol AND L $\neq$ eof) \ra
monitor(Ls,[R$|$Rs],left))\\[2mm]
\> + ask(State=right) \ra ask(char(R)) \ra \ $\ldots$ \ ) \em \
\ \ \ \% { analogously for the right stream}
\end{programss}

According to Definition \ref{def:required}, the agent $\sf
reader(left,Ls)$ requires the variable $\sf Ls$ and $\sf
reader(right,Rs)$ requires the variable $\sf Rs$. By Remark
\ref{rem:distribution} it is possible for us to apply twice the
distribution operation\footnote{Remark
  \ref{rem:distributionrestricted}, guarantees also in both cases it
  is a restricted distribution operation.} and bring them
inside the \textsf{ask} constructs. The result is the following
program.

\begin{programss}
handle\_two(L,R,State)\la\\
\> \PS ask(State=idle) \ra \em \ \ \ \ \% { waiting for an input}\\
\> \> \PS \> ask(char(L)) \ra reader(left,Ls) \Par
reader(right,Rs)\Par
\\
 \> \> \> \>  \> \>    monitor([L$|$Ls],[R$|$Rs],left)
\\
\> \> + \> ask(char(R)) \ra
 reader(left,Ls) \Par    reader(right,Rs)\Par
 \\
    \> \> \> \> \> \>  monitor([L$|$Ls],[R$|$Rs],right))
 \\[2mm]
\> + ask(State=left) \ra ask(char(L)) \ra write(L) \Par   \em \ \ \ \ \%
{ processing the left stream}\\
\> \> \PS \> ask(L=eof) \ra reader(left,Ls) \Par    reader(right,Rs)
\Par    tell(Ls=[\ ])
\\ \> \> \> \>  \> \> \Par    onestream([R$|$Rs]) \\
\>\> + \> ask(L=eol) \ra reader(left,Ls) \Par    reader(right,Rs) \Par
monitor(Ls,[R$|$Rs],idle)\\
\>\> + \> ask(L$\neq$eol AND L$\neq$eof) \ra reader(left,Ls)
\Par reader(right,Rs)\Par
\\
\> \> \> \>  \> \>   monitor(Ls,[R$|$Rs],left))\\[2mm]
\> + ask(State=right) \ra ask(char(R)) \ra  \ $\ldots$ \ ) \em \
\ \ \ \% { analogously for the right stream}
\end{programss}
In this program we can now eliminate $\sf tell(Ls=[\ ])$ in the agent
$\sf reader(left,Ls) \Par$ $\sf reader(right,Rs)$ $\sf \Par tell(Ls=[\
]) \Par onestream([R|Rs])$ thus obtaining\footnote{Again, it is true
  that the variable $\sf Ls$ here occurs also elsewhere in the
  definition, but since it occurs only on choice-branches different
  than the one on which the considered agent lies, we can assume it to
  be renamed.}:
\begin{programss}
  handle\_two(L,R,State)\la\\
  \> \PS ask(State=idle) \ra \em \ \ \ \ \% { waiting for an input}\\
  \> \> \PS \> ask(char(L)) \ra reader(left,Ls) \Par    reader(right,Rs)\Par
\\ \> \> \> \>  \> \>   monitor([L$|$Ls],[R$|$Rs],left) \\
  \> \> + \> ask(char(R)) \ra reader(left,Ls) \Par    reader(right,Rs)\Par
\\ \> \> \> \>  \> \>    monitor([L$|$Ls],[R$|$Rs],right))\\[2mm]
  \> + ask(State=left) \ra ask(char(L)) \ra write(L) \Par  \em \ \ \ \
\% { processing the left stream}\\
  \> \> \PS \> ask(L=eof) \ra reader(left,[\ ]) \Par    reader(right,Rs)
\Par    onestream([R$|$Rs]) \\
  \>\> + \> ask(L=eol) \ra reader(left,Ls) \Par    reader(right,Rs)
\Par    monitor(Ls,[R$|$Rs],idle)\\
  \>\> + \> ask(L$\neq$eol AND L$\neq$eof) \ra reader(left,Ls) \Par
reader(right,Rs) \Par
\\ \> \> \> \>  \> \> monitor(Ls,[R$|$Rs],left))\\[2mm]
  \> + ask(State=right) \ra ask(char(R)) \ra \ $\ldots$\ ) \em \ \ \
  \ \% { analogously for the right stream}
\end{programss}
In this program, the unfolding of the agent $\sf reader(left,[\ ])$
yields as result the agent
\begin{programss}
\>  \>  ask($\exists_{\sf X,Xs'}$ [\ ]=[X$|$Xs']) \ra tell([\
]=[X$|$Xs'])\Par read(Channel,X)
\\ \> \> \> \> \> \> \Par
 reader(Channel,Xs')\\
\> + \> ask([\ ]=[\ ]) \ra  stop .
\end{programss}
By (trivial) guard simplification, this can become
\begin{programs}
  \> \> ask(false) \ra tell([\ ]=[X$|$Xs'])\Par read(Channel,X) \Par
  reader(Channel,Xs')\\
  \> + \> ask(true) \ra stop.
\end{programs}
Now, by using branch elimination we can eliminate the first branch and
by applying the conservative ask elimination we can transform the
second branch into {\sf stop}.  The application of these operations
yields:

\begin{programss}
handle\_two(L,R,State)\la \\
\> \PS ask(State=idle) \ra \em \ \ \ \ \% { waiting for an input}\\
\> \> \PS \> ask(char(L)) \ra reader(left,Ls) \Par    reader(right,Rs)
\Par
 \\ \> \> \> \> \> \>            monitor([L$|$Ls],[R$|$Rs],left) \\
\> \> + \> ask(char(R)) \ra reader(left,Ls) \Par  reader(right,Rs) \Par
  \\ \> \> \> \> \> \>           monitor([L$|$Ls],[R$|$Rs],right))\\[2mm]
\> + ask(State=left) \ra ask(char(L)) \ra write(L) \Par  \em \ \ \ \ \%
{ processing the left stream}\\
\> \> \PS \> ask(L=eof) \ra  reader(right,Rs) \Par    onestream([R$|$Rs])
\\
\>\> + \> ask(L=eol) \ra reader(left,Ls) \Par    reader(right,Rs) \Par
             monitor(Ls,[R$|$Rs],idle)\\
\>\> + \> ask(L$\neq$eol AND L$\neq$eof) \ra reader(left,Ls) \Par
reader(right,Rs) \Par
  \\ \> \> \> \> \> \>           monitor(Ls,[R$|$Rs],left))\\[2mm]
\> + ask(State=right) \ra ask(char(R)) \ra \ $\ldots$ \ ) \em \
\ \ \ \% {analogously for the right stream}
\end{programss}
We now apply the backward instantiation operation to
\textsf{monitor(Ls,[R$|$Rs],idle)} and to
\textsf{monitor(Ls,[R$|$Rs],left)}.  By cleaning up the result
with a tell elimination\footnote{This is the first operation in
this example
  that is not restricted.}, this amounts to instantiating \textsf{Ls}
to \textsf{[L'$|$Ls']}. Therefore, we have obtained.
\begin{programss}
handle\_two(L,R,State)\la \\
\> \PS ask(State=idle) \ra \em \ \ \ \ \% { waiting for an input}\\
\> \> \PS \> ask(char(L)) \ra reader(left,Ls) \Par
reader(right,Rs)\Par
\\ \> \> \> \> \> \>   monitor([L$|$Ls],[R$|$Rs],left) \\
\> \> + \> ask(char(R)) \ra reader(left,Ls) \Par
reader(right,Rs)\Par
\\ \> \> \> \> \> \>    monitor([L$|$Ls],[R$|$Rs],right))\\[2mm]
\> + ask(State=left) \ra ask(char(L)) \ra write(L) \Par  \em \ \ \ \ \%
{ processing the left stream}\\
\> \> \PS \> ask(L=eof) \ra  reader(right,Rs) \Par    onestream([R$|$Rs])
\\
\>\> + \> ask(L=eol) \ra reader(left,[L'$|$Ls']) \Par
reader(right,Rs)\Par
\\ \> \> \> \> \> \>   monitor([L'$|$Ls'],[R$|$Rs],idle)\\
\>\> + \> ask(L$\neq$eol AND L$\neq$eof) \ra reader(left,[L'$|$Ls'])
\Par
reader(right,Rs)\Par \\ \> \> \> \> \> \>    monitor([L'$|$Ls'],[R$|$Rs],left))\\[2mm]
\> + ask(State=right) \ra ask(char(R)) \ra  \ $\ldots$ \ ) \em \
\ \ \ \% {analogously for the right stream}
\end{programss}

In order to prepare the program for the folding operation we need one
more clean up phase: using the unfolding and some simplification
operations, we can replace each call $\sf reader(left,[L'|Ls'])$ with
$\sf read(left,L')\Par reader(left,Ls')$.  The result of these
operations is the program:

\begin{programss}
handle\_two(L,R,State)\la \\
\> \PS ask(State=idle) \ra \em \ \ \ \ \% { waiting for an input}\\
\> \> \PS \> ask(char(L)) \ra reader(left,Ls) \Par
reader(right,Rs)\Par
\\ \> \> \> \> \> \>    monitor([L$|$Ls],[R$|$Rs],left) \\
\> \> + \> ask(char(R)) \ra reader(left,Ls) \Par
reader(right,Rs)\Par
\\ \> \> \> \> \> \>    monitor([L$|$Ls],[R$|$Rs],right))\\[2mm]
\> + ask(State=left) \ra ask(char(L)) \ra write(L)  \Par  \em \ \ \ \ \%
{ processing the left stream}\\
\> \> \PS \> ask(L=eof) \ra  reader(right,Rs) \Par    onestream([R$|$Rs])
\\
\>\> + \> ask(L=eol) \ra read(left,L') \Par    reader(left,Ls') \Par
reader(right,Rs) \Par \\ \> \> \> \> \> \>   monitor([L'$|$Ls'],[R$|$Rs],idle)\\
\>\> + \> ask(L$\neq$eol AND L$\neq$eof) \ra read(left,L') \Par
reader(left,Ls') \Par    reader(right,Rs) \Par
\\ \> \> \> \> \> \>monitor([L'$|$Ls'],[R$|$Rs],left))\\[2mm]
\> + ask(State=right) \ra ask(char(R)) \ra  \ $\ldots$ \ ) \em \
\ \ \ \% {analogously for the right stream}
\end{programss}
We can now apply twice the \emph{extended folding} operation.  The
first folding allows us to replace \textsf{reader(left,Ls) \Par
  reader(right,Rs) \Par monitor([L$|$Ls],[R$|$Rs],left)} with
\textsf{handle\_two(L,R,left)}. With the second one we replace
\textsf{reader(left,Ls) \Par reader(right,Rs) \Par
  monitor([L$|$Ls],[R$|$Rs],right)} with
\textsf{handle\_two(L,R,right)}. Recall that the extended folding
operation, as described in Subsection~\ref{subs:folding}, occurs when
the replaced agent coincides with a non-trivial
\emph{instance} of the body of the folding definition; as
already explained in the discussion after Definition \ref{def:folding}
this is only a shorthand for a sequence of tell introduction,
folding and tell elimination, as described in Subsection~\ref{subs:folding}.
The resulting program after these two operations
is:
\begin{programss}
handle\_two(L,R,State)\la \\
\> \PS ask(State=idle) \ra \em \ \ \ \ \% { waiting for an input}\\\
\> \> \PS \> ask(char(L)) \ra handle\_two(L,R,left)\\
\> \> + \> ask(char(R)) \ra  handle\_two(L,R,right)  \\[2mm]
\> + ask(State=left) \ra ask(char(L)) \ra write(L) \Par  \em \ \ \ \ \%
{ processing the left stream}\\
\> \> \PS \> ask(L=eof) \ra  reader(right,Rs) \Par    onestream([R$|$Rs])
\\
\>\> +  \> ask(L=eol) \ra read(left,L') \Par    reader(left,Ls')
             \Par    reader(right,Rs) \Par
\\ \>\> \>\> \>\>monitor([L'$|$Ls'],[R$|$Rs],idle)\\
\>\> + \> ask(L$\neq$eol AND L$\neq$eof) \ra read(left,L') \Par
reader(left,Ls') \Par    reader(right,Rs) \Par
\\ \>\> \>\> \>\>monitor([L'$|$Ls'],[R$|$Rs],left))\\[2mm]
\> + ask(State=right) \ra ask(char(R)) \ra \ $\ldots$ \ ) \em \
\ \ \ \% {analogously for the right stream}
\end{programss}

Then, we perform two more extended foldings: with the first one we
replace the agent \textsf{reader(left,Ls') \Par reader(right,Rs) \Par
  monitor([L'$|$Ls'],[R$|$Rs],idle)} with the agent
\mbox{\textsf{handle\_two(L',R,idle)}}, with the latter we replace
\mbox{\textsf{reader(left,Ls')}} \Par \mbox{\textsf{reader(right,Rs)}}
\Par \mbox{\textsf{monitor([L'$|$Ls'],[R$|$Rs],left)}} with
\mbox{\textsf{handle\_two(L',R,left)}}. The resulting program is
\begin{programss}
  handle\_two(L,R,State)\la \\
  \> ask(State=idle) \ra \em \ \ \ \ \% { waiting for an input}\\
  \> \> \PS \> ask(char(L)) \ra handle\_two(L,R,left)\\
  \> \> + \> ask(char(R)) \ra  handle\_two(L,R,right)  \\[2mm]
  \> +\ ask(State=left) \ra ask(char(L)) \ra write(L) \Par  \em \ \ \ \
\% { processing the left stream}\\
  \> \> \PS \> ask(L=eof) \ra  reader(right,Rs) \Par
onestream([R$|$Rs]) \\
  \>\> +  \> ask(L=eol) \ra read(left,L') \Par    handle\_two(L',R,idle)\\
  \>\> + \> ask(L$\neq$eol AND L$\neq$eof) \ra read(left,L') \Par
handle\_two(L',R,left)\\[2mm]
  \> +\ ask(State=right) \ra ask(char(R)) \ra \ $\ldots$ \ ) \em \ \
  \ \ \% {analogously for the right stream}
\end{programss}
Notice that now the definition of {\sf handle\_two} is recursive.
Moreover, the above program is almost completely independent from the
definition of {\sf reader}.  In order to eliminate the atom {\sf
  reader(right,Rs)} as well, we use an unfold/fold transformation
similar to (but simpler than) the previous one. This transformation
starts with the new definition\footnote{This definition is presented
  here for the sake of clarity; however recall that we assume that it
  is added to the original program at the beginning of the
  transformation.}:

\begin{programss}
handle\_one(X, Channel) \la\ reader(Channel,Xs) \Par
onestream([X$|$Xs])
\end{programss}
After the transformation, we end up with the definition:
\begin{programss}
handle\_one(X, Channel)\la\ ask(char(X)) \ra \\
 \> \PS ask(X=eof) \ra stop\\
 \> + ask(X$\neq$eof) \ra
  write(X) \Par read(Channel,X') \Par handle\_one(X',Channel))
\end{programss}

Also in this case the folding operation allows us to save
computational space. In fact, the parallel composition of
\textsf{reader} and of \textsf{onestream} in the original definition
leads to the construction of a list containing all the data read so
far. In a concurrent setting this list could easily be of unbounded
size and monotonically increasing. The initial definition employs a
computational space which is linear in the input. After the
transformation we have a definition which does not build the list any
longer, and which could be optimized to employ only
constant space (this could be achieved by a using a garbage
collection mechanism which allows one to re-use the
space allocated for local variables).

We now continue with the last steps of our example. By
folding {\sf handle\_one} into the last definition of {\sf
  handle\_two}, we obtain

\begin{programss}
handle\_two(L,R,State)\la \em \\
  \> \PS ask(State=idle) \ra \em \ \ \ \ \% { waiting for an input}\\
  \> \> \PS \> ask(char(L)) \ra handle\_two(L,R,left)\\
  \> \> + \> ask(char(R)) \ra  handle\_two(L,R,right)  \\[2mm]
  \> +\ ask(State=left) \ra ask(char(L)) \ra write(L) \Par   \em \ \ \ \
\% { processing the left stream}\\
  \> \> \PS \> ask(L=eof) \ra  handle\_one(R,right)\\
  \>\> +  \> ask(L=eol) \ra read(left,L') \Par
                   handle\_two(L',R,idle)\\
  \>\> + \> ask(L$\neq$eol AND L$\neq$eof) \ra read(left,L') \Par
             handle\_two(L',R,left)\\[2mm]
  \> +\ ask(State=right) \ra ask(char(R)) \ra  \ $\ldots$ \ ) \em \ \ \
\ \% { analogously for the right stream}
\end{programss}
We now want to let {\sf streamer} benefit from the improvements we
have obtained via this transformation. First, we transform its
definition by applying the backward instantiation to {\sf
  monitor(Ls,Rs,idle)}, and obtain:
\begin{programss}
streamer\la\
  reader(left,[L$|$Ls]) \Par
 reader(right,[R$|$Rs]) \Par
 monitor([L$|$Ls],[R$|$Rs],idle).
\end{programss}
Next, we unfold the two {\sf reader} atoms, and eliminate the
redundant {\sf ask} and \textsf{tell} guards.
\begin{programss}
 streamer\la\ \= read(left,L) \Par reader(left,Ls) \Par \\
 \>  read(right,R)
  \Par reader(right,Rs)\Par monitor([L$|$Ls],[R$|$Rs],idle).
\end{programss}

\NI We can now fold {\sf handle\_two} in it (via an extended folding
operation), obtaining:
\begin{programss}
 streamer\la\ read(left,L) \Par read(right,R) \Par
  handle\_two(L,R,idle).
\end{programss}
Note that this last folding operation is applied to a
non-guarding context. As discussed in Remark
\ref{rem:propagation}, we can apply the folding also in this
case because the definition of streamer is never modified nor
used by the transformation. So we can simply assume that the
original definition of streamer contained a dummy ask guard as in
\begin{programss}
\sf streamer\la\ ask(true) \ra
(\ \=  \sf  read(left,L) \Par reader(left,Ls)\Par \sf read(right,R) \Par\\
\> 
\sf reader(right,Rs) \Par monitor([L$|$Ls],[R$|$Rs],idle))
\end{programss}
Then we assume that the folding operation is applied to this
definition, and that the guard $\sf ask(true)$ will eventually
be removed by an ask elimination operation.

In the final program, we only need the definitions of
\textsf{streamer} and of \textsf{handle\_two} together with the ones
of the built-it predicates.  Observe that the definition of {\sf
  streamer} is much more efficient than the original one.  Firstly, it
now benefits from a straightforward left-to-right dataflow. In the
initial program the variables \textsf{Ls} and \textsf{Rs} are employed
as bidirectional communication channels, in fact there exist two
agents (\textsf{reader} and \textsf{monitor}) which alternate in
``instantiating'' them further. This is not the case in the final
program, where for each variable it is clear which is the agent that
is supposed to ``instantiate'' it (i.e.\ to progressively add
information to the store about it).  This fact implies that on the
final program are possible a number of powerful compile-time
(low-level) optimizations which in the first program are not possible.

Secondly, the number of suspension points is dramatically reduced: in
the original program \textsf{reader} had to suspend and awaken itself
at each input token. In the final one \textsf{streamer} is independent
from \textsf{reader} and has to suspend less often.

Last but certainly not least, as previously mentioned
\textsf{streamer} now does not construct the list and could be
optimized to employ a constant computational space, while in its
initial version it employed a space linear in the input, that is,
possibly unbounded. It is worth remarking that in a concurrent
setting processes are often not meant to end their computation,
in which case it is of vital importance that the computational
space remains bounded in size; thus in this context a space gain
like the one obtained in the above example makes the difference
between a viable and a non-viable definition.
\end{example}

\begin{example} This is a variation on a standard example for
unfold/fold transformations: a program computing the sum and the
length of the elements in a list. The variation consists in the fact
that we consider only the elements of the list which are larger than
the given parameter \textsf{Limit}. We assume here that the
constraint system being used incorporates some arithmetic domain.
Therefore, in the following
program we use also arithmetic constraints, with the obvious intended
meaning.

\begin{programss}
  sumlen(Xs,Limit,S,L) \la\ sum(Xs,Limit,S) \Par len(Xs,Limit,L)
  \\[2mm]
  sum(Xs,Limit,S) \la\ \\
  \> ( \> ask(Xs=[\ ]) \ra  tell(S=0)\\
  \> + \> ask($\exists_{\sf Y,Ys}$ (Xs=[Y$|$Ys] \A Y $\leq$ Limit)) \ra
tell(Xs=[Y$|$Ys]) \Par\\
\> \> \>   sum(Ys,Limit,S)\\
  \> + \> ask($\exists_{\sf Y,Ys}$ (Xs=[Y$|$Ys] \A Y $>$ Limit)) \ra tell (Xs=
  [Y$|$Ys]) \Par \\
\> \> \> sum(Ys,Limit,S') \Par \\
\> \> \> tell(S=S'+ Y))
  \\[2mm]
  len(Xs,Limit,L) \la\ \\
  \> ( \> ask(Xs=[\ ]) \ra  tell(L=0) \\
  \> + \> ask($\exists_{\sf Y,Ys}$( Xs=[Y$|$Ys] \A Y $\leq$  Limit)) \ra
tell (Xs=[Y$|$Ys]) \Par \\
  \> \> \>              len(Ys,Limit,L)\\
  \> + \> ask($\exists_{\sf Y,Ys}$ (Xs=[Y$|$Ys] \A Y $>$ Limit)) \ra  tell
(Xs=[Y$|$Ys]) \Par \\
  \> \> \>              len(Ys,Limit,L') \Par \\
  \> \> \>              tell(L=L'+ 1))
\end{programss}
With two unfoldings we obtain:
\begin{programss}
  sumlen(Xs,Limit,S,L) \la\  \\
  \> ( \> ask(Xs=[\ ]) \ra  tell(S=0) \\
  \> + \> ask($\exists_{\sf Y,Ys}$ (Xs=[Y$|$Ys] \A Y $\leq$  Limit)) \ra
tell (Xs=[Y$|$Ys]) \Par \\
  \> \> \>              sum(Ys,Limit,S)\\
  \> + \> ask($\exists_{\sf Y,Ys}$ (Xs=[Y$|$Ys] \A Y $>$ Limit)) \ra  tell
(Xs=[Y$|$Ys]) \Par \\
  \> \> \>              sum(Ys,Limit,S') \Par \\
  \> \> \>              tell(S=S'+ Y))\\
  \>    \Par \\
  \> ( \> ask(Xs=[\ ]) \ra  tell(L=0)\\
  \> + \> ask($\exists_{\sf Y',Ys'}$ (Xs=[Y'$|$Ys'] \A Y' $\leq$  Limit))
\ra tell (Xs=[Y'$|$Ys']) \Par \\
  \> \> \>              len(Ys',Limit,L)\\
  \> + \> ask($\exists_{\sf Y',Ys'}$ (Xs=[Y'$|$Ys'] \A Y' $>$ Limit))\ra
tell (Xs=[Y'$|$Ys']) \Par \\
  \> \> \>              len(Ys',Limit,L') \Par \\
  \> \> \>              tell(L=L'+ 1))
\end{programss}
We now apply the (restricted) distribution operation; in practice, we now
bring one choice inside the other one.
\begin{programss}
  sumlen(Xs,Limit,S,L) \la\  \\
  \> ( \> ask(Xs=[\ ]) \ra  tell(S=0) \Par \\
  \>   \>  \> ( \> ask(Xs=[\ ]) \ra  tell(L=0) \\
  \>   \>  \> + \> ask($\exists_{\sf Y',Ys'}$ (Xs=[Y'$|$Ys'] \A Y' $\leq$
Limit)) \ra tell (Xs=[Y'$|$Ys']) \Par \\
  \> \> \> \> \> \>             len(Ys',Limit,L)\\
  \> \> \>    + \> ask($\exists_{\sf Y',Ys'}$ ( Xs=[Y'$|$Ys'] and Y' $>$ Limit))
\ra
tell (Xs=[Y'$|$Ys']) \Par \\
  \> \> \> \> \> \>             len(Ys',Limit,L') \Par \\
  \> \> \> \> \> \>             tell(L=L'+ 1))\\
  \> + \> ask($\exists_{\sf Y,Ys}$ (Xs=[Y$|$Ys] \A Y $\leq$  Limit)) \ra
tell (Xs=[Y$|$Ys]) \Par \\
  \> \> \>              sum(Ys,Limit,S) \Par \\
  \> \> \>              ( \> ask(Xs=[\ ]) \ra  tell(L=0)\\
  \> \> \>              + \> ask($\exists_{\sf Y',Ys'}$ (Xs=[Y'$|$Ys'] and Y'
$\leq$  Limit)) \ra tell (Xs=[Y'$|$Ys']) \Par \\
  \> \> \> \> \> \>             len(Ys',Limit,L)\\
  \> \> \>              + \> ask($\exists_{\sf Y',Ys'}$ (Xs=[Y'$|$Ys'] and
Y' $>$ Limit)) \ra  tell (Xs=[Y'$|$Ys']) \Par \\
  \> \> \> \> \> \>                     len(Ys',Limit,L') \Par \\
  \> \> \> \> \> \>             tell(L=L'+ 1))\\
  \> + \> ask($\exists_{\sf Y,Ys}$ (Xs=[Y$|$Ys] \A Y $>$ Limit) )\ra  tell
(Xs=[Y$|$Ys]) \Par \\
  \> \> \>              sum(Ys,Limit,S') \Par \\
  \> \> \>              tell(S=S'+ Y) \Par \\
  \> \> \>              ( \> ask(Xs=[\ ]) \ra tell(L=0)\\
  \> \> \>                + \> ask($\exists_{\sf Y',Ys'}$ (Xs=[Y'$|$Ys'] \A
Y' $\leq$  Limit)) \ra  tell (Xs=[Y'$|$Ys']) \Par \\
  \> \> \> \> \> \>             len(Ys',Limit,L)\\
  \> \> \>              + \> ask($\exists_{\sf Y',Ys'}$( Xs=[Y'$|$Ys'] \A Y'
$>$ Limit)) \ra tell (Xs=[Y'$|$Ys']) \Par \\
  \> \> \> \> \> \>             len(Ys',Limit,L') \Par \\
  \> \> \> \> \> \>             tell(L=L'+ 1)))
\end{programss}
It is worth noticing that the applicability conditions of
Definition \ref{def:distribution} are trivially satisfied thanks
to the fact that both choices depend on the same variable
\textsf{Xs}. Notice also that in this case we cannot apply Remark
\ref{rem:distribution}, in fact this is an example of a
distribution operation which is not possible with the tools of
\cite{EGM98}.

By using the ask simplification followed by a branch elimination and by a
conservative ask elimination we obtain the following program.
Notice that the ask simplification is possible here because
we can take arithmetic constraints into account.

\begin{programss}
sumlen(Xs,Limit,S,L) \la\  \\
        \> ( \> ask(Xs=[\ ]) \ra  tell(S=0) \Par  tell(L=0)\\
        \> + \> ask($\exists_{\sf Y,Ys}$ (Xs=[Y$|$Ys] \A Y $\leq$  Limit))
\ra  tell (Xs=[Y$|$Ys]) \Par \\
\> \> \>                sum(Ys,Limit,S) \Par \\
\> \> \>                tell (Xs=[Y'$|$Ys']) \Par \\
\> \> \>                len(Ys',Limit,L)\\
        \> + \> ask($\exists_{\sf Y,Ys}$ (Xs=[Y$|$Ys] \A Y $>$ Limit)) \ra
tell (Xs=[Y$|$Ys]) \Par \\
\> \> \>                sum(Ys,Limit,S') \Par \\
\> \> \>                tell(S=S'+ Y) \Par \\
\> \> \>                tell (Xs=[Y'$|$Ys']) \Par \\
\> \> \>                len(Ys',Limit,L') \Par \\
\> \> \>                tell(L=L'+ 1))
\end{programss}
Via a tell simplification (first and last non-restricted operation of this
example), we can transform $\sf tell (Xs=[Y'|Ys'])$ into $\sf tell
([Y|Ys]=[Y'|Ys'])$, and subsequently apply a tell elimination we
obtain:
\begin{programss}
sumlen(Xs,Limit,S,L) \la\  \\
        \> ( \> ask(Xs=[\ ]) \ra  tell(S=0) \Par  tell(L=0)\\
        \> + \> ask($\exists_{\sf Y,Ys}$ (Xs=[Y$|$Ys] \A Y $\leq$  Limit))
\ra tell (Xs=[Y$|$Ys]) \Par \\
\> \> \>                sum(Ys,Limit,S) \Par \\
\> \> \>                len(Ys,Limit,L)\\
        \> + \> ask($\exists_{\sf Y,Ys}$ (Xs=[Y$|$Ys] \A Y $>$ Limit)) \ra
tell (Xs=[Y$|$Ys]) \Par \\
\> \> \>                sum(Ys,Limit,S') \Par \\
\> \> \>                tell(S=S'+ Y) \Par \\
\> \> \>                len(Ys,Limit,L') \Par \\
\> \> \>                tell(L=L'+ 1))
\end{programss}
We can now apply the folding operation.
\begin{programss}
sumlen(Xs,Limit,S,L) \la\  \\
        \> ( \> ask(Xs=[\ ]) \ra  tell(S=0) \Par  tell(L=0)\\
        \> + \> ask($\exists_{\sf Y,Ys}$ (Xs=[Y$|$Ys] \A Y $\leq$  Limit))
\ra tell (Xs=[Y$|$Ys]) \Par \\
\> \> \>                sumlen(Ys,Limit,S,L)\\
        \> + \> ask($\exists_{\sf Y,Ys}$ (Xs=[Y$|$Ys] \A Y $>$ Limit)) \ra
tell (Xs=[Y$|$Ys]) \Par \\
\> \> \>                sumlen(Ys,Limit,S',L') \Par \\
\> \> \>                tell(S=S'+ Y) \Par \\
\> \> \>                tell(L=L'+ 1))
\end{programss}
Again, we have reached a point in which the main definition is
directly recursive. Moreover, the number of choice-points
encountered while traversing a list is now half of what it was
initially.
\end{example}

\section{Related Work}
\label{sec:related}

As mentioned in the introduction, this is one of the few attempts to
apply fold/unfold techniques in the field of concurrent languages. In
fact, in the literature we find only three papers which are relatively
closely related to the present one: Ueda and Furukawa \citeyear{UF88}
defined transformation systems for the concurrent logic language GHC
\cite{Ueda86}, Sahlin \citeyear{Sah95} defined a partial evaluator for
AKL, while de Francesco and Santone in \citeyear{DFS96} presented a
transformation system for CCS \cite{Miln89}.
\medskip

The transformation system we are proposing builds on the systems
defined in the papers above and can be considered an extension of
them. Differently from the previous cases, our system is defined for a
generic (concurrent) constraint language. Thus, together with some new
transformations such as the distribution, the backward instantiation
and the branch elimination, we introduce also specific operations
which allow constraint simplification and elimination (though, some
constraint simplification is done in \cite{Sah95} as well).

It is interesting and not straightforward to compare our system
with the one of Ueda and Furukawa \citeyear{UF88}. This is specific
for the GHC language, which has a different syntactic structure
from CCP and uses the Herbrand universe as computational domain.
Also because of this, \cite{UF88} employs operations which are
completely different from ours.  In particular, our operation of
unfolding is replaced by immediate execution and case splitting
in \cite{UF88}. Our unfolding is a weaker operation which has a
broader applicability than case splitting, since the latter
operation involves the moving of synchronization points and
therefore requires suitable applicability conditions.
Furthermore, the distribution operation is not present in
\cite{UF88}, as it would not be possible in the syntactic
structure of GHC. However, in many cases the effect of
distribution can be achieved in \cite{UF88} by introduction of a
new clause followed by case splitting. In order to clarify this,
below we report how the transformations of the Example
\ref{exa:read-write} could be mimicked in GHC by using the
  operations of \cite{UF88}. The transformation in the following example was provided
  by a reviewer of this paper.

\begin{example} The initial program \textsf{collect\_deliver} considered in
Example \ref{exa:read-write}, in terms of the GHC syntax is
\begin{verbatim}
1: collect_deliver :- | collect(Xs), deliver(Xs).

2: collect([X|Xs]) :- | get_token(X), collect(Xs).
3: collect([])     :- | true.

4: deliver(Ys0) :- | Ys0=[Y|Ys], deliver_2(Y,Ys).
5: deliver_2(eof,Ys) :- | Ys=[].
6: deliver_2(Y,Ys) :- Y\=eof | deliver_token(Y), deliver(Ys).
\end{verbatim}
The presence of \verb&deliver_2& is due to the fact that GHC
does not allow nested guards. The first operation to be used is
an \emph{immediate execution}, applied to clause (1). The result
is
\begin{verbatim}
7: collect_deliver :- | collect(Xs), Xs=[Y|Ys], deliver_2(Y,Ys).
\end{verbatim}
By normalizing this, we obtain
\begin{verbatim}
8: collect_deliver :- | collect([Y|Ys]), deliver_2(Y,Ys).
\end{verbatim}
Another \emph{immediate execution} operation yields
\begin{verbatim}
9: collect_deliver :- | get_token(Y), collect(Ys), deliver_2(Y,Ys).
\end{verbatim}
Now, we need to \emph{introduce a new definition}.
\begin{verbatim}
10: collect_deliver_2(Y) :- | collect(Ys), deliver_2(Y,Ys).
\end{verbatim}
By applying to this the \emph{case splitting} operation, we obtain
\begin{verbatim}
11: collect_deliver_2(eof) :- | collect(Ys), Ys=[].
\end{verbatim}
\begin{verbatim}
12: collect_deliver_2(Y) :- Y\=eof | collect(Ys), deliver_token(Y),
                            deliver(Ys).
\end{verbatim}
By normalizing clause 11, and subsequently applying an \emph{immediate
execution} operation, we obtain
\begin{verbatim}
13: collect_deliver_2(eof) :- | true.
\end{verbatim}
To (12) and (9) we can apply the folding operation,
and the resulting program is thus
\begin{verbatim}
collect_deliver :- | get_token(Y), collect_deliver_2(Y).
collect_deliver_2(eof) :- | true.
collect_deliver_2(Y) :- Y\=eof | deliver_token(Y), collect_deliver.
\end{verbatim}
It is worth noting  how it is possible to achieve a resulting
program which is basically identical to the one of Example
\ref{exa:read-write}, despite the completely different nature of
the operation used.
\end{example}

Compared to \cite{UF88} we also provide a more flexible definition
for the folding operation which allows the folding clause to be
recursive (which is really a step forward in the context of
folding operations which are themselves capable of introducing
recursion) and frees the \emph{initial program} from having to be
partitioned in $P_{new}$ and $P_{old}$. In fact, as opposed to
virtually all fold operations which enable to introduce recursion
presented so far (the only exception being \cite{DFS96}), the
applicability of the folding operation does not depend on the
transformation history, (which has always been one of the
``obscure sides'' of it) but it relies on plain syntactic
criteria. The idea of using a guarded folding in order to obtain
applicability conditions independent of the transformation
history was first introduced by de Francesco and Santone
\citeyear{DFS96} in the CCS setting. However, their technical
development is rather different from ours, in particular our
correctness results and proofs are completely different from
those sketched in \cite{DFS96}.

As previously mentioned, differently from our case in
\cite{Sah95} it is considered a definition of \emph{ask
elimination} which allows us to remove potentially selectable
branches; the consequence is that the resulting transformation
system is only \emph{partially} (thus not totally) correct. We
should mention that in \cite{Sah95} two preliminary assumptions
on the ``scheduling'' are made in such a way that this limitation
is actually less constraining than it might appear.

\section{Conclusions}
\label{sec:conclusions}

We have introduced an unfold/fold transformation system for CCP and we
have proved its total correctness w.r.t.\ the input/output semantics
defined by the observables $\cal O$, which takes into account also the
termination modes. This semantics corresponds (modulo irrelevant
differences due to the treatment of failure and of local variables) to
that one proposed in \cite{BP91}.  This is one of the two fully
abstract ``standard'' semantics for CCP, the other being that one
defined in \cite{SRP91}.  (Actually, these two semantic models have
been proved to be isomorphic (\cite{BP92}), provided that the
termination mode and the consistency checks are eliminated.)

We have also shown that the proposed transformation system
preserves another, stronger semantics which takes into account
the intermediate results of computations up to logical
implication (Theorem \ref{thm:correctness2}). We argued that this
result should be strong enough for transforming also programs
which might not terminate, in particular for transforming
reactive programs.  Nevertheless, in addition to this we have
presented a restricted transformation system, obtained from the
initial one by adding some (relatively mild) restrictions on some
operations. We have shown that this second system preserves the
trace semantics of programs (up to simulation, Theorem
\ref{thm:correctnessstrong}) and therefore it is totally correct
w.r.t.\ the semantics $\sf {\cal O}_i$ which takes into account
all the intermediate results (Corollary \ref{cor:intermediate}).
We have also proved that this system preserves active infinite
computations and we claim that, more generally, this system does
not introduce in the transformed program any new infinite
computation which was not present in the original one.

As shown by the examples, this system can be used for the
optimization of concurrent constraint programs both in terms of
time and of space. In fact, it allows us to eliminate unnecessary
suspension points (and therefore to reduce sequentiality), to
reduce the number of communication channels and to avoid the
construction of some global data structures.  The system can also
be used to simplify the dynamic behavior of a program, thus
allowing us to prove directly absence of deadlock.

\medskip

Concerning future work, there exist other techniques for proving
deadlock freeness for CCP programs, notably in \cite{CFM94} a
methodology based on abstract interpretation has been defined.
It could be interesting to investigate an integration of our
methodology with abstract interpretation tools. We are also
considering a formal comparison of some different transformation
systems (in particular our system and that one of \cite{UF88}) to
assess their relative strength. This task is not immediate, since
the target languages are different.

\appendix


\section{Detailed Proofs}

\appendixhead{Appendix A is }{toplas}{2039}

\begin{acks}

We would like to thank the reviewers for their precise and helpful
comments.

\end{acks}

\bibliographystyle{acmtrans}
\bibliography{sandro}


\newpage

\begin{received} Received 1/4/99; Revised 12/14/00; Accepted 4/5/01.
\end{received}

\elecappendix{\it 2001}{http://www.acm.org/toplas, ACM Transactions on Programming Languages and Systems, Vol. , No. , 2001, Pages \pages}

\setcounter{section}{1}

\medskip

In this Appendix we provide the detailed proofs for the results
which ensure that the transformation system we have defined is
totally correct. In particular, we provide the detailed proofs
for Theorems~\ref{thm:correctness} and
\ref{thm:correctnessstrong}. In order to obtain a self contained
Appendix some technical Lemmata contained also in the paper are
repeated here. In what follows, we are going to refer to a fixed
\emph{transformation sequence} $\sf D_0, \ldots, D_n$.

\begin{lemma}
        Assume that there exists a derivation
        $\sf \langle D.C[A],c\rangle \rightarrow^*
        \langle D.C'[A],c'\rangle $ where $\sf c$ is a satisfiable constraint
        and the context $\sf C'[\ ]$ has the form
        $$\sf A_{1}\Par \ldots\Par \bar C[\ ]\Par \ldots \Par A_{n}$$ and,
        for each $j\in [1,n]$,
        $\sf A_{j}$ is either a choice agent, or a procedure call
        or the agent $\sf Stop$. Then
        ${\cal D} \models
        \sf \ (pc(\bar C[\ ]) \And c') \rightarrow pc(C[\ ]) $
        holds and in case $\sf \bar C[\ ]$ is the empty context also
        ${\cal D} \models \sf \ c' \rightarrow pc(C[\ ]) $ holds.
        \label{lem:proppc}
\end{lemma}
\begin{proof}
By a straightforward inductive argument it follows that if there exists a
derivation
$\sf \langle D.C[A],c\rangle \rightarrow^*
\langle D.C'[A],c'\rangle $, then
${\cal D} \models \sf \ (pc(C'[\ ]) \And c')
\rightarrow pc(C[\ ])$. Now, if $\sf C'[\ ]$ has the form
$\sf A_{1}\Par \ldots\Par \bar C[\ ]\Par \ldots \Par A_{n}$,
where each $\sf A_{j}$ is either a choice agent or a procedure call or
$\sf Stop$, then
$\sf pc(C'[\ ]) =
pc(\bar C[\ ])$ which implies ${\cal D} \models
        \sf \ (pc(\bar C[\ ]) \And c') \rightarrow pc(C[\ ])$.
Obviously if $\sf \bar C[\ ]$ is the empty context then
$\sf pc(\bar C [\ ]) = true$, from which the second part of the Lemma follows.
\end{proof}

We prove now Proposition~\ref{pro:partial}.

\begin{proposition}[\ref{pro:partial} (Partial Correctness)] If, for
each agent $\sf A$, $\sf {\cal O}(D_0.A) = {\cal O}(D_i.A)$
then, for each agent $\sf A$, $\sf {\cal O}(D_i.A) \supseteq
{\cal O}(D_{i+1}.A)$.
\end{proposition}

\NI \begin{proof} We now show that given an agent $\sf A$ and a
satisfiable constraint $\sf c_{I}$, if there exists a derivation
$\sf \xi = \langle D_{i+1}.A,c_{I}\rangle \rightarrow^* \langle
D_{i+1}.B,c_{F}\rangle$, with $\sf m(B,c_{F}) \in \{ss, dd,
ff\}$, then there exists also a derivation $\sf \xi' = \langle
D_{i}.A,c_{I}\rangle \rightarrow^* \langle
D_{i}.B',c_{F}'\rangle$ with $\sf \exists_{-\Var(A,c_{I})}c_{F}'
= \exists_{-\Var(A,c_{I})}c_{F}$ and ${\sf m(B',c'_{F})}={\sf
m(B,c_{F})}$.  By Definition~\ref{def:semantics}, this will
imply the thesis.  The proof is by induction on the length $l$
of the derivation.  \II

$(l=0)$. In this case $\sf \xi = \langle D_{i+1}.A,c_{I}\rangle$. By the
definition $\sf \langle D_{i}.A,c_{I}\rangle$ is also a
derivation of length $0$ and then the thesis holds.  \II

$(l>0)$. If the first step of derivation $\xi$ does not use rule
${\bf R4}$, then the proof follows from the inductive hypothesis:
In fact, if
$\sf \xi = \langle D_{i+1}.A,c_{I}\rangle \rightarrow
\langle D_{i+1}.A_{1},c_{1}\rangle \rightarrow^*
\langle D_{i+1}.B,c_{F}\rangle$
then by the inductive hypothesis, there exists a derivation
$$\sf \xi''= \langle D_{i}.A_{1},c_{1}\rangle \rightarrow^*
\langle D_{i}.B',c_{F}'\rangle$$ with $\sf
\exists_{-\Var(A_{1},c_{1})}c_{F}' =
\exists_{-\Var(A_{1},c_{1})}c_{F}$ and ${\sf m(B',c'_{F})}={\sf
m(B,c_{F})}$. We can assume, without loss of generality, that
$\sf \Var(A,c_{I}) \cap \Var(\xi'') \subseteq \Var(A_{1},c_{1})$.
Therefore, there exists a derivation $\sf \xi'=\langle
D_{i}.A,c_{I}\rangle \rightarrow^* \langle
D_{i}.B',c_{F}'\rangle$. Now, to prove the thesis it is
sufficient to observe that, by the hypothesis on the variables,
$\sf \exists_{-\Var(A,c_{I})}c_{F}' =
\exists_{-\Var(A,c_{I})}(c_{1} \wedge
\exists_{-\Var(A_{1},c_{1})}c_{F}') =
\exists_{-\Var(A,c_{I})}(c_{1} \wedge
\exists_{-\Var(A_{1},c_{1})}c_{F})
=\exists_{-\Var(A,c_{I})}c_{F}$.

Now, assume that the first step of derivation $\sf \xi$ uses rule
${\bf R4}$ and let $\sf d' \in D_{i+1}$ be the declaration used
in the first step of $\sf \xi$.
If $\sf d'$ was not modified in the transformation
step from $\sf D_i$ to $\sf D_{i+1}$ (that is, $\sf d' \in D_{i}$),
then the result follows from the inductive hypothesis. We assume then
that $\sf d' \not\in D_{i}$, $\sf d'$ is then the result of the
transformation operation applied to obtain $\sf D_{i+1}$, and we now
distinguish various cases according to the operation itself.
\II

\NI {\bf Case 1:} $\sf d'$ is the result of an unfolding operation.\\
In this case the proof is straightforward.  \II

\NI {\bf Case 2:} $\sf d'$ is the result of a tell elimination
or of a tell introduction.\\
In this case the thesis follows from a straightforward analysis of the
possible derivations which use \textsf{d} or \textsf{d'}.
First, observe that for any derivation which uses a declaration
$\sf H \la C[tell(\ol{s}=\ol{t}) \Par B]$,
we can construct another derivation such that the agent
$\sf tell(\ol{s}=\ol{t})$ is evaluated before $\sf B$.
Moreover for any constraint $\sf c$ such that
$\sf \exists_{dom(\sigma)} c =
\exists_{dom(\sigma)} c \sigma$,
(where $\sigma$ is a relevant most general unifier of
$\sf \ol{s}$ and $\sf \ol{t}$), there exists a derivation step
$\sf \langle D_{i}.B_{1}\sigma,c \sigma \rangle
\rightarrow \langle D_{i}.B_{2}\sigma,c'\rangle$
if and only if there exists a derivation step
$\sf \langle D_{i}.B_{1},c \wedge (\ol{s}=\ol{t}) \rangle
\rightarrow \langle D_{i}.B_{2},c''\rangle$,
where, for some constraint \textsf{e},
$\sf c'=e \sigma $, $\sf c''= e \wedge (\ol{s}=\ol{t})$
and therefore
$\sf c'  =\exists_{dom(\sigma)} c''$.
Finally, since by definition $\sigma$ is idempotent
and the variables in the domain of $\sigma$ do not occur
neither in $\sf C[\ ]$ nor in $\sf H$, for any constraint
$\sf e$  we have that
$\sf \exists_{-\Var(A,c_{I})}e \sigma=
\exists_{-\Var(A,c_{I})} (e \wedge (\ol{s}=\ol{t}))$.
\II

\NI {\bf Case 3:} $\sf d'$ is the result of a backward instantiation.\\
Let $\sf d$ be the corresponding declaration in $\sf D_i$.  The
situation is the following:

- $\sf d: \ q(\ol r) \la C[p(\ol t)]$

- $\sf d': \ q(\ol r) \la C[p(\ol t) \Par tell(b)\Par
tell(\ol t = \ol s)]$\\
where  $\sf f: \ p(\ol s) \la tell(b)\Par  H \in  D_i$  has
no variable in common with $\sf d$
(the case  $\sf d': \
q(\ol r) \la C[p(\ol t)\Par tell(\ol t = \ol s)]$
is analogous and hence omitted). In this case
\[
\begin{array}[t]{ll}
\sf \xi = &\sf \langle D_{i+1}.C_{I}[q (\ol{v})],c_{I}\rangle
\rightarrow \langle D_{i+1}.C_{I}[C[p(\ol t) \Par tell(b)\Par
tell(\ol t = \ol s)] \Par tell(\ol{v}=\ol{r})],c_{I}\rangle
\\
& \sf \rightarrow^* \langle D_{i+1}.B,c_{F}\rangle.
\end{array}
\]
By the inductive hypothesis, there exists a derivation
\[
\sf \chi =
\langle D_{i}.C_{I}[C[p(\ol t) \Par tell(b)\Par tell(\ol t = \ol s)]
\Par tell(\ol{v}=\ol{r})], c_{I}\rangle \rightarrow^*
\langle D_{i}.B'',c_{F}''\rangle,
\]
with \[
\begin{array}[t]{ll} \sf \exists_{-\Var(C_{I}[C[p(\ol t) \Par tell(b)\Par
tell(\ol t = \ol s)] \Par tell(\ol{v}=\ol{r})],c_{I})}  c_{F}'' =
\\
\sf \exists_{-\Var(C_{I}[C[p(\ol t) \Par tell(b)\Par tell(\ol t =
\ol s)] \Par tell(\ol{v}=\ol{r})],c_{I})}c_{F}
\end{array}
\]
 and
\begin{equation}
        {\sf m(B'',c_{F}'')} ={\sf m (B,c_{F})}.
        \label{eq:app6ott3}
\end{equation}
Moreover, since $\sf \Var(C_{I}[q (\ol{v})],c_{I}) \subseteq
\Var(C_{I}[C[p(\ol t) \Par tell(b)\Par tell(\ol t = \ol s)]
\Par tell(\ol{v}=\ol{r})],c_{I})$, we have that
\begin{equation}
        \sf \exists_{-\Var(C_{I}[q (\ol{v})],c_{I})}  c_{F}'' =
        \exists_{-\Var(C_{I}[q (\ol{v})],c_{I})}c_{F}.
        \label{eq:app29sett1}
\end{equation}
If $\sf p(\ol t) $ is not evaluated in $\sf \chi$, then the proof is
immediate. Otherwise, by the definition of $\chi$ and
since $\sf f \in D_i$, there exists also a derivation
\[
\sf \chi'=\langle D_{i}.C_{I}[C[p(\ol t)] \Par tell(\ol{v}=\ol{r})],
c_{I}\rangle \rightarrow^*
\langle D_{i}.B',c_{F}'\rangle
\]
such that
${\sf \exists_{-\Var(C_{I}[C[p(\ol t)] \Par
tell(\ol{v}=\ol{r})],c_{I})}c_{F}'=
\exists_{-\Var(C_{I}[C[p(\ol t)] \Par
tell(\ol{v}=\ol{r})],c_{I})}c''_{F}}$
and ${\sf m(B',c_{F}')}$ = ${\sf m (B'',c''_{F})}$.
Therefore, by (\ref{eq:app29sett1}) and  (\ref{eq:app6ott3})
\begin{equation}
        {\sf \exists_{-\Var(C_{I}[q (\ol{v})],c_{I})}c_{F}'=
        \exists_{-\Var(C_{I}[q (\ol{v})],c_{I})}c_{F}}
        \mbox{ and } {\sf m(B',c_{F}')} = {\sf m (B,c_{F})}.
        \label{eq:app28genn2}
\end{equation}
By the definition of $\sf \chi'$,
$\sf \Var(C_{I}[q (\ol{v})],c_{I}) \cap \Var(\chi') \subseteq
\Var(C_{I}[C[p(\ol t)] \Par tell(\ol{v}=\ol{r})],c_{I}).$
Then, by the definition of derivation and since $\sf d \in D_{i}$,
\[
\sf \langle D_{i}.C_{I}[q (\ol{v})],c_{I}\rangle \rightarrow
\langle D_{i}.C_{I}[C[p(\ol t) ] \Par tell(\ol{v}=\ol{r})],c_{I}\rangle
\rightarrow^* \langle D_{i}.B',c_{F}'\rangle
\]
and then the thesis follows from (\ref{eq:app28genn2}).
\II

\NI {\bf Case 4:} $\sf d'$ is obtained from $\sf d$ by either
an ask simplification or a tell simplification.\\
We consider only the first case (the proof of the other one is
analogous and hence it is omitted).  Let

- $\sf d':\ q(\ol r) \la C[\sum_{j=1}^{n} ask(c'_j)
\rightarrow A_j]$, and

- $\sf d:\ q(\ol r) \la C[\sum_{j=1}^{n} ask(c_j) \rightarrow A_j]$, \\
where for $j \in [1,n]$,
${\cal D} \models \sf \exists_{- \Var(q(\ol r),C,A_j)}\
(pc(C[\ ]) \And c_j)$
$\sf \lra (pc(C[\ ]) \And c'_j)$.
According to the definition of $\sf pc$ and by Lemma~\ref{lem:proppc}, for
any derivation $\chi$ for
$$\sf \langle D_{i}.C_{I}[C[\sum_{j=1}^{n} ask(c'_j) \rightarrow A_j]
\Par tell(\ol{v}=\ol{r})], c_{I}\rangle$$ there exists a
derivation $\chi'$ for $$\sf \langle D_{i}.C_{I}[C[\sum_{j=1}^{n}
ask(c_j) \rightarrow A_j] \Par tell(\ol{v}=\ol{r})],
c_{I}\rangle$$ which performs the same steps of $\chi$ (possibly
in a different order) and such that whenever the choice agent
inside $\sf C[\ ]$ is evaluated the current store implies $\sf
pc(C[\ ])$.  Therefore the thesis follows from the above
equivalence.  \II

\NI {\bf Case 5:} $\sf d'$ is the result of a branch elimination
or of a conservative ask elimination.\\
The proof is straightforward by noting that:
(a) according to Definition~\ref{def:semantics} we consider also
inconsistent stores  resulting from non-terminated computations;
(b) an ask action of the form $\sf ask(true)$ always succeeds.
\II

\NI {\bf Case 6:} $\sf d'$ is the result of a distribution operation.
Let

- $\sf d:\ q(\ol r) \la C[H \Par \sum_{j=1}^{n} ask(c_j) \rightarrow B_j]
\in D_{i}$

- $\sf d':\ q(\ol r) \la C[\sum _{j=1}^{n} ask(c_j) \rightarrow (H
\Par B_j)] \in D_{i+1}$  \\

where $\sf e = pc(C[\ ])$ and for every
constraint $\sf c$ such that $\sf \Var(c)
\cap \Var(d) \subseteq \Var(q(\ol r),C)$, if
$\sf  \langle D_{i}.H, c \wedge e \rangle$ is productive
then both the following conditions hold:
\begin{itemize}
        \item
        there exists at least one $j\in [1,n]$ such that
        $\sf {\cal D}\models (c \wedge e ) \rightarrow  c_j$

        \item  for each $j\in [1,n]$, either
        $\sf {\cal D}\models (c \wedge e ) \rightarrow  c_j$
    or $\sf {\cal D}\models (c \wedge e ) \rightarrow  \neg c_j$.
\end{itemize}
In this case
$\sf \xi = \langle D_{i+1}.C_{I}[q (\ol{v})],c_{I}\rangle \rightarrow
\langle D_{i+1}.C_{I}[C[ \sum _{j=1}^{n} ask(c_j) \rightarrow (H \Par B_j)]
\Par tell(\ol{v}=\ol{r})],c_{I}\rangle
\rightarrow^* \langle D_{i+1}.B,c_{F}\rangle$.
By the inductive hypothesis, there exists a derivation
\[
\sf \chi =
\langle D_{i}.C_{I}[C[ \sum _{j=1}^{n} ask(c_j) \rightarrow (H \Par B_j)]
\Par tell(\ol{v}=\ol{r})], c_{I}\rangle \rightarrow^*
\langle D_{i}.B'',c_{F}''\rangle
\]
with
\[\begin{array}[t]{ll}
\sf \exists_{-\Var(C_{I}[C[\sum _{j=1}^{n} ask(c_j) \rightarrow
(H \Par B_j)] \Par tell(\ol{v}=\ol{r})],c_{I})} c_{F}'' =
\\
\sf \exists_{-\Var(C_{I}[C[\sum _{j=1}^{n} ask(c_j) \rightarrow
(H \Par B_j)] \Par tell(\ol{v}=\ol{r})],c_{I})}c_{F}
\end{array}
\]
and
\begin{equation}
        {\sf  m(B'',c_{F}'')} = {\sf m(B,c_{F})}.
        \label{eq:app4nov1}
\end{equation}
Moreover, since $\sf \Var(C_{I}[q (\ol{v})],c_{I}) \subseteq
\Var(C_{I}[C[\sum _{j=1}^{n} ask(c_j) \rightarrow (H \Par B_j)]
\Par tell(\ol{v}=\ol{r})],c_{I})$, we have that
\begin{equation}
        \sf \exists_{-\Var(C_{I}[q (\ol{v})],c_{I})}c_{F}'' =
        \exists_{-\Var(C_{I}[q (\ol{v})],c_{I})} c_{F}.
        \label{eq:app4nov2}
\end{equation}
Now, we distinguish two cases:

\emph{1)}\ \  $\sf \sum _{j=1}^{n} ask(c_j) \rightarrow (H \Par B_j)$ is
not evaluated in $\sf \chi$. In this case the proof is obvious.

\emph{2)}\ \ $\sf \sum_{j=1}^{n} ask(c_j) \rightarrow (H \Par B_j)$ is
evaluated in $\sf \chi$. We have two more possibilities:

\emph{2a)}\ \  There exists $h \in [1,n]$, such that
\[
\begin{array}{ll}
    \sf \chi =
    & \sf \langle D_{i}.C_{I}[C[ \sum_{j=1}^{n} ask(c_j)
    \rightarrow (H \Par B_j)]
    \Par tell(\ol{v}=\ol{r})], c_{I}\rangle \rightarrow^*\\
    & \sf  \langle D_{i}.C_{m}[ \sum_{j=1}^{n}  ask(c_j)
    \rightarrow (H \Par B_j)], c_{m}\rangle \rightarrow
    \langle D_{i}.C_{m}[H \Par B_h], c_{m}\rangle \rightarrow^*
    \langle D_{i}.B'',c_{F}''\rangle
\end{array}
\]
where $\sf {\cal D} \models c_{m} \ra c_{h}$.  In this case
the thesis follows immediately, since using $\sf d$ one can
obtain the agent $\sf C_{m}[H \Par B_h]$ after having
evaluated the choice agent in $\sf C[\ ]$.

\emph{2b)} \ \ There is no $h \in [1,n]$, such that
$\sf {\cal D} \models c_{F}'' \ra c_{h}$.  In this case
\begin{equation}
        {\sf c_{F}''} \mbox{ is satisfiable, }
        {\sf  m(B'',c_{F}'') = dd},
        \label{eq:app18nov199}
\end{equation}
$\sf B''$ is the agent $\sf C_{F}[\sum_{j=1}^{n} ask(c_j) \rightarrow (H
\Par
B_j)]$
and
\[
\begin{array}[t]{ll}
\sf \chi = &\sf \langle D_{i}.C_{I}[C[ \sum_{j=1}^{n} ask(c_j)
\rightarrow (H \Par  B_j)] \Par tell(\ol{v}=\ol{r})],
c_{I}\rangle \rightarrow^*
\\
&\sf \langle D_{i}.C_{F}[\sum_{j=1}^{n} ask(c_j) \rightarrow (H
\Par B_j)], c_{F}''\rangle \not \rightarrow.
\end{array}
 \]



>From the definition of derivation, the definition
of $\sf B''$ and the hypothesis that
$\sf \sum_{j=1}^{n} ask(c_j) \rightarrow (H \Par B_j)$ is
evaluated in $\sf \chi$, it follows that
$\sf C_{F}[\sum_{j=1}^{n} ask(c_j) \rightarrow (H \Par B_j)]$
is of the form
$\sf A_{1}\Par \ldots\Par
\sum_{j=1}^{n} ask(c_j) \rightarrow (H \Par B_j)\Par
\ldots \Par A_{l}$, where either $\sf A_{k}$ is a
choice agent or $\sf A_{k}= Stop$.
By Lemma~\ref{lem:proppc},
${\cal D} \models \sf \ c_{F}'' \rightarrow pc(C[\ ])$
and by definition of derivation
$\sf \Var(c_{F}'') \cap \Var(d) \subseteq \Var(q(\ol r),C)$.
Then, since there is no $j\in [1,n]$ such that
$\sf {\cal D}\models c_{F}'' \rightarrow  c_j$,
by definition of distribution,
$\sf \langle D_{i}.H, c_{F}'' \rangle$ is not productive.
Then, by definition,
$\sf \langle D_{i}.H, c_{F}''\rangle$
has at least one finite derivation
$\sf \chi_{1}= \langle D_{i}.H, c_{F}''\rangle \rightarrow^*
\langle D_{i}.H', c_{F}'\rangle \not \rightarrow$ such that
$\sf {\cal D}\models \exists_{-\ol{Z}}\ c''_{F}
\leftrightarrow \exists_{-\ol{Z}}\ c'_{F}$, where
$\sf \ol{Z} = \Var(H)$.
Moreover, since in a derivation we can add to the store only
constraints on the variables occurring in the agents,
$\sf c''_{F} =\exists_{-\Var(H,c_{F}'')}\ c''_{F} =
\exists_{-\Var(H,c_{F}'')}\ c'_{F}$ holds.

Without loss of generality, we can assume that
$\sf \Var (\chi_{1}) \cap \Var (\chi) \subseteq \Var(H,c_{F}'')$.
Therefore, by the previous observation,
\begin{equation}
        \sf  \exists_{-\Var(C_{F}[\sum_{j=1}^{n} ask(c_j)
        \rightarrow (H \Par B_j)],c_{F}'')}\ c'_{F} =
        c''_{F}
        \label{eq:app4nov4}
\end{equation}
and since $\sf \langle D_{i}.C_{F}[\sum_{j=1}^{n} ask(c_j)
\rightarrow (H \Par B_j)],c_{F}''\rangle \not \rightarrow$ and
$\sf \langle D_{i}.H',c_{F}'\rangle \not \rightarrow$, there
exists a derivation
\[
\begin{array}{ll}
    \sf \chi' =
    & \sf  \langle D_{i}.C_{I}[C[H \Par
    \sum_{j=1}^{n} ask(c_j) \rightarrow B_j]
    \Par tell(\ol{v}=\ol{r})], c_{I}\rangle \rightarrow^* \\
    & \sf \langle D_{i}.C_{F}[H \Par\sum_{j=1}^{n} ask(c_j)
    \rightarrow  B_j], c_{F}''\rangle \rightarrow^*
    \langle D_{i}.C_{F}[H' \Par
    \sum_{j=1}^{n} ask(c_j) \rightarrow  B_j],
    c_{F}'\rangle  \not \rightarrow.
\end{array}
\]
Moreover, since
$\sf d \in D_{i}$, there exists a derivation
\[
\begin{array}{ll}
     \sf \xi' =
     & \sf \langle D_{i}.C_{I}[q(\ol{v})], c_{I}\rangle
     \rightarrow \langle D_{i}.C_{I}[C[H \Par
     \sum_{j=1}^{n} ask(c_j) \rightarrow B_j]
     \Par tell(\ol{v}=\ol{r})], c_{I}\rangle \rightarrow^* \\
     & \sf \langle D_{i}.C_{F}[H \Par
     \sum_{j=1}^{n} ask(c_j) \rightarrow  B_j], c_{F}''\rangle
     \rightarrow^*
     \langle D_{i}.C_{F}[H' \Par\sum_{j=1}^{n} ask(c_j)
     \rightarrow  B_j],  c_{F}'\rangle  \not \rightarrow.
     \end{array}
\]
Finally, to prove the thesis it is sufficient to observe that
from (\ref{eq:app4nov1}), (\ref{eq:app18nov199}), (\ref{eq:app4nov4})
and from the definition of
$\sf  B'=C_{F}[H' \Par\sum_{j=1}^{n} ask(c_j) \rightarrow  B_j]$
it follows that
${\sf  m(B',c_{F}')} = {\sf m(B,c_{F}) =dd}$. Moreover
\[\begin{array}{lll}
      \sf \exists_{-\Var(C_I[q(\ol v)],c_I)} c'_{F} & = &
      \mbox{(by construction)}  \\
      \sf \exists_{-\Var(C_I[q(\ol v)],c_I)} (c''_{F} \wedge
      \exists_{-\Var(C_{F}[H \Par\sum_{j=1}^{n} ask(c_j)
      \rightarrow   B_j],c_{F}')} c'_{F}) & = &
      \mbox{(by (\ref{eq:app4nov4}))}  \\
      \sf \exists_{-\Var(C_I[q(\ol v)],c_I)} c''_{F}& = &
      \mbox{(by (\ref{eq:app4nov2}))}  \\
      \sf \exists_{-\Var(C_I[q(\ol v)],c_I)} c_{F}
\end{array}
\]
which concludes the proof of this case.  \II

\NI
{\bf Case 7:} $\sf d'$ is the result of a folding.\\
Let

- $\sf d: \ q(\ol r) \la C[H]$ be the folded declaration
($\sf \in D_i$),

- $\sf f: \ p(\ol X) \la H$ be the folding declaration ($\in \sf D_{0}$),

- $\sf d': \ q(\ol r) \la C[p(\ol X)]$ be the result of the folding
operation ($\in \sf D_{i+1}$)
\\
where, by hypothesis, $\sf \Var(d) \cap \Var(\ol X)\subseteq \Var(H)$
and $\sf \Var(H) \cap (\Var(\ol r)\cup \Var(C))\subseteq \Var(\ol X)$.
In this case
$\sf \xi = \langle D_{i+1}.C_{I}[q (\ol{v})],c_{I}\rangle
\rightarrow \langle D_{i+1}.C_{I}[C[p(\ol X)] \Par
tell(\ol v = \ol r)],c_{I}\rangle \rightarrow^*
\langle D_{i+1}.B,c_{F}\rangle$
and we can assume, without loss of generality, that $\sf
\Var (C_{I}[q (\ol{v})],c_{I}) \cap \Var(H)=\emptyset$.

By the inductive hypothesis, there exists a derivation
\[
\sf \chi =
\langle D_{i}.C_{I}[C[p(\ol X)] \Par
tell(\ol v = \ol r)],c_{I}\rangle \rightarrow^*
\langle D_{i}.B'',c_{F}''\rangle,
\]
with
$\sf \exists_{-\Var(C_{I}[C[p(\ol X)] \Par
tell(\ol v = \ol r)],c_{I})} c_{F}'' =
\exists_{-\Var(C_{I}[C[p(\ol X)] \Par
tell(\ol v = \ol r)],c_{I})} c_{F}$ and
\begin{equation}
        {\sf m(B'',c''_{F})= m(B,c_{F})}.
        \label{eq:app25nov199}
\end{equation}
Since $\sf \Var(C_{I}[q (\ol{v})],c_{I}) \subseteq
\Var(C_{I}[C[p(\ol X)] \Par tell(\ol v = \ol r)],c_{I})$,
we have that
\begin{equation}
        \sf \exists_{-\Var(C_{I}[q (\ol{v})],c_{I})} c_{F}'' =
    \exists_{-\Var(C_{I}[q (\ol{v})],c_{I})} c_{F}.
        \label{eq:app25nov299}
\end{equation}
Since by hypothesis for any agent $\sf A'$,
$\sf {\cal O}(D_0.A') = {\cal O}(D_i.A')$,
there exists a derivation
\[\sf \xi_{0}=\langle D_{0}.C_{I}[C[p(\ol X)] \Par
tell(\ol v = \ol r)],c_{I}\rangle
\rightarrow^* \langle D_{0}.B_{0}, c_{0}\rangle
\]
such that
$\sf \exists_{-\Var(C_{I}[C[p(\ol X)] \Par
tell(\ol v = \ol r)],c_{I})}c_{0} =
\exists_{-\Var(C_{I}[C[p(\ol X)] \Par
tell(\ol v = \ol r)],c_{I})} c_{F}''$
and ${\sf m( B_{0}, c_{0})} = {\sf m(B'',c''_{F})}$.
By (\ref{eq:app25nov199}), (\ref{eq:app25nov299}) and since
$\sf \Var(C_{I}[q (\ol{v})],c_{I}) \subseteq
\Var(C_{I}[C[p(\ol X)] \Par tell(\ol v = \ol r)],c_{I})$,
we have that
\begin{equation}
        {\sf \exists_{-\Var(C_{I}[q (\ol{v})],c_{I})} c_0 =
    \exists_{-\Var(C_{I}[q (\ol{v})],c_{I})} c_{F}}
    \mbox{ and } {\sf m( B_{0}, c_{0})} = {\sf m(B,c_{F})}.
        \label{eq:app25nov1099}
\end{equation}
Let $\sf f': \ p(\ol X') \la H'$ be an appropriate
renaming of $\sf f$, which renames only the variables in $\sf \ol X$,
such that $\sf \Var(d) \cap \Var(f')= \emptyset$ (note that this
is possible, since $\sf \Var(H) \cap (\Var(\ol r)\cup
\Var(C))\subseteq \Var(\ol X)$). Moreover by hypothesis,
$\sf \Var (C_{I}[q (\ol{v})],c_{I}) \cap \Var(H)=\emptyset$.
Then, without loss of generality we can assume that
$\sf \Var(\xi_{0}) \cap \Var (f') \neq \emptyset$
if and only if the procedure call $ \sf p(\ol X)$ is evaluated,
in which case declaration $\sf f'$ is used.

Thus there exists a derivation
\[\sf \langle D_{0}.
C_{I}[ C [H'\Par tell(\ol X = \ol X') ]
\Par tell(\ol v = \ol r)],c_{I}\rangle \rightarrow^*
\langle D_{0}. B'_{0}, c_{0}\rangle,
\]
where ${\sf m(B'_{0}, c_{0})} = {\sf m(B_{0}, c_{0})}$.
By (\ref{eq:app25nov1099}) we have
\begin{equation}
    {\sf m(B'_{0}, c_{0})} = {\sf m(B,c_{F})}.
    \label{eq:app1ott2}
\end{equation}
We show now that we can substitute
$\sf  H$ for $\sf H' \Par tell(\ol X = \ol X') $
in the previous derivation. Since
$\sf f': \ p(\ol X') \la H'$ is a renaming of
$\sf f: \ p(\ol X) \la H$, the equality
$\sf \ol X = \ol X'$ is a conjunction
of equations involving only distinct variables.
Then, by replacing
$\sf \ol X$ with $\sf \ol X'$ and vice versa in the previous
derivation we obtain the derivation
$\sf \chi_{0}= \sf \langle D_{0}.
C_{I}[C [H \Par tell(\ol X' = \ol X) ]
\Par tell(\ol v = \ol r)],c_{I}\rangle \rightarrow^*
\langle D_{0}. B''_{0}, c_{0}'\rangle$ where
\[
\begin{array}[t]{l}
\sf \exists_{-\Var(C_{I}[C[H \Par tell(\ol X' = \ol X)] \Par
tell(\ol v = \ol r)],c_{I})} c_{0}' = \sf  \exists_{-\Var(C_{I}[C [H \Par tell (\ol X' = \ol X) ] \Par
tell(\ol v = \ol r)],c_{I})} c_{0}\\
\mbox{and } {\sf m(B''_{0},
c'_{0})} = {\sf m(B'_{0}, c_{0})}.
\end{array}
\]
>From (\ref{eq:app1ott2}) it follows that
\begin{equation}
        {\sf m(B''_{0}, c'_{0})} = {\sf m(B,c_{F}). }
        \label{eq:app1ott3}
\end{equation}
Then, from (\ref{eq:app25nov1099}) and since
$\sf \Var(C_{I}[q(\ol v)],c_{I}) \subseteq
\Var(C_{I}[C [H \Par tell (\ol X' = \ol X) ]
\Par tell(\ol v = \ol r)],c_{I})$ we obtain
\begin{equation}
        \sf \exists_{-\Var(C_{I}[q(\ol v)],c_{I})} c_{0}' =
       \exists_{-\Var(C_{I}[q(\ol v)],c_{I})}c_{F}.
        \label{eq:app28genn10}
\end{equation}
Moreover, we can drop the constraint $\sf tell(\ol X' = \ol X)$,
since the declarations used in the derivation are renamed apart
and, by construction,
$\sf \Var(C_{I}[C [H] \Par tell(\ol r = \ol v)], c_{I})
\cap \Var(\ol X') =\emptyset$.
Therefore there exists a derivation
$\sf \langle D_{0}.
C_{I}[C [H] \Par tell(\ol v = \ol r)],c_{I}\rangle
\rightarrow^*  \langle D_{0}.\bar B_{0},\bar c_{0}\rangle$
which performs exactly the same steps of $\sf \chi_{0}$,
(possibly) except for the evaluation of
$\sf tell(\ol X' = \ol X)$, and such that
$\sf \exists_{-\Var(C_{I}[C [H] \Par
tell(\ol v = \ol r)],c_{I})} \bar c_{0} =
\exists_{-\Var(C_{I}[C [H] \Par
tell(\ol v = \ol r)],c_{I})} c_{0}'$ and
${\sf m(\bar B_{0},\bar c_{0})}= {\sf m(B''_{0}, c'_{0})}$.
>From (\ref{eq:app1ott3}),  (\ref{eq:app28genn10}) and since
$\sf \Var(C_{I}[q(\ol v)],c_{I}) \subseteq
\Var(C_{I}[C [H] \Par tell(\ol v = \ol r)],c_{I})$, it follows that
\begin{equation}
    {\sf m(\bar B_{0},\bar c_{0})= m (B,c_{F})}
    \mbox{ and }
    {\sf \exists_{-\Var(C_{I}[q(\ol v)],c_{I})}\bar c_{0} =
    \exists_{-\Var(C_{I}[q(\ol v)],c_{I})}c_{F}}.
    \label{eq:app1ott4}
\end{equation}
Since $\sf {\cal O}(D_0.A') = {\cal O}(D_i.A')$ holds
by hypothesis for any agent $\sf A'$,
there exists a derivation
\[
\sf \langle D_{i}.
C_{I}[C [H] \Par tell(\ol v = \ol r)],c_{I}\rangle
\rightarrow^*  \langle D_{i}.B',c_{F}'\rangle
\]
where
$$ \sf \exists_{-\Var(C_{I}[C [H] \Par
tell(\ol v = \ol r)],c_{I})} c_{F}'= \exists_{-\Var(C_{I}[C [H]
\Par tell(\ol v = \ol r)],c_{I})} \bar c_{0}$$ and ${\sf
m(B',c'_{F})} = {\sf m(\bar B_{0},\bar c_{0})}$.  From
(\ref{eq:app1ott4}) and since $\sf \Var(C_{I}[q(\ol v)],c_{I})
\subseteq \Var(C_{I}[C [H] \Par tell(\ol v = \ol r)],c_{I})$, we
obtain
\begin{equation}
     {\sf m(B',c'_{F})} = {\sf m (B,c_{F})}
     \mbox{ and }
     {\sf \exists_{-\Var(C_{I}[q(\ol v)],c_{I})}c_{F}'=
     \exists_{-\Var(C_{I}[q(\ol v)],c_{I})}c_{F}}.
     \label{eq:app1ott5}
\end{equation}
Finally, since $\sf d: \ q(\ol r) \la C[H]
\in \sf D_i$, there exists a derivation
\[
\sf \xi' =
\langle D_{i}.C_{I}[q ( \ol v)],c_{I}\rangle \rightarrow
\langle D_{i}.C_{I}[C [H] \Par tell(\ol v = \ol r)],c_{I}\rangle
\rightarrow^* \langle D_{i}.B',c_{F}'\rangle
\]
and then the thesis follows from (\ref{eq:app1ott5}).
\end{proof}

Before proving the total correctness result
we need some technical lemmata.
Here and in the following we use the notation $\sf w_t$
(with $\sf t \in \{ss,dd,ff\}$) as a shorthand for indicating
the success weight $\sf w_{ss}$, the deadlock weight $\sf w_{dd}$
and the failure weight $\sf w_{ff}$.

\begin{lemma}
    \label{lem:apppesoclausola}
    Let $ \sf q (\ol{r})\la H \in D_{0}$, $\sf t \in \{ss,dd,ff\}$ and
    let $\sf C[\ ]$ be context. For any satisfiable
    constraint $\sf c$ and for any constraint $\sf c'$, such that
    $\sf \Var(C[q(\ol{t})],c) \cap \Var(\ol{r}) =\emptyset$ and
    $\sf w_t( C[q (\ol{t}) ],c,c')$ is defined,
    there exists a constraint $\sf d'$ such that 
    $\sf w_t( C[q (\ol{r}) \Par tell(\ol{t}=\ol{r})],c,d')\leq
    w_t( C[q (\ol{t}) ],c,c')$ and
    $\sf \exists_{-\Var( C[q (\ol{t}) ],c)}d' =
    \exists_{-\Var( C[q (\ol{t}) ],c)}c'$.
    \end{lemma}
 \begin{proof}
Immediate. \end{proof}

\begin{lemma}
    \label{lem:apppesiind0}
    Let $\sf q (\ol{r})\la H \in D_{0}$ and $\sf t \in \{ss,dd,ff\}$.
    For any context $\sf C_{I}[\ ]$, any satisfiable
    constraint $\sf c$ and for any constraint $\sf c'$, the following
holds.
    \begin{enumerate}
        \item\label{pt:apppesiind01}
        If $\sf \Var(H) \cap \Var(C_{I},c) \subseteq \Var(\ol r)$
        and $\sf w_t( C_{I}[q (\ol{r}) ],c,c')$ is defined, then there
        exists a constraint $\sf d'$, such that
        $\sf \Var(d') \subseteq \Var(C_{I}[H],c)$,
        $\sf w_t( C_{I}[H],c,d') \leq  w_t( C_{I}[q (\ol{r}) ],c,c')$
        and $\sf \exists_{-\Var( C_{I}[q (\ol{r}) ],c)}d' =
        \exists_{-\Var( C_{I}[q (\ol{r}) ],c)}c'$.

        \item\label{pt:apppesiind02}
        If $\sf \Var(H) \cap \Var(C_{I},c) \subseteq \Var(\ol r)$,
        $\sf \Var(c') \cap \Var(\ol r) \subseteq \Var(C_{I}[H],c)$
        and $\sf w_t( C_{I}[H],c,c')$ is defined, then there exists a
        constraint $\sf d'$, such that \\
        $\sf w_t( C_{I}[q (\ol{r}) ],c,d') \leq  w_t( C_{I}[H],c,c')$
        and $\sf \exists_{-\Var( C_{I}[q (\ol{r}) ],c)}d' =
        \exists_{-\Var( C_{I}[q (\ol{r}) ],c)}c'$.
    \end{enumerate}
\end{lemma}
\begin{proof}
Immediate.
\end{proof}
The following Lemma is crucial in the proof of
completeness.

\begin{lemma}
    \label{lem:apppesiindi}
    Let $0 \leq i \leq n$, $\sf t \in \{ss,dd,ff\}$,
    $ \sf cl:\  q (\ol{r})\la H \in D_{i}$,
    and let $ \sf cl':\  q (\ol{r})\la H' $ be the corresponding
    declaration in $\sf D_{i+1}$ (in the case $i<n$).
    For any context $\sf C_{I}[\ ]$ and any satisfiable
    constraint $\sf c$ and for any constraint $\sf c'$ the following holds:
    \begin{enumerate}
       \item\label{pt:apppesiindi1}
       If $\sf \Var(H) \cap \Var(C_{I},c) \subseteq \Var(\ol r)$
       and $\sf w_t( C_{I}[q (\ol{r}) ],c,c')$ is defined,
       then there exists a constraint $\sf d'$, such that
       $\sf \Var(d') \subseteq \Var(C_{I}[H],c)$,
       $\sf w_t( C_{I}[H],c,d') \leq
       w_t( C_{I}[q (\ol{r}) ],c,c')$
       and $\sf \exists_{-\Var( C_{I}[q (\ol{r}) ],c)}d' =
       \exists_{-\Var( C_{I}[q (\ol{r}) ],c)}c'$;
       \item\label{pt:apppesiindi2}
       If $\sf \Var(H,H') \cap \Var(C_{I},c) \subseteq \Var(\ol r)$,
       $\sf \Var(c') \cap \Var(\ol r)\subseteq \Var(C_{I}[H],c)$ and
       $\sf w_t( C_{I}[H],c,c')$ is defined, then there exists a
       constraint $\sf d'$, such that $\sf \Var(d') \subseteq
       \Var(C_{I}[H'],c)$,
       $\sf w_t( C_{I}[H'],c,d') \leq  w_t( C_{I}[H],c,c')$
       and
       \\
       $\sf \exists_{-\Var( C_{I}[q (\ol{r}) ],c)}d' =
       \exists_{-\Var( C_{I}[q (\ol{r}) ],c)}c'$.
    \end{enumerate}
\end{lemma}
\begin{proof}
Observe that, for $i=0$, the proof of 1
follows from the first part of Lemma~\ref{lem:apppesiind0}.
We prove here that, for each $i \geq 0$,

{\bf a)} if 1 holds for $i$ then 2 holds for $i$;

{\bf b)} if 1 and 2 hold for i then 1 holds for $i+1$. \\
The proof of the Lemma then follows from straightforward inductive
argument.

{\bf a)} If $\sf cl$ was not affected by the transformation step
from $\sf D_i$ to $\sf D_{i+1}$ then the result is obvious by choosing
$ \sf d'= \exists_{-\Var( C_{I}[H],c)} c'$.
Assume then that $\sf cl$ is
affected when transforming $\sf D_i$ to $\sf D_{i+1}$ and
let us distinguish various cases.  \II

\NI
{\bf Case 1:} $\sf cl' \in  D_{i+1}$ was obtained from $\sf D_i$ by
unfolding.\\
In this case, the situation is the following:

- $\sf cl:\ q (\ol{r}) \la  C[p(\ol{t})] \in D_{i}$

- $\sf u:\ p(\ol{s})\la  B  \in D_{i}$

- $\sf cl':\ q (\ol{r}) \la C[B \Par tell(\ol{t}=\ol{s})] \in D_{i+1}$
\\
where \textsf{cl} and \textsf{u} are assumed to be renamed
so that they do not share variables.
Let $\sf n = w_t(C_{I}[C[p(\ol{t})]],c,c')$.
By the definition of transformation sequence, there exists a
declaration $\sf p(\ol{s})\la B_{0} \in D_{0}$.
Moreover, by the hypothesis on the variables,
$\sf \Var(C[p(\ol{t})],C[B \Par tell(\ol{t}=\ol{s})]) \cap
\Var(C_{I},c) \subseteq \Var(\ol r)$ and then
$\sf \Var(C_{I}[C[p(\ol{t})]],c) \cap \Var(\ol s)=\emptyset$.
Therefore, by Lemma~\ref{lem:apppesoclausola},
there exists a constraint $\sf d_{1}$, such that
\begin{equation}
        \sf w_t( C_{I}[C[p(\ol{s})\Par
        tell(\ol{t}=\ol{s})]],c,d_{1})
        \leq w_t( C_{I}[C[p(\ol{t})]],c,c')=n
        \label{eq:app14ott1}
\end{equation}
and
\begin{equation}
        \sf \exists_{-\Var( C_{I}[C[p(\ol{t})]],c)}d_{1} =
        \exists_{-\Var( C_{I}[C[p(\ol{t})]],c)}c'.
        \label{eq:app14ott2}
\end{equation}
By the hypothesis on the variables and since \textsf{u} is renamed
apart from \textsf{cl},
$\sf \Var(B) \cap \Var(C_{I},C,\ol{t},c) = \emptyset$ and therefore
$\sf \Var(B) \cap \Var(C_{I}[C[\ ]\Par
tell(\ol{t}=\ol{s})],c) \subseteq \Var(\ol s)$.
Then, by Point 1, there
exists a constraint $\sf d'$, such that
$\sf \Var(d') \subseteq
\Var(C_{I}[C[B\Par tell(\ol{t}=\ol{s})]],c)$,
$\sf w_t(C_{I}[C[B\Par tell(\ol{t}=\ol{s})]],c,d') \leq
w_t(C_{I}[C[p(\ol{s})\Par tell(\ol{t}=\ol{s})]],c,d_{1})$ and
\\
$\sf \exists_{-\Var( C_{I}[C[p(\ol{s})\Par
tell(\ol{t}=\ol{s})]],c)}d' =
\exists_{-\Var( C_{I}[C[p(\ol{s})\Par
tell(\ol{t}=\ol{s})]],c)}d_{1}$.

By (\ref{eq:app14ott1}),
$\sf w_t( C_{I}[C[B\Par tell(\ol{t}=\ol{s})]],c,d')\leq n$.

Furthermore, by hypothesis and construction,
$\sf \Var(c',d') \cap \Var(\ol r) \subseteq
\Var(C_{I}[C[p(\ol{t})]],c)$ and, without loss of generality, we can
assume that $\sf \Var(d_{1}) \cap \Var(\ol r) \subseteq
\Var(C_{I}[C[p(\ol{t})]],c)$.

Then, by (\ref{eq:app14ott2}) and since
$\sf \Var( C_{I}[C[p(\ol{t})]],c) \subseteq
\Var( C_{I}[C[p(\ol{s})\Par tell(\ol{t}=\ol{s})]],c)$, we have that
$\sf \exists_{-\Var( C_{I}[q(\ol{r})],c)}d' =
\exists_{-\Var( C_{I}[q(\ol{r})],c)}c'$ and this completes the proof.
\II

\NI {\bf Case 2:} $\sf cl'$ is the result of a tell elimination
or introduction.\\
The proof is analogous to that one given for Case 2 of
Proposition~\ref{pro:partial} and  it is omitted.

%
%
%
%
%
\II

\NI {\bf Case 3:} $\sf cl'$ is the result of a backward instantiation.\\
Let $\sf cl$ be the corresponding declaration in $\sf D_i$.
The situation is then the
following:

- $\sf cl: \ q(\ol r) \la C[p(\ol t)]$

- $\sf cl': \ q(\ol r) \la C[p(\ol t) \Par tell(b)
 \Par tell(\ol t = \ol s)]$\\
where $\sf f: \ p(\ol s) \la tell(b)\Par H \in D_i$ has no variable
in common with $\sf cl$ (the case
$\sf cl': \ q(\ol r) \la C[p(\ol t) \Par tell(\ol t = \ol s)]$
is analogous and hence omitted). By the hypothesis, $\sf \Var(C[p(\ol t)],
C[p(\ol t) \Par tell(b)\Par tell(\ol t = \ol s)]) \cap \Var(C_{I},c)
\subseteq \Var(\ol r)$, $\sf \Var(c') \cap \Var(\ol r) \subseteq
\Var(C_{I}[C[p(\ol t)]],c)$ and there exists $n$ such that
$\sf w_t( C_{I}[C[p(\ol t)]],c,c')=n$.
Then
$\sf \Var(C_{I}[C[p(\ol t)]],c) \cap \Var(\ol s)= \emptyset$ and,
without loss of generality, we can assume that
$\sf \Var(H) \cap\Var(C_{I},c) = \emptyset$.

Moreover, by the definition of transformation sequence,
there exists a declaration $\sf p(\ol{s})\la B_{0} \in D_{0}$
and then, by Lemma~\ref{lem:apppesoclausola},
there exists a constraint $\sf d_{1}$ such that
\begin{equation}
        \sf w_t( C_{I}
        [C[p(\ol{s})\Par tell(\ol{t}=\ol{s})]],c,d_{1})
        \leq w_t( C_{I}[C[p(\ol{t})]],c,c')=n
        \label{eq:app14ott3}
\end{equation}
and
\begin{equation}
        \sf \exists_{-\Var( C_{I}[C[p(\ol{t})]],c)}d_{1} =
        \exists_{-\Var( C_{I}[C[p(\ol{t})]],c)}c'.
        \label{eq:app14ott4}
\end{equation}

Using the hypothesis on the variables and since \textsf{f}
is renamed apart from $\sf \Var(\ol r)$, we have that
\[
\sf \Var(tell(b)\Par H) \cap
\Var(C_{I}[C[\ \Par tell(\ol{t}=\ol{s})]],c) \subseteq \Var(\ol{s}).
\]
Then, from Point 1 of the Lemma (assumed as hypothesis)
and (\ref{eq:app14ott3}) it follows that there exists a
constraint $\sf d_{2}$ such that
\begin{equation}
    \sf w_t( C_{I}[C[tell(b)\Par  H\Par tell(\ol{t}=\ol{s})]],c,d_{2})
    \leq  w_t( C_{I}[C[p(\ol{s})\Par tell(\ol{t}=\ol{s})]],c,d_{1})
    \leq n
    \label{eq:app18nov1}
\end{equation}
and
\begin{equation}
     \sf \exists_{-\Var( C_{I}[C[p(\ol{s})\Par
     tell(\ol{t}=\ol{s})]],c)}d_{2} =
     \exists_{-\Var( C_{I}[C[p(\ol{s})\Par
     tell(\ol{t}=\ol{s})]],c)}d_{1}
         \label{eq:app14ott5}
\end{equation}
hold. By definition of weight, we can assume that
$\sf \Var(d_{1}) \subseteq
\Var(C_{I}[C[p(\ol{s})\Par tell(\ol{t}=\ol{s})]],c)$ and
therefore, we have that
$\sf \Var(b)
\cap \Var(C_{I}[C[p(\ol{s})\Par tell(\ol{t}=\ol{s})]],c,d_{1})
\subseteq \Var(\ol{s})$.

We have now two cases:

\emph{1)}\ \ $\sf {\cal D} \models \exists_{-\Var(\ol{s})} d_{1} \ra
\exists_{-\Var(\ol{s})} b$.  In this case, by (\ref{eq:app14ott3}),
there exists a derivation
\[
\sf \xi =
\langle D_{0}.C_{I}[C[p(\ol{s})\Par tell(\ol{t}=\ol{s})]],c\rangle
\rightarrow^*
\langle D_{0}.B_{F},c_{F}\rangle,
\]
such that $\sf m(B_{F},c_{F})=t$, $wh(\xi)\leq \sf n$ and $$\sf
\exists_{-\Var( C_{I}[C[p(\ol{s})\Par
tell(\ol{t}=\ol{s})]],c)}c_{F}= \exists_{-\Var(
C_{I}[C[p(\ol{s})\Par tell(\ol{t}=\ol{s})]],c)}d_{1}.$$ By the
hypothesis on the variables, we can build a derivation
\[
\sf \chi = \langle D_{0}.
C_{I}[C[p(\ol{t})\Par tell(b) \Par tell(\ol{t}=\ol{s})]],c\rangle
\rightarrow^* \langle D_{0}.B'_{F},d_{3}\rangle
\]
which performs exactly the same steps of $\xi$, plus possibly a tell
action, such that $wh(\chi) \leq \sf n$,
${\sf m(B'_{F},d_{3})} = {\sf m(B_{F},c_{F})}$ and
\begin{equation}
        \sf \exists_{-\Var( C_{I}[C[p(\ol{s})\Par
        tell(\ol{t}=\ol{s})]],c)}d_{3}=
    \exists_{-\Var( C_{I}[C[p(\ol{s})\Par
        tell(\ol{t}=\ol{s})]],c)}d_{1}.
        \label{eq:app41nov99}
\end{equation}
Let $\sf d' = \exists_{-\Var(  C_{I}[C[p(\ol t) \Par tell(b)
\Par tell(\ol t = \ol s)]],c)} d_{3}$.
By the previous result and by definition of weight $ \sf w_t(
C_{I}[C[p(\ol{t})\Par tell(b) \Par tell(\ol{t}=\ol{s})]],c,d') \leq
n$.

Moreover, by hypothesis, $\sf \Var(c', d') \cap \Var(\ol r) \subseteq
\Var(C_{I}[C[p(\ol{t})]],c)$ and we can assume, without loss of
generality, that $\sf \Var(d_{1},d_{2}) \cap \Var(\ol r)\subseteq
\Var(C_{I}[C[p(\ol{t})]],c)$.  Then, by (\ref{eq:app14ott4}),
(\ref{eq:app41nov99}) and by definition of $\sf d'$,
it follows that
$\sf \exists_{-\Var(C_{I}[q(\ol{r})],c)}d' =
\exists_{-\Var( C_{I}[q(\ol{r})],c)}c'$ and then the thesis holds.\\

\emph{2)}\ \  $\sf {\cal D} \not \models \exists_{-\Var(\ol{s})} d_{1} \ra
\exists_{-\Var(\ol{s})} b$.
In this case, by (\ref{eq:app14ott5}),
$\sf {\cal D} \not \models \exists_{-\Var(\ol{s})} d_{2} \ra
\exists_{-\Var(\ol{s})} b$. By (\ref{eq:app18nov1})
this means that there exists a
derivation
\[
\sf \xi =
\langle D_{0}.C_{I}[C[tell(b)\Par  H\Par
tell(\ol{t}=\ol{s})]],c\rangle
\rightarrow^*
\langle D_{0}.B_{F},c_{F}\rangle
\not \rightarrow
\]
such that $\sf tell(b)\Par  H\Par
tell(\ol{t}=\ol{s})$ is not evaluated in $\xi$,
${\sf m(B_{F},c_{F}) =t}$, $wh(\xi)\leq \sf n$ and
\\
$\sf \exists_{-\Var( C_{I}[C[tell(b)\Par  H\Par
tell(\ol{t}=\ol{s})]],c)}c_{F} =
\exists_{-\Var( C_{I}[C[tell(b)\Par  H\Par tell(\ol{t}=\ol{s})]],c)}d_{2}$.
By definition, we can construct another derivation
\[
\sf \chi =
\langle D_{0}. C_{I}[C[p(\ol{t})\Par tell(b)\Par tell(\ol{t}=\ol{s})]],
c\rangle \rightarrow^*
\langle D_{0}.B'_{F},c_{F}\rangle \not \rightarrow
\]
which performs exactly the same steps of $\xi$ (and therefore
$wh(\chi)\leq \sf n$) and such that ${\sf m(B_{F},c_{F})
  =m(B'_{F},c_{F})}$.  Let $\sf d' = \exists_{-\Var( C_{I}[C[p(\ol t)
  \Par tell(b)
\Par tell(\ol t = \ol s)]],c)} c_{F}$.
By definition of derivation $$\sf \Var(c_{F}) \cap \Var(
C_{I}[C[tell(b)\Par H\Par tell(\ol{t}=\ol{s})]],c)\subseteq
\Var(C_{I}, C, c)$$
and therefore ${\sf \exists_{-\Var(
    C_{I}[C[tell(b)\Par H\Par tell(\ol{t}=\ol{s})]],c)}d'}$ = ${\sf
  \exists_{-\Var( C_{I}[C[tell(b)\Par H \Par
    tell(\ol{t}=\ol{s})]],c)}d_{2}}$. The remainder of the proof is
now analogous to that one of the previous case. \II

\NI {\bf Case 4:} Either $\sf cl'$ is the result of an ask simplification
or $\sf cl''$ is the result of a tell simplification.
The proof is analogous to that one given for
Case 4 of Proposition~\ref{pro:partial} and hence it is omitted.
\II

\NI
{\bf Case 5:} $\sf cl'$ is the result of a branch elimination or
of a conservative ask elimination.\\
The proof is straightforward by noting that:
(a) according to Definition~\ref{def:semantics} we consider also
inconsistent stores resulting from non-terminated computations;
(b) an ask action of the form $\sf
ask(true)$ always succeeds; (c) if we delete an
$\sf ask(true)$ action we obtain a derivation whose weight is smaller.
\II

\NI
{\bf Case 6:} $\sf cl'$ is the result of a distribution.\\
Let

- $\sf cl:\ q(\ol r) \la C[H \Par \sum_{j=1}^{n} ask(c_j) \rightarrow B_j]
\in D_{i}$

- $\sf cl':\   q(\ol r) \la  C[ \sum_{j=1}^{n} ask(c_j)
\rightarrow (H \Par B_j)]\in D_{i+1}$  \\

where $\sf e = pc(C[\ ])$
and for every constraint $\sf e'$ such that
$\sf \Var(e') \cap \Var(cl) \subseteq \Var(q(\ol r), C)$, if
$\sf  \langle D_{i}.H, e' \wedge e \rangle$ is productive
then both the following conditions hold:
\begin{itemize}
        \item
        there exists at least one $j\in [1,n]$ such that
        $\sf {\cal D}\models (e' \wedge e ) \rightarrow  c_j$

        \item  for each $j\in [1,n]$, either
        $\sf {\cal D}\models (e' \wedge e ) \rightarrow  c_j$
    or $\sf {\cal D}\models (e' \wedge e ) \rightarrow  \neg c_j$
\end{itemize}
We prove that, for any derivation
$$\sf \xi=\langle D_0.  C_{I}[C[H \Par \sum_{j=1}^{n} ask(c_j)
\rightarrow B_j]],c\rangle \rightarrow^* \langle D_0.B,d\rangle$$
with
${\sf m(B,d) \in \{ss,dd,ff\}}$, there exists a derivation
$$\sf\xi'= \langle D_0.C_{I}[C[\sum_{j=1}^{n} ask(c_j) \rightarrow (H
\Par B_j)]],c\rangle \rightarrow^* \langle D_0.B',d'\rangle$$
such that 
\[\sf \exists_{-\Var(C_{I} [C[H \Par \sum_{j=1}^{n} ask(c_j)
  \rightarrow B_j]],c)} d' = \exists_{-\Var( C_{I} [C[H \Par
  \sum_{j=1}^{n} ask(c_j) \rightarrow B_j]],c)} d
\] where also
$wh(\xi')\leq wh(\xi)$, and ${\sf m(B',d')}= {\sf m(B,d)}$. This
together with the definition of weight implies the thesis.

If $\sf H \Par \sum_{j=1}^{n}
ask(c_j) \rightarrow B_j$ is not evaluated in $\xi$,
then the proof is immediate.
Otherwise we have to distinguish two cases:

\emph{1)}\ \ There exists an $h \in [1,n]$, such that
\[
\begin{array}{ll}
   \sf \xi =
   & \sf \langle D_{0}.C_{I}[C[H \Par \sum_{j=1}^{n} ask(c_j)
   \rightarrow B_j]], c\rangle \rightarrow^*
   \langle D_{0}.C_{m}[H \Par \sum_{j=1}^{n}  ask(c_j)
   \rightarrow B_j], d_{m}\rangle \\
   & \sf \rightarrow  \langle D_{0}.C_{m}[H \Par B_h], d_{m}\rangle
   \rightarrow^* \langle D_{0}.B,d\rangle
\end{array}
\]
and $\sf {\cal D} \models d_{m} \ra c_{h}$.
In this case we can construct the derivation
\[
\begin{array}{ll}
   \sf \chi =
   & \sf \langle D_{0}.C_{I}[C[\sum_{j=1}^{n} ask(c_j)
   \rightarrow (H \Par B_j)]], c\rangle \\
   & \sf \rightarrow^*
   \langle D_{0}.C_{m}[\sum_{j=1}^{n}  ask(c_j)
   \rightarrow (H \Par B_j)], d_{m}\rangle \\
   & \sf \rightarrow \langle D_{0}.C_{m}[H \Par B_h], d_{m}\rangle
   \rightarrow^* \langle D_{0}.B,d\rangle
\end{array}
\]
which performs exactly the same steps of $\xi$ and then the thesis holds.

\emph{2)}\ \ $\xi$ is of the form
\[
\begin{array}{ll}
   \sf \xi =
   & \sf \langle D_{0}.C_{I}[C[H \Par \sum_{j=1}^{n} ask(c_j)
   \rightarrow B_j]], c\rangle \rightarrow^*
   \langle D_{0}.C_{m}[H \Par \sum_{j=1}^{n}  ask(c_j)
   \rightarrow B_j], d_{m}\rangle  \\
   & \rightarrow \sf \langle D_{0}.C_{m}[H' \Par \sum_{j=1}^{n}  ask(c_j)
   \rightarrow B_j], d_{m+1}\rangle \rightarrow^*
   \langle D_{0}.B,d\rangle.
\end{array}
\]
By Lemma~\ref{lem:proppc} and by definition of $\sf pc$,
we can construct another derivation
\[\begin{array}{ll}
    \sf \chi=
    &\sf \langle D_{0}.C_{I}[C[H \Par \sum_{j=1}^{n} ask(c_j)
    \rightarrow B_j]], c\rangle \rightarrow^*
    \langle D_{0}.C_{m}[H \Par \sum_{j=1}^{n}
    ask(c_j) \rightarrow B_j],d_{m}\rangle \\
    & \sf  \rightarrow^*
    \langle D_{0}.C_k[H  \Par \sum_{j=1}^{n} ask(c_j)
    \rightarrow B_j], d_k\rangle\rightarrow^*
    \langle D_{0}.B,d\rangle
\end{array}\]
which performs the same steps of $\xi$ (possibly in a different order)
and
such that the the agent $\sf H$ is not evaluated in the first $k$ steps,
where $\sf \Var(d_{k}) \cap \Var(cl) \subseteq \Var(q(\ol r), C)$
and $\sf {\cal D} \models d_{k} \ra e (= pc(C[\ ]))$.
Let $\sf \chi_{1}=
\langle D_{0}.C_k[H  \Par \sum_{j=1}^{n} ask(c_j)
\rightarrow B_j], d_k\rangle\rightarrow^*
\langle D_{0}.B,d\rangle$.
Now, if  $\sf  \langle D_{0}.H, d_{k} \rangle$ is not productive,
the proof is analogous to that one of Case 6 of
Proposition~\ref{pro:partial} and hence it is omitted.
Then assume that $\sf  \langle D_{0}.H, d_{k} \rangle$ is productive.
By definition of distribution there exists at least one $j\in [1,n]$
such that $\sf {\cal D}\models d_{k} \rightarrow  c_j$ and
for each $j\in [1,n]$, either
$\sf {\cal D}\models d_{k} \rightarrow  c_j$
or $\sf {\cal D}\models d_{k} \rightarrow  \neg c_j$.
Then, by definition, there exists a derivation
$\sf \xi_{1}=\langle D_{0}.C_k[\sum_{j=1}^{n} ask(c_j)
\rightarrow (H  \Par B_j)], d_k\rangle\rightarrow^*
\langle D_{0}.B,d\rangle$,
which performs the same steps of $\chi_{1}$ (possibly in a different
order).

Therefore there exists a derivation
\[\begin{array}{ll}
    \sf \xi'=
    &\sf \langle D_{0}.C_{I}[C[\sum_{j=1}^{n} ask(c_j)
    \rightarrow (H \Par B_j)]], c\rangle \rightarrow^*
    \langle D_{0}.C_{m}[\sum_{j=1}^{n}
    ask(c_j) \rightarrow (H \Par B_j)],d_{m}\rangle  \\
    & \rightarrow^* \sf \langle D_{0}.C_k[\sum_{j=1}^{n} ask(c_j)
    \rightarrow (H  \Par B_j)], d_k\rangle\rightarrow^*
    \langle D_{0}.B,d\rangle
\end{array}\]
which performs the same steps of $\chi$ (in a different order).
By construction $wh(\xi')=wh(\chi)=wh(\xi)$ and then the thesis
holds.
\II

\NI{\bf Case 7:} $\sf cl'$ is the result of a folding.\\
Let

- $\sf cl: \ q(\ol r) \la C[B]$ be the folded declaration ($\in \sf D_i$),

- $\sf f: \ p(\ol X) \la B$ be the folding declaration ($\in \sf
D_{0}$),

- $\sf cl': \ q(\ol r) \la C[p(\ol X)]$ be the result of the folding
operation $(\sf \in D_{i+1})$,
\\
where, by hypothesis, $\sf \Var(cl) \cap \Var(\ol X) \subseteq \Var(B)$,
$\sf \Var(B) \cap \Var(\ol r,C) \subseteq \Var(\ol X)$, $\sf
\Var(C[B],C[p(\ol X)]) \cap \Var(C_{I},c) \subseteq \Var(\ol r)$, $\sf
\Var(c') \cap \Var(\ol r) \subseteq \Var(C_{I}[C[B]],c)$ and there
exists $n$ such that $\sf w_t( C_{I}[C[B]],c,c')=n$.  Then,
\begin{equation}
        \sf \Var(B)  \cap \Var(C_{I}[C[\ ]],c)  \subseteq
        \Var(B)  \cap \Var(\ol r,C) \subseteq \Var(\ol X)
        \label{eq:app18nov3}
\end{equation}
and
\begin{equation}
        \sf \Var(c') \cap \Var(\ol r) \subseteq \Var(C_{I}[C[B]],c)
        \cap \Var(\ol r) \subseteq
        \Var(C_{I}[C[p(\ol X)]],c)
        \label{eq:app14ott10}
\end{equation}
hold. Moreover, we can assume without loss of generality that
$\sf \Var(c') \cap \Var(\ol X) \subseteq \Var(C_{I}[C[B]],c)$.\\
Since $\sf f \in \sf D_{0}$, from (\ref{eq:app18nov3})
and Point 2 of Lemma~\ref{lem:apppesiind0}
it follows that there exists a constraint $\sf d'$  such that
$\sf w_t( C_{I}[C [p(\ol X)]],c,d') \leq w_t( C_{I}[C[B]],c,c')$
and
\begin{equation}
        \sf \exists_{-\Var( C_{I}[C [p(\ol X)]],c)}d' =
        \exists_{-\Var( C_{I}[C [p(\ol X)]],c)}c'.
        \label{eq:app14ott11}
\end{equation}
We can assume, without loss of generality, that $\sf \Var(d')
\subseteq \Var(C_{I}[C[p(\ol X)]],c)$.
Then by using (\ref{eq:app14ott10}) and (\ref{eq:app14ott11}) we obtain that
$\sf \exists_{-\Var( C_{I}[q(\ol r)],c)} d' =
\exists_{-\Var( C_{I}[q(\ol r)],c)}c'$
which concludes the proof of {\bf a)}.\\

{\bf b)} Assume that the parts 1 and 2 of this Lemma hold for $i \geq 0$.
We prove that 1 holds for $i+1>0$.\\
Let $ \sf cl:\  q (\ol{r})\la H \in D_{i+1}$,
and let $ \sf \bar{cl}:\  q (\ol{r})\la \bar{H} $ be the corresponding
declaration in $\sf D_{i}$. Moreover let $\sf C_{I}[\ ]$ be a context,
$\sf c$ a satisfiable constraint and let $\sf c'$ be a constraint,
such that $\sf \Var(H) \cap \Var(C_{I},c) \subseteq \Var(\ol r)$
and  $\sf w_t( C_{I}[q (\ol{r}) ],c,c')$ is defined.
Without loss of generality, we can assume that
$\sf \Var(\bar{H}) \cap \Var(C_{I},c) \subseteq \Var(\ol r)$.
Then, since by inductive hypothesis, part 1 holds for $i$,
there exists a constraint $\sf d_{1}$ such that
$\sf \Var(d_{1}) \subseteq \Var(C_{I}[\bar{H}],c)$,
\begin{equation}
        {\sf w_t( C_{I}[\bar{H}],c,d_{1}) \leq  w_t( C_{I}[q (\ol{r})
],c,c')}
        \mbox{ and }
        {\sf \exists_{-\Var( C_{I}[q (\ol{r}) ],c)}d_{1} =
       \exists_{-\Var( C_{I}[q (\ol{r}) ],c)}c'}.
        \label{eq:app42nov99}
\end{equation}
Since by inductive hypothesis part 2 holds for $i$,
there exists a  constraint $\sf d'$, such that
$\sf \Var(d') \subseteq \Var(C_{I}[H],c)$,
$\sf w_t( C_{I}[H],c,d') \leq  w_t( C_{I}[\bar{H}],c,d_{1})$
and $\sf \exists_{-\Var( C_{I}[q (\ol{r}) ],c)}d' =
\exists_{-\Var( C_{I}[q (\ol{r}) ],c)}d_{1}$.\\
By (\ref{eq:app42nov99}) we obtain $\sf w_t( C_{I}[H],c,d') \leq
w_t( C_{I}[q (\ol{r}) ],c,c')$ and $$\sf \exists_{-\Var( C_{I}[q
(\ol{r}) ],c)}d' = \exists_{-\Var( C_{I}[q (\ol{r}) ],c)}c'$$ and
then the thesis holds.\end{proof}

\begin{lemma}
   \label{lem:apppesideri}
   Let $0 \leq i \leq n$,  $\sf c_1, c_{m}$ satisfiable constraints,
   $\sf c_{k}$ a constraint and assume that there exists a derivation
   $\xi:\ \sf \langle D_i.A_1,c_1\rangle \rightarrow^{*}
   \langle D_i.A_m,c_m\rangle \rightarrow ^{*} \langle
   D_i.A_{k},c_{k}\rangle$, such that
   \begin{description}
      \item[i)] in the first $m-1$ steps of $\xi$ rule ${\bf R2}$
      is used only for evaluating agents of the
      form $\sf ask(c)\rightarrow B$,
      \item[ii)] $\sf w_t(A_{1},c_{1},c_{k})$ is defined
      (for ${\sf t} ={\sf m(A_{k},c_{k})\in \{ss,dd,ff\}}$).
   \end{description}
   Then there exists a constraint $\sf c'$ such that
   $\sf \Var(c') \subseteq \Var(A_{m},c_{m})$,
   $\sf \exists_{-\Var(A_{1},c_{1})}c_{k}
   =\exists_{-\Var(A_{1},c_{1})}c'$ and
   $\sf w_t( A_{m},c_{m},c') \leq w_t(A_{1},c_{1},c_{k})$.
\end{lemma}
\begin{proof} We prove the thesis for one derivation step.
Then the proof of the Lemma follows by using a straightforward
inductive argument.  Assume that $\sf c_1, c_{2}$ are satisfiable
constraints, $\sf c_{k}$ is a constraint
and that there exists a derivation
\[
\sf \langle D_i.A_{1},c_{1}\rangle \rightarrow
\langle  D_i.A_{2},c_{2}\rangle\rightarrow ^{*}
\langle D_i.A_{k},c_{k}\rangle
\]
such that ${\sf m(A_{k},c_{k})\in \{ss,dd,ff\}}$ and the first step
can use rule ${\bf R2}$ only for evaluating agents
of the form $\sf ask(c)\rightarrow B$.
By the definition of derivation we have $\sf A_{1} =C_{1}[A]$,
where $\sf C_{1}[\ ]$ is not a guarding context.
We have now three cases:

\emph{1)}\ \  $\sf A =tell(c)$. In this case
\[
\sf \langle D_i.C_{1}[tell(c)],c_{1}\rangle \rightarrow
\langle  D_i.C_{1}[Stop],c_{1} \wedge c\rangle\rightarrow ^{*}
\langle D_i.A_{k},c_{k}\rangle.
\]
Since $\sf C_{1}[\ ]$ is not a guarding context the definition of weight
implies that
$$\sf w_t( C_{1}[Stop],c_{1} \wedge c,
\exists_{-\Var(C_{1}[Stop],c_{1} \wedge c)}c_{k}) =
w_t(C_{1}[tell(c)],c_{1},c_{k})$$
where ${\sf t} = {\sf
  m(A_{k},c_{k})}$. Then the thesis holds

\emph{2)}\ \  $\sf A =q(\ol v)$ and there exists a declaration
$\sf cl: \ q(\ol r) \la B \in D_i$.  In this case
\[
\sf \langle D_i.C_{1}[q(\ol v)],c_{1}\rangle \rightarrow
\langle  D_i.C_{1}[B \Par tell(\ol v = \ol r)],c_{1} \rangle
\rightarrow ^{*} \langle D_i.A_{k},c_{k}\rangle.
\]
>From the definition of derivation it follows that
$\sf \Var(C_{1}[q (\ol{v})],c_{1})
\cap \Var(q (\ol{r})) =\emptyset$.
Furthermore, by definition of transformation sequence,
there exists a declaration $\sf q (\ol{r})\la H \in D_{0}$.
Since $\sf w_t( C_{1}[q (\ol{v})],c_{1},c_{k})$ is defined by
hypothesis (where ${\sf t} = {\sf m(A_{k},c_{k})}$),
from Lemma~\ref{lem:apppesoclausola} it follows that
there exists a constraint $\sf d'$ such that
$\sf w_t( C_{1}[q (\ol{r}) \Par tell(\ol{v}=\ol{r})],c_{1},d')
\leq w_t( C_{1}[q (\ol{v})],c_{1},c_{k})$ and 
$\sf \exists_{-\Var( C_{1}[q (\ol{v}) ],c_{1})}d'
$ $=$ $\sf\exists_{-\Var( C_{1}[q (\ol{v}) ],c_{1})}c_{k}$.

>From the definition of derivation it follows that
$\sf \Var(B) \cap \Var(C_{1}[\  \Par
tell(\ol{v}=\ol{r})],c_{1}) \subseteq \Var(\ol r)$.
Part~\ref{pt:apppesiindi1} of Lemma~\ref{lem:apppesiindi} implies that there
exists a constraint $\sf c'$ such that
${\sf \Var(c') \subseteq
\Var(C_{1}[B\Par tell(\ol{v}=\ol{r})],c_1)}$,
$\sf w_t(C_{1}[B\Par tell(\ol{v}=\ol{r})],c_{1},c') \leq
w_t( C_{1}[q (\ol{r}) \Par tell(\ol{v}=\ol{r})],c_{1},d')$
and
\[\sf \exists_{-\Var( C_{1}[q (\ol{r}) \Par
tell(\ol{v}=\ol{r})],c_{1})}c' =
\sf \exists_{-\Var( C_{1}[q (\ol{r}) \Par tell(\ol{v}=\ol{r})],c_{1})}d'.
\]
These results together with the inclusion $\sf \Var( C_{1}[q
(\ol{v}) ],c_{1}) \subseteq \Var( C_{1}[q (\ol{r}) \Par
tell(\ol{v}=\ol{r})],c_{1})$ imply that $\sf w_t(C_{1}[B\Par
tell(\ol{v}=\ol{r})],c_{1},c') \leq w_t( C_{1}[q (\ol{v})
],c_{1},c_{k})$ and $$\sf \exists_{-\Var( C_{1}[q (\ol{v})
],c_{1})}c' = \exists_{-\Var( C_{1}[q (\ol{v}) ],c_{1})}c_{k},$$
thus concluding the proof for this case.

\emph{3)}\ \
$\sf A = ask(c)\rightarrow B$ and $\sf {\cal D} \models c_{1} \ra c$.
In this case
\[
\sf \langle D_i.C_{1}[ask(c)\rightarrow B],c_{1}\rangle \rightarrow
\langle  D_i.C_{1}[B],c_{1} \rangle\rightarrow ^{*}
\langle D_i.A_{k},c_{k}\rangle.
\]
Since $\sf C_{1}[\ ]$ is not a guarding context and $\sf {\cal
D} \models c_{1} \ra c$ we obtain $$\sf w_t( C_{1}[B],c_{1},
\exists_{-\Var(C_{1}[B],c_{1})}c_{k}) \leq
w_t(C_{1}[ask(c)\rightarrow B],c_{1},c_{k})$$ where ${\sf t} =
{\sf m(A_{k},c_{k})}$, which concludes the proof.
\end{proof}

We need one last lemma.

\begin{lemma}
    Let $\sf c$ be a satisfiable constraint, $\sf A$ be the agent
    $\sf A_{1}\Par \ldots \Par A_{l}$,
    where for any $j \in [1,l]$ either $\sf A_{j}$ is a choice agent
    or $\sf A_{j}= Stop$ and assume there exists a split
    derivation $\sf \nu $ in $\sf D_0$,
    \[
    \sf \nu = \langle D_{0}.A, c\rangle \rightarrow
    \langle D_{0}.A', c'\rangle \rightarrow^*
    \langle D_{0}.B,d \rangle,
    \]
    where ${\sf m(B,d)} \in {\sf \{ss,dd,ff\}}$.
    Then
    $\sf \langle D_{i}.A, c\rangle \rightarrow
    \langle D_{0}.A', c'\rangle \rightarrow^*
    \langle D_{0}.B,d  \rangle$ is a split derivation in
    $\sf D_{i} \cup D_0$.
    \label{lem:appdasplitasplit}
\end{lemma}
\begin{proof}

The proof is straightforward, by observing that by the
hypothesis on $\sf A$ the first step of $\sf \nu $ uses the rule
${\bf R2}$ (in case such a step exists) and therefore, by
definition of split derivation, $\sf w_{t}(A,c,d) >
w_{t}(A',c',d)$, where ${\sf t} = {\sf m(B,d)}$. Then by
definition, $\sf \langle D_{i}.A, c\rangle \rightarrow \langle
D_{0}.A', c'\rangle \rightarrow^* \langle D_{0}.B,d  \rangle$ is
a split derivation in $\sf D_{i} \cup D_0$.
\end{proof}

We can now prove our main theorem.
\begin{theorem}[\ref{thm:correctness} (Total Correctness)] Let
$\sf D_0, \ldots, D_n$ be a transformation sequence. Then, for
any agent $\sf A$,
\begin{itemize}
\item
$\sf {\cal O}(D_0.A) = {\cal O}(D_n.A)$.
\end{itemize}
\end{theorem}

\begin{proof}  The proof proceeds by showing simultaneously,
by induction on $i$, that for $i \in [0,n]$:
\begin{enumerate}
    \item for any agent $\sf A$, $\sf {\cal O}(D_0.A) = {\cal O}(D_i.A)$;
    \item $\sf D_i$ is weight complete.
\end{enumerate}

\NI \emph{Base case}. We just need to prove that $\sf D_0$ is
weight complete.  Assume that there exists a derivation
$\sf \langle D_0.A,c_{I}\rangle \rightarrow^*
\langle D_0.B,c_{F}\rangle$,
where $\sf c_{I}$ is a satisfiable constraint and
${\sf m(B,c_{F})} \in {\sf \{ss,dd,ff\}}$.
Then there exists a derivation
$\sf \xi:\ \langle D_0.A,c_{I}\rangle \rightarrow^*
\langle D_0.B',c'_{F}\rangle $,
such that ${\sf m(B',c'_{F})} = {\sf m(B,c_{F})}$,
whose weight is minimal and where
$\sf \exists_{-\Var(A,c_{I})} c'_{F} =
\exists_{-\Var(A,c_{I})} c_{F}$.
It follows from Definition~\ref{def:splitder}
that $\xi$ is a split derivation.
\II

\NI \emph{Induction step}.

\NI By the inductive hypothesis for any agent $\sf A$, $\sf {\cal
  O}(D_0.A) = {\cal O}(D_{i-1}.A)$ and $\sf D_{i-1}$ is weight
complete.  From propositions~\ref{pro:partial} and \ref{pro:total1} it
follows that if $\sf D_{i}$ is weight complete then for any agent $\sf
A$, $\sf {\cal O}(D_0.A) = {\cal O}(D_{i}.A)$.  So, in order to prove
parts 1 and 2, we only have to show that $\sf D_{i}$ is weight
complete.

Assume then that there exists a derivation
$\sf \langle D_0.A,c_{I}\rangle \rightarrow^*
\langle D_0.B,c_{F}\rangle$ such that
$\sf c_{I}$ is a satisfiable constraint and
${\sf m(B,c_{F})} \in {\sf \{ss,dd,ff\}}$.
>From the inductive hypothesis it follows that there exists a split
derivation
\[\sf \chi=
\langle D_{i-1}.A,c_{I}\rangle \rightarrow^*
\langle D_{i-1}.A_m,c_m\rangle \rightarrow
\langle D_{0}.A_{m+1},c_{m+1}\rangle \rightarrow^*
\langle D_0.B'',c''_{F} \rangle
\]
where
\begin{equation}
        {\sf \exists_{-\Var(A,c_{I})}c''_{F} =
        \exists_{-\Var(A,c_{I})}c_{F}}
        \mbox{ and }
        {\sf m(B'',c''_{F})} = {\sf m(B,c_{F})}.
        \label{eq:app7ott6}
\end{equation}
Let $\sf d \in D_{i-1} \backslash D_{i}$ be the modified clause in the
transformation step from $\sf D_{i-1}$ to $\sf D_{i}$.

If in the
first $m$ steps of $\chi$ there is no procedure call which uses $\sf d$
then clearly there exists a split derivation $\xi$ in $\sf D_{i} \U D_{0}$,

$\sf \xi=\langle D_{i}.A,c_{I}\rangle \rightarrow^*
\langle D_{i}.A_m,c_m\rangle \rightarrow
\langle D_{0}.A_{m+1},c_{m+1}\rangle
\rightarrow^* \langle D_0.B'',c''_{F} \rangle$
which performs the same steps of $\chi$ and then the thesis holds.

Otherwise, assume without loss of generality that ${\bf R4}$
is the rule used in the first step of derivation $\chi$ and that
$\sf d$ is the clause employed in the first step of $\chi$.
We also assume that the declaration $\sf d$ is used only once in $\chi$,
since
the extension to the general case is immediate.

We have to distinguish various cases according to what happens to the
clause $\sf d$ when moving from  $\sf D_{i-1}$ to $\sf D_{i}$.
\II

\NI
{\bf Case 1:} $\sf d$ is unfolded.\\
Let $\sf d'$ be the corresponding declaration in $\sf D_i$.
The situation is the following:

- $\sf d: \  \sf q (\ol{r})\la C[p(\ol{t})] \in D_{i-1}$,

- $\sf u: \   \sf p (\ol{s}) \la  H \in D_{i-1}$, and

- $\sf d': \  \sf q (\ol{r})\la C[H\Par tell(\ol t = \ol s)]
\in  D_{i}$,\\
where $\sf d$ and $\sf u$ are assumed to be renamed apart.  By the
definition of split derivation, $\chi$ has the form
\[\begin{array}{ll}
      \sf \langle D_{i-1}.C_I[q(\ol v)],c_I\rangle \rightarrow
      \langle D_{i-1}.C_I[C[p(\ol t)]\Par tell(\ol{v}=\ol{r})],
      c_I\rangle \rightarrow^*
      \langle D_{i-1}.A_m,c_m\rangle \rightarrow \\
      \sf \langle D_0.A_{m+1},c_{m+1}\rangle \rightarrow^*
      \langle D_0.B'',c''_F\rangle.
\end{array}\]
Without loss of generality, we can assume that $\sf \Var(\chi)\cap
\Var(u) \neq \emptyset$ if and only if $\sf p(\ol t)$ is evaluated in
the first $m$ steps of $\chi$, in which case $\sf u$ is used for
evaluating it.  We have to distinguish two cases.

\emph{1)}\ \ There exists $k<m$ such that the $k$-th derivation step
of $\chi$ is the procedure call $\sf p(\ol t)$.  In this case $\chi$
has the form
\[\begin{array}{ll}
      \sf \langle D_{i-1}.C_I[q(\ol v)],c_I\rangle \rightarrow
      \langle D_{i-1}.C_I[C[p(\ol t)]\Par tell(\ol{v}=\ol{r})],
      c_I\rangle \rightarrow^*
      \langle D_{i-1}.C_k[p(\ol t)], c_k\rangle\rightarrow \\
      \sf \langle D_{i-1}.C_k[H\Par tell(\ol t = \ol s)],
      c_k\rangle
      \rightarrow^* \langle D_{i-1}.A_m,c_m\rangle \rightarrow
      \langle D_0.A_{m+1},c_{m+1}\rangle \rightarrow^*
      \langle D_0.B'',c''_F\rangle.
\end{array}\]
Then there exists a corresponding derivation in
$\sf D_{i} \U  D_{0}$
\[\begin{array}{ll}
      \sf \xi=& \sf
      \langle D_{i}.C_I[q(\ol v)],c_I\rangle \rightarrow
      \langle D_{i}.C_I[C[H\Par tell(\ol t = \ol s)]\Par
      tell(\ol{v}=\ol{r})], c_I\rangle \rightarrow^* \\
     &\sf  \langle D_{i}.C_k[H\Par tell(\ol t = \ol s)], c_k\rangle
      \rightarrow^*\\
      &\sf \langle D_{i}.A_m,c_m\rangle \rightarrow
      \langle D_0.A_{m+1},c_{m+1}\rangle \rightarrow^*
      \langle D_0.B'',c''_F\rangle,
\end{array}\]
which performs exactly the same steps of $\chi$ except for a
procedure call to $\sf p(\ol t)$.
In this case the proof follows by observing that, since by the
inductive hypothesis  $\chi$ is a split derivation,
the same holds for $\xi$.

\emph{2)}\ \
There is no procedure call to $\sf p(\ol t)$ in the first $m$
steps. Therefore $\chi$ has the form
\[\begin{array}{ll}
      \sf \langle D_{i-1}.C_I[q(\ol v)],c_I\rangle \rightarrow
      \langle D_{i-1}.C_I[C[p(\ol t)]\Par tell(\ol{v}=\ol{r})],
      c_I\rangle  \rightarrow^*
      \langle D_{i-1}.C_m[p(\ol t)], c_m\rangle\rightarrow\\
      \sf \langle D_{0}.C_{m+1}[p(\ol t)], c_m\rangle\rightarrow^*
      \langle D_0.B'',c''_F\rangle.
\end{array}\]
Then, by the definition of $\sf D_{i}$, there exists a derivation
\[\begin{array}{ll}
     \sf \xi_{0}= & \sf \langle D_{i}.C_I[q(\ol v)],c_I\rangle
     \rightarrow
     \langle D_{i}.C_I[C[H\Par tell(\ol t = \ol s)]\Par
     tell(\ol{v}=\ol{r})], c_I\rangle \rightarrow^* \\
     &\sf\langle D_{i}.C_m[H\Par tell(\ol t = \ol s)],
     c_m\rangle\rightarrow
     \langle D_{0}.C_{m+1}[H\Par tell(\ol t = \ol s)], c_m\rangle.
\end{array}\]

Observe that from the derivation
$\sf \langle D_{0}.C_{m+1}[p(\ol t)], c_m\rangle\rightarrow^*
\langle D_0.B'',c''_F\rangle$ and (\ref{eq:app7ott6}) it follows that
\begin{equation}
        {\sf w_t(C_{m+1}[p(\ol t)],c_{m},c''_F)}
        \mbox{ is defined, where }
        {\sf t} ={\sf m(B,c_{F})}.
        \label{eq:app23nov199}
\end{equation}
The hypothesis on the variables implies that $\sf \Var
(C_{m+1}[p(\ol t)], c_m) \cap \Var(u) =\emptyset$. Then, by the
definition of transformation sequence and since $\sf u \in
D_{i-1}$, there exists a declaration $\sf p (\ol{s}) \la H_{0}
\in D_{0}$.  By Lemma~\ref{lem:apppesoclausola} and part 1 of
Lemma~\ref{lem:apppesiindi} it follows that there exists a constraint $\sf
d_{F}$
such that
\begin{equation}
    \sf w_t(C_{m+1}[H\Par tell(\ol t = \ol s)],c_{m},d_F) \leq
        w_t(C_{m+1}[p(\ol t)],c_{m},c''_F)
    \label{eq:app7ott3}
\end{equation}
and
\begin{equation}
    \sf\exists_{-\Var(C_{m+1}[p(\ol t)],c_{m})} d_F =
    \exists_{-\Var(C_{m+1}[p(\ol t)], c_{m})} c''_F .
        \label{eq:app7ott1}
\end{equation}
Therefore, by the definition of $\sf w_t$, by (\ref{eq:app7ott3}) and
since  $\sf w_t(C_{m+1}[p(\ol t)],c_{m},c''_F)$ is defined,
there exists a derivation
\[
\sf \xi_{1} =
\langle D_0.C_{m+1}[H\Par tell(\ol t = \ol s)],c_{m}\rangle
\rightarrow^* \langle D_0.B',c'_F\rangle,
\]
where
$\sf \exists_{-\Var(C_{m+1}[H\Par
tell(\ol t = \ol s)], c_{m})} c'_{F} =
\exists_{-\Var(C_{m+1}[H\Par
tell(\ol t = \ol s)], c_{m})} d_{F}$
and, by (\ref{eq:app23nov199}),
\begin{equation}
        {\sf  m(B',c'_F)} ={\sf m(B,c_{F})}.
        \label{eq:app23nov299}
\end{equation}
By (\ref{eq:app7ott1})
\begin{equation}
      {\sf \exists_{-\Var(C_{m+1}, c_{m})} c'_{F} =
      \exists_{-\Var(C_{m+1}, c_{m})} c''_{F}}
      \label{eq:app7ott7}
\end{equation}
holds and, by definition of weight, we obtain
\begin{equation}
     \sf w_t(C_{m+1}[H\Par tell(\ol t = \ol s)], c_{m},c'_{F})
     =  w_{t}(C_{m+1}[H\Par tell(\ol t = \ol s)], c_{m},  d_{F}) .
     \label{eq:app7ott2}
\end{equation}
Moreover, we can assume without loss of generality that
$\sf \Var(\xi_{0}) \cap \Var(\xi_{1})=
\Var(C_{m+1}[H\Par tell(\ol t = \ol s)],c_{m})$.
Then, by the definition of procedure call
\begin{equation}
    \sf \Var(C_I[q(\ol v)],c_I) \cap (\Var(c'_{F})
    \cup \Var(c''_{F}))
    \subseteq \Var(C_{m+1},c_{m})
    \label{eq:app7ott8}
\end{equation}
and there exists a derivation
\[\begin{array}{ll}
     \sf \xi=&\sf \langle D_{i}.C_I[q(\ol v)],c_I\rangle
     \rightarrow \langle D_{i}.C_I[C[H\Par tell(\ol t = \ol s)]
     \Par tell(\ol{v}=\ol{r})],
     c_I\rangle \rightarrow^*\\
     &\sf \langle D_{i}.C_m[H\Par tell(\ol t = \ol s)],
     c_m\rangle\rightarrow
     \langle D_{0}.C_{m+1}[H\Par tell(\ol t = \ol s)], c_m\rangle
     \rightarrow^* \langle D_0.B',c'_F\rangle
\end{array}\]
such that the first $m-1$ derivation steps do not use rule ${\bf R2}$
and the $m$-th derivation step uses the rule ${\bf R2}$.
Now, we have the following equalities
\[\begin{array}{lll}
     \sf \exists_{-\Var(C_I[q(\ol v)],c_I)} c'_{F} & = &
     \mbox{(by (\ref{eq:app7ott8}) and by construction)}  \\
     \sf \exists_{-\Var(C_I[q(\ol v)],c_I)} (c_{m} \wedge
     \exists_{-\Var(C_{m+1},c_{m})}c'_{F}) & = &
     \mbox{(by (\ref{eq:app7ott7}))}  \\
     \sf \exists_{-\Var(C_I[q(\ol v)],c_I)} (c_{m} \wedge
     \exists_{-\Var(C_{m+1},c_{m})}c''_{F}) & = &
     \mbox{(by (\ref{eq:app7ott8}) and by construction)}  \\
     \sf \exists_{-\Var(C_I[q(\ol v)],c_I)} c''_{F} & = &
     \mbox{(by the first statement in (\ref{eq:app7ott6}))}\\
     \sf \exists_{-\Var(C_I[q(\ol v)],c_I)} c_{F}.
\end{array}
\]
By the definition of weight,
$\sf w_t(C_I[q(\ol v)],c_I,c'_F)=w_t(C_I[q(\ol v)],c_I,c''_F)$,
by (\ref{eq:app7ott2}) and (\ref{eq:app7ott3}),
$\sf w_t(C_{m+1}[H\Par tell(\ol t = \ol s)],c_{m},c'_F) \leq
w_t(C_{m+1}[p(\ol t)],c_{m},c''_F)$
and $\sf w_t(C_{m+1}[p(\ol t)],c_{m},c''_F) <
w_t(C_I[q(\ol v)],c_I,c''_F)$, since $\chi$ is a split derivation.
Therefore $\sf w_t(C_{m+1}[H\Par tell(\ol t = \ol s)],c_{m},c'_F) <
w_t(C_I[q(\ol v)],c_I,c'_F)$ and then,
by definition, $\xi$ is a split derivation in $\sf D_{i} \cup
D_{0}$. This, together with (\ref{eq:app23nov299}), implies the thesis.
\II

\NI {\bf Case 2:} A tell constraint in $\sf d$ is eliminated or
introduced.
\\
In the first case, let $\sf d'$ be the corresponding declaration in
$\sf D_i$.  Therefore the situation is the following:

- $\sf d: \  \sf q (\ol{r})\la C[tell(\ol{s}=\ol{t}) \Par H]$

- $\sf d': \   \sf q (\ol{r}) \la  C[H\sigma]$\\
where $\sigma$ is a relevant most general unifier of $\sf s$ and
$\sf t$ and the variables in the domain of $\sigma$ do not occur neither
in $\sf C[\ ]$ nor in $\sf \ q(\ol r)$.  Observe that for any derivation
which uses the declaration $\sf d$, we can construct another derivation
such that the agent $\sf tell(\ol{s}=\ol{t})$ is evaluated
before $\sf H$.  Then the thesis follows from Lemma~\ref{lem:apppesiindi}
and from the argument used in the proof of Case 2 of
Proposition~\ref{pro:partial}.
The proof for the tell introduction is analogous and hence it is omitted.
\II

\NI {\bf Case 3:} $\sf d$ is  backward instantiated.\\
Let $\sf d'$ be the corresponding declaration in $\sf D_i$. The
situation is the following:

- $\sf d: \ q(\ol r) \la C[p(\ol t)] \in D_{i-1}$,

- $\sf d': \ q(\ol r) \la C[p(\ol t) \Par tell(b)\Par tell(\ol
t = \ol s)] \in D_i$,\\
where $\sf c: \ p(\ol s) \la tell(b)\Par B \in D_{i-1}$ has
no variable in common with $\sf d$ (the case  $\sf d': \
q(\ol r) \la C[p(\ol t) \Par tell(\ol t = \ol s)]$
is analogous and hence omitted).
We distinguish two cases:

\emph{1)}\ \ There is no procedure call to $\sf p(\ol t)$
in the first $m$ steps. Therefore $\chi$ has the form
\[\begin{array}{ll}
     \sf \langle D_{i-1}.C_I[q(\ol v)],c_I\rangle \rightarrow
     \langle D_{i-1}.C_I[C[p(\ol t)]\Par tell(\ol{v}=\ol{r})], c_I\rangle
     \rightarrow^*
     \langle D_{i-1}.C_m[p(\ol t)], c_m\rangle\rightarrow\\
     \sf \langle D_{0}.C_{m+1}[p(\ol t)], c_m\rangle\rightarrow^*
     \langle D_0.B'',c''_F\rangle.
\end{array}\]
Without loss of generality, we can assume that $\sf \Var(\chi)\cap
\Var(p(\ol t) \Par tell(b)\Par tell(\ol t = \ol s))=\Var(\ol t)$.
Then, by the definition of $\sf D_{i}$, there exists a derivation
corresponding to $\chi$,
\[\begin{array}{ll}
     \sf \xi_{0}= & \sf \langle D_{i}.C_I[q(\ol v)],c_I\rangle
     \rightarrow  \langle D_{i}.C_I[C[p(\ol t) \Par tell(b)
     \Par tell(\ol t = \ol s)] \Par tell(\ol{v}=\ol{r})], c_I\rangle
     \rightarrow^*\\
     &\sf\langle D_{i}.C_m[p(\ol t) \Par tell(b)\Par
     tell(\ol t = \ol s)], c_m\rangle\rightarrow
     \\
     &\sf\langle D_{0}.C_{m+1}[p(\ol t) \Par tell(b)\Par
     tell(\ol t = \ol s)], c_m\rangle.
\end{array}\]
Following the same reasoning as in Case 3 of Lemma~\ref{lem:apppesiindi},
we can prove that there exists a constraint $\sf d_{F}$ such that
\[
\sf w_t(C_{m+1}[p(\ol t) \Par tell(b)\Par tell(\ol t = \ol
s)],c_{m},d_{F})
\leq  w_t(C_{m+1}[p(\ol t)],c_{m},c''_F)
\]
where $\sf \exists_{-\Var(C_{m+1}[p(\ol t)], c_{m})} d_{F} =
\exists_{-\Var(C_{m+1}[p(\ol t)],c_{m})}c''_F $ and
$\sf t=m(B'',c''_F)$.  The rest of the proof is
analogous to Case 1 (unfolding) and hence it is omitted.

\emph{2)}\ \ There is exists $k<m$ such that the $k$-th derivation
step of $\chi$ is the procedure call $\sf p(\ol t)$.  We
distinguish two more cases:

\emph{2a)}\ \ $\sf p \neq q$.  In this case we can assume, without
loss of generality, that $\chi$ has the form
\[\begin{array}{ll}
      \sf \langle D_{i-1}.C_I[q(\ol v)],c_I\rangle \rightarrow
      \langle D_{i-1}.C_I[C[p(\ol t)]\Par tell(\ol{v}=\ol{r})], c_I\rangle
      \rightarrow^*
      \langle D_{i-1}.C_k[p(\ol t)], c_k\rangle
      \rightarrow
      \\
      \sf \langle D_{i-1}.C_k[tell(\bar b)\Par  \bar B
      \Par tell(\ol t = \ol s')], c_k\rangle \rightarrow^*
      \langle D_{i-1}.A_m, c_m\rangle\rightarrow
      \\
      \sf \langle D_{0}.A_{m+1}, c_m\rangle\rightarrow^*
      \langle D_0.B'',c''_F\rangle
\end{array}\]
where $\sf c'= p(\ol s') \la tell(\bar b)\Par \bar B$ is a renaming of
$\sf c$ such that $\sf \Var(c') \cap \Var(d')=\emptyset$.  In this
case there exists a derivation
\[\begin{array}{ll}
     \sf \langle D_{i}.C_I[q(\ol v)],c_I\rangle \rightarrow
     \langle D_{i}.C_I[C[p(\ol t) \Par tell(b)\Par
     tell(\ol t = \ol s)] \Par tell(\ol{v}=\ol{r})], c_I\rangle
     \rightarrow^*\\
     \sf\langle D_{i}.C_k[tell(\bar b)\Par  \bar B \Par
     tell(\ol t = \ol s') \Par tell(b)\Par
     tell(\ol t = \ol s)], c_k \rangle.
\end{array}\]
Observe now that, given any set of declarations,
if there exists a derivation $\sf \chi'$ for the configuration
$\sf \langle C'[tell(\bar b)\Par \bar B \Par tell(\ol t = \ol s')], c'
\rangle$ where $\sf c'$ is satisfiable and
$\sf \Var(C', c') \cap \Var(b, \ol s) = \emptyset$,
then there exists a derivation for
$\sf \langle C'[tell(\bar b)\Par \bar B \Par tell(\ol t = \ol s')
\Par tell(b)\Par tell(\ol t = \ol s)], c'\rangle$
which performs the same steps of $\chi'$ plus (possibly) two steps
corresponding to the evaluation of $\sf tell(b)$ and $\sf tell(\ol t =
\ol s)$.  Since $\sf (\ol t = \ol s') \wedge (\ol t = \ol s)$ is
logically equivalent to $\sf (\ol t = \ol s') \wedge (\ol s' = \ol
s)$, we can substitute $\sf tell(\ol t = \ol s') \Par tell(\ol t = \ol
s)$ for $\sf tell(\ol t = \ol s') \Par tell(\ol s' = \ol s)$.
Moreover, since $\sf p(\ol s') \la tell(\bar b)\Par \bar B$ is a
renaming of $\sf c$ and therefore $\sf {\cal D} \models (\bar b \wedge
(\ol s' = \ol s)) \ra b$ holds, we can drop the agent $\sf tell(b)$.

Finally, observe that $\sf \ol s' = \ol s$ can be reduced to a
conjunction of equations of the form $\sf \ol X = \ol Y$,
where $\sf \ol X \subseteq \Var(\ol s)$ and
$\sf \ol Y \subseteq \Var(\ol s')$ are distinct variables.
Therefore, we can drop the constraint
$\sf tell(\ol s' = \ol s)$, since the declarations used in
the derivation are renamed apart and
$\sf \Var(C'[tell(\bar b)\Par \bar B \Par
tell(\ol t = \ol s')], c') \cap \Var(\ol s) = \emptyset$.
Then the thesis holds for this case.

\emph{2b)}\ \ $\sf p = q$.  In this case, the situation is the
following:

- $\sf d: \ p(\ol r) \la tell(b') \Par C''[p(\ol t)] \in D_{i-1}$,

- $\sf d': \ p(\ol r) \la tell(b' ) \Par
C''[p(\ol t) \Par tell(b)\Par tell(\ol t = \ol s)] \in D_i$,\\
where $\sf c: \ p(\ol s) \la tell(b)\Par C'[p(\ol u)] $ is a
renaming of $\sf d$ which has no variables in common with $\sf d$.
Let $\sf c'= p(\ol s') \la tell(\bar b)\Par \bar C[p(\ol u')]$
be a renaming of $\sf c$ such that
$\sf \Var(c') \cap \Var(d')=\emptyset$.
Now the proof is analogous to the previous one by observing that,
for any set of declarations, if there exists a derivation
$\sf \chi'$  for
$\sf \langle \bar C'[tell(\bar b)\Par \bar
C[p(\ol u')]\Par tell(\ol t = \ol s')], c'\rangle $ where $\sf c'$
is satisfiable and
$\sf \Var(\bar C', c') \cap \Var(b, \ol s) = \emptyset$,
then there exists a derivation for
$\sf \langle \bar C'[tell(\bar b)\Par
\bar C[p(\ol u')] \Par tell(\ol t = \ol s')
\Par tell(b)\Par tell(\ol t = \ol s)], c'\rangle $
which performs the same steps of $\chi'$, plus some
tell actions (analogously to the previous
case, we can drop the tell agents $\sf  tell(b)$ and
$\sf tell(\ol t = \ol s)$).
This concludes the proof of this case.
\II

\NI {\bf Case 4:} An ask guard in $\sf d$ is simplified. Let

- $\sf d:\ q(\ol r) \la C[\sum_{j=1}^{n} ask(c_j) \rightarrow B_j]$,

- $\sf d':\ q(\ol r) \la C[\sum_{j=1}^{n} ask(c'_j) \rightarrow B_j] \in
D_{i}$,
\\
where for $j \in [1,n]$, ${\cal D} \models \sf
\exists_{- \Var(q(\ol r),C,B_j)}\
(pc(C[\ ]) \And c_j)$
$\sf \lra (pc(C[\ ]) \And c'_j)$ and
$\sf d \in D_{i-1}$ is
the declaration to which the guard simplification was applied.

By the definition of split derivation $\chi$ has the form
\[\begin{array}{ll}
      \sf \chi= &\sf
      \langle D_{i-1}.C_{I}[q(\ol v)],c_{I}\rangle \rightarrow
      \langle D_{i-1}.C_{I}[C[\sum_{j=1}^{n} ask(c_j) \rightarrow B_j]
      \Par tell(\ol{v}=\ol{r})],c_{I}\rangle \rightarrow^* \\
      & \sf \langle D_{i-1}.C_m[\sum_{j=1}^{n}
      ask(c_j) \rightarrow B_j], c_m\rangle \rightarrow
      \langle D_{0}.A_{m+1},c_{m}\rangle \rightarrow^*
      \langle D_0.B'',c''_{F} \rangle.
\end{array}\]
Since by the inductive hypothesis for any agent $\sf A$,
$\sf {\cal O}(D_0.A) = {\cal O}(D_{i-1}.A)$,
it is easy to check that there exists a derivation
\[\begin{array}{ll}
      \sf \chi'= &\sf
      \langle D_{i-1}.C_{I}[q(\ol v)],c_{I}\rangle \rightarrow
      \langle D_{i-1}.C_{I}[C[\sum_{j=1}^{n} ask(c_j) \rightarrow B_j]
      \Par tell(\ol{v}=\ol{r})],c_{I}\rangle \rightarrow^* \\
      & \sf \langle D_{i-1}.C_m[\sum_{j=1}^{n} ask(c_j) \rightarrow B_j],
      c_m\rangle \rightarrow ^*
      \langle D_{i-1}.C_{m+h}[\sum_{j=1}^{n} ask(c_j) \rightarrow B_j],
      c_{m+h}\rangle
      \\
      & \sf \rightarrow^*
      \langle D_{i-1}.\bar B ,\bar c_{F} \rangle
\end{array}\]
such that
${\sf \exists_{-\Var(C_{I}[q(\ol v)],c_{I})} \bar c_{F}=
\exists_{-\Var(C_{I}[q(\ol v)],c_{I})} c_{F}''}$ and
$\sf m(\bar B ,\bar c_{F} )$ = $\sf m(B'',c''_{F})$.
>From (\ref{eq:app7ott6}) it follows that
\begin{equation}
   {\sf
   \exists_{-\Var(C_{I}[q(\ol v)],c_{I})} \bar c_{F}=
   \exists_{-\Var(C_{I}[q(\ol v)],c_{I})} c_{F}}
   \mbox{ and }
   {\sf m(\bar B ,\bar c_{F} )} = {\sf m(B,c_{F})}.
   \label{eq:app6nov1}
\end{equation}

Without loss of generality, we can assume that $\sf \chi'$ is chosen
in such a way that the first $m+h$ steps of $\sf \chi'$ do not use
rule ${\bf R2}$ and that $h$ is maximal, in the sense that either $\sf
c_{m+h}$ is not satisfiable or in the $m+h+1$-th step we can only use
rule ${\bf R2}$.

In the first case, let $\sf C'_{m+h}$
be the context obtained from
$\sf C_{m+h}$ as follows:
any (renamed) occurrence of the agent
$\sf \sum_{j=1}^{n} ask(c_j) \rightarrow B_j$ in $\sf C_{m+h}[\ ]$,
introduced in $\chi_{0}$ by a procedure call of the form
$\sf q(\ol s)$, is replaced by  a (suitably renamed) occurrence of the
agent
$\sf \sum_{j=1}^{n} ask(c'_j) \rightarrow B_j$.
Then, by definition of $\sf D_{i}$,  we have that
\[\begin{array}{ll}
      \sf \xi= &\sf
      \langle D_{i}.C_{I}[q(\ol v)],c_{I}\rangle \rightarrow
      \langle D_{i}.C_{I}[C[\sum_{j=1}^{n} ask(c'_j) \rightarrow B_j]
      \Par tell(\ol{v}=\ol{r})],c_{I}\rangle \rightarrow^* \\
      & \sf
      \langle D_{i}.C'_{m+h}[\sum_{j=1}^{n} ask(c'_j) \rightarrow B_j],
      c_{m+h}\rangle
\end{array}\]
is a derivation in $\sf D_{i}$ which does not
use rule ${\bf R2}$ and such that
$${\sf m(C'_{m+h}[\sum_{j=1}^{n} ask(c'_j)
\rightarrow B_j], c_{m+h})} = {\sf m(B,c_{F})} = \sf ff.$$ Then
the thesis follows by definition of split derivation.

Now assume that $\sf c_{m+h}$ is satisfiable.
By Lemma~\ref{lem:apppesideri} and (\ref{eq:app6nov1}), there exists a
constraint $\sf \bar d$, such that
$\sf \Var(\bar d) \subseteq
\Var(C_{m+h}[\sum_{j=1}^{n} ask(c_j)
\rightarrow  B_j],c_{m+h})$ and
\begin{equation}
        \sf w_t(C_{m+h}[\sum_{j=1}^{n} ask(c_j) \rightarrow
        B_j],c_{m+h},\bar d) \leq
        w_t(C_{I}[q(\ol v)],c_{I},c_{F})
        \label{eq:app6nov4}
\end{equation}
where
\begin{equation}
        {\sf \exists_{-\Var(C_{I}[q(\ol v)],c_{I})}\bar d=
        \exists_{-\Var(C_{I}[q(\ol v)],c_{I})}c_{F}}
        \mbox{ and } {\sf t} = {\sf m(B,c_{F})}.
        \label{eq:app6nov5}
\end{equation}
By definition of weight, by (\ref{eq:app6nov4}) and since
$\sf \Var(\bar d) \subseteq
\Var(C_{m+h}[\sum_{j=1}^{n} ask(c_j)
\rightarrow  B_j],c_{m+h})$, there exists a derivation
\[
\sf \langle D_{0}.
C_{m+h}[\sum_{j=1}^{n} ask(c_j) \rightarrow B_j],c_{m+h}\rangle
\rightarrow^* \langle D_{0}.\bar B',\bar d' \rangle
\]
such that $\sf \exists_{-\Var(C_{m+h}[\sum_{j=1}^{n} ask(c_j)
  \rightarrow B_j], c_{m+h})} \bar d' = \bar d$ and $\sf m(\bar
B',\bar d') = t$.  Then, by the definition of weight and by
(\ref{eq:app6nov4}),
\begin{equation}
        \sf w_t(C_{m+h}[\sum_{j=1}^{n} ask(c_j) \rightarrow
        B_j],c_{m+h},\bar d') \leq
        w_t(C_{I}[q(\ol v)],c_{I},c_{F})
        \label{eq:app18nov5}
\end{equation}
holds. Without loss of generality, we can assume that
$ \sf \Var(\bar d') \cap \Var(C_{I}[q(\ol v)],c_{I})
\subseteq \Var(C_{m+h}, c_{m+h})$.
Therefore, from (\ref{eq:app6nov5}) it follows that
\begin{equation}
        {\sf \exists_{-\Var(C_{I}[q(\ol v)],c_{I})}\bar d' =
        \exists_{-\Var(C_{I}[q(\ol v)],c_{I})}c_{F}}
        \mbox{ and }
        {\sf m(\bar B',\bar d')} = {\sf m(B,c_{F})}.
        \label{eq:app6nov8}
\end{equation}
Let $\sf \tilde {B}'= C'_{m+h}[\sum_{j=1}^{n} ask(c'_j) \rightarrow
B_j]$ be the agent obtained from
$$\sf \tilde {B} =
C_{m+h}[\sum_{j=1}^{n} ask(c_j) \rightarrow B_j]$$ as follows:
any (renamed) occurrence of the agent $\sf \sum_{j=1}^{n}
ask(c_j) \rightarrow B_j$ in $\sf C_{m+h}[\ ]$, introduced in
$\chi_{0}$ by a procedure call of the form $\sf q(\ol s)$, is
replaced by  a (suitably renamed) occurrence of the agent $\sf
\sum_{j=1}^{n} ask(c'_j) \rightarrow B_j$. By the definition of
$\sf D_{i}$ and since $\sf \langle D_{i-1}.C_{I}[q(\ol
v)],c_{I}\rangle \rightarrow ^* \langle
D_{i-1}.C_{m+h}[\sum_{j=1}^{n} ask(c_j) \rightarrow B_j],
c_{m+h}\rangle $, there exists a derivation
\[\sf \xi_{0}=
\langle D_{i}.C_{I}[q(\ol v)],c_{I}\rangle \rightarrow^*
\langle D_{i}.C'_{m+h}[\sum_{j=1}^{n} ask(c'_j) \rightarrow
B_j],c_{m+h}\rangle,
\]
which does not use rule ${\bf R2}$.  Observe that, by construction,
$\sf\tilde{B}$ has the form $\sf A_{1}\Par \ldots \Par A_{l}$, where
$\sf A_{j}$ is either a choice agent or $\sf Stop$ for each $j\in
[1,l]$.  Moreover, since the first $m+h$ steps of $\sf \chi_{0}$ do
not use rule ${\bf R2}$ (and therefore, it is not possible evaluate a
procedure call of the form $\sf q(\ol s)$ inside a guarding context),
$\sf \tilde{B}'$ has the form $\sf A'_{1}\Par \ldots \Par A'_{l}$,
where either $\sf A'_{j}=A_{j}$ or $\sf A_{j}$ is a (renamed)
occurrence of the agent $\sf \sum_{j=1}^{n} ask(c_j) \rightarrow B_j$
while $\sf A'_{j}$ is a (suitably renamed) occurrence of the agent
$\sf \sum_{j=1}^{n} ask(c'_j) \rightarrow B_j$.  By
Lemma~\ref{lem:proppc}, ${\cal D} \models \sf \ c_{m+h} \rightarrow
pc(C'[\ ]) $, where $\sf C'[\ ]$ is a renamed version of the context
$\sf C [\ ]$ in $\sf\tilde{B}$, which was introduced in $\chi_{0}$ by
a procedure call of the form $\sf q(\ol s)$.

Now from the definition of derivation and of ask simplification it
follows that, if $\sf ask(\tilde{c}_j) \rightarrow \tilde{B}_j$ is a
choice branch in $\sf\tilde{B}$ and $\sf ask(\tilde{c}'_j) \rightarrow
\tilde{B}_j$ is the corresponding choice branch in $\sf\tilde{B}'$,
then
\[
\sf{\cal D} \models
\exists_{- \Var(\tilde{B}_j,c_{m+h})}\
(c_{m+h} \And \tilde{c}_j)
\lra (c_{m+h} \And \tilde{c}'_j)
\]
holds. Therefore, by using the same arguments as in
Case 4 of Proposition~\ref{pro:partial}, since
(by inductive hypothesis) $\sf D_0$ is weight complete
and $\sf \langle D_{0}.C_{m+h}[\sum_{j=1}^{n}
ask(c_j) \rightarrow B_j], c_{m+h}\rangle \rightarrow^*
\langle D_{0}.\bar B',\bar d' \rangle$, we obtain that
there exists a split derivation in $\sf D_0$ of the form
\[
\sf \nu=  \sf
\langle D_{0}.C'_{m+h}[\sum_{j=1}^{n} ask(c'_j)
\rightarrow B_j], c_{m+h}\rangle \rightarrow
\langle D_{0}.B_{m+h+1},c_{m+h}\rangle \rightarrow^*
\langle D_{0}.B ',c'_{F} \rangle
\]
such that
$\sf \exists_{-\Var(C'_{m+h}[\sum_{j=1}^{n} ask(c'_j)
\rightarrow B_j],c_{m+h})} c'_{F} =
\exists_{-\Var(C'_{m+h}[\sum_{j=1}^{n} ask(c'_j)
\rightarrow B_j],c_{m+h})} \bar d'$ and
${\sf m(B ',c'_{F})}$ = ${\sf m(\bar B',\bar d')}$.

Then, by using the same arguments as in Case 4 of
Proposition~\ref{pro:partial}, from the definition of
weight and from (\ref{eq:app18nov5}) it follows that
\begin{equation}
\label{eq:app6nov10}
\begin{array}{ll}
      \sf w_t(C'_{m+h}[\sum_{j=1}^{n} ask(c'_j)
      \rightarrow B_j],c_{m+h},c'_{F}) & = \\
      \sf  w_t(C'_{m+h}[\sum_{j=1}^{n} ask(c'_j)
      \rightarrow B_j], c_{m+h},\bar d') & = \\
      \sf w_t(C_{m+h}[\sum_{j=1}^{n} ask(c_j)
      \rightarrow B_j], c_{m+h}, \bar d')
      \leq w_t(C_{I}[q(\ol v)],c_{I},c_{F}),
\end{array}
\end{equation}
where ${\sf t}={\sf m(B ',c'_{F})}$.  Moreover, we can
assume without loss of generality that
\[
\sf \Var(\xi_{0}) \cap \Var(\nu) =
\Var(C'_{m+h}[\sum_{j=1}^{n} ask(c'_j) \rightarrow B_j],c_{m+h}).
\]
Then by (\ref{eq:app6nov8}) we obtain
\begin{equation}
        {\sf \exists_{-\Var(C_{I}[q(\ol v)],c_{I})}c'_{F} =
        \exists_{-\Var(C_{I}[q(\ol v)],c_{I})}c_{F}}
        \mbox{ and }
        {\sf m(B ',c'_{F})} = {\sf m(B,c_{F})}
        \label{eq:app126ott3}
\end{equation}
and therefore, by definition of weight,
\begin{equation}
        \sf w_t(C_{I}[q(\ol v)],c_{I},c'_{F})=
        w_t(C_{I}[q(\ol v)],c_{I},c_{F})
        \label{eq:app12nov1}
\end{equation}
holds. By Lemma~\ref{lem:appdasplitasplit} and by construction of
$\sf C'_{m+h}[\sum_{j=1}^{n} ask(c'_j) \rightarrow B_j]$
\[
\sf \xi_{1}=
\langle D_{i}.C'_{m+h}[\sum_{j=1}^{n} ask(c'_j) \rightarrow B_j],
c_{m+h}\rangle \rightarrow
\langle D_{0}.B_{m+h+1},c_{m+h}\rangle \rightarrow^*
\langle D_{0}.B ',c'_{F} \rangle
\]
is a split derivation in $\sf D_{i}\cup D_0$.  By the definition of
split derivation $\sf w_t(B_{m+h+1},c_{m+h}, c_{F}') <
w_t(C'_{m+h}[\sum_{j=1}^{n} ask(c'_j) \rightarrow
B_j],c_{m+h},c'_{F})$, where ${\sf t}={\sf m(B ',c'_{F})}$. Then, by
(\ref{eq:app12nov1}) and (\ref{eq:app6nov10}), we have that
\begin{equation}
        \sf w_t(B_{m+h+1},c_{m+h},c'_{F}) <
        w_t(C_{I}[q(\ol v)],c_{I},c'_{F}).
 \label{eq:app12nov2}
\end{equation}
Finally,
$$
\begin{array}[t]{ll}
\sf \xi= \langle D_{i}.C_{I}[q(\ol v)],c_{I}\rangle \rightarrow^*
\langle D_{i}.C'_{m+h}[\sum_{j=1}^{n} ask(c'_j)
\rightarrow B_j],c_{m+h}\rangle \rightarrow
\\
\sf \langle D_{0}.B_{m+h+1},c_{m+h}\rangle \rightarrow^* \langle
D_{0}. B ',c'_{F} \rangle \end{array}
$$ is a derivation in $\sf
D_{i}\cup D_0$. By construction the first $m+h$ steps of $\xi$
do not use rule ${\bf R2}$, the $m+h+1$-th step uses rule ${\bf
R2}$. Thus the thesis follows from (\ref{eq:app12nov2}) and
(\ref{eq:app126ott3}). \II

\NI {\bf Case 5:} $\sf d$ is the declaration to which either a branch
elimination or an ask elimination was applied.  In the case of branch
elimination the proof follows immediately from the fact that we
consider also the inconsistent results of non-terminated computations.
As for the ask elimination case, let us assume that

- $\sf d:\ q(\ol r) \la C[ask(true) \rightarrow H] \in D_{i-1}$ and

- $\sf d':\ q(\ol r) \la C[ H]\in D_{i}$.
\\
We show, by induction on the weight
$\sf w_t(C_{I}[q(\ol v)],c_{I},c''_{F} )$,
where $\sf t = {\sf m(B,c_{F})}$,
that there exists a split derivation
$\sf \xi=
\langle D_{i}.C_{I}[q(\ol v)],c_{I}\rangle \rightarrow^*
\langle D_{0}.B', c'_{F} \rangle$ in
$\sf D_{i} \cup D_0$, such that
${\sf \exists_{-\Var(C_{I}[q(\ol v)],c_{I})} c'_{F}=
\exists_{-\Var(C_{I}[q(\ol v)],c_{I})}  c_{F}'' }$
and ${\sf m(B',c'_{F})}  = {\sf  m(B'',c''_{F})}$.
Then the proof follows by (\ref{eq:app7ott6}).

\NI \emph{Base case}. In this case
$\sf w_t(C_{I}[q(\ol v)],c_{I},c''_{F} )=0$ and
by definition of split derivation,
$\sf B''=C_k[ask(true) \rightarrow H]$,
$\chi$ has the form
\[\begin{array}{ll}
    \sf \chi= &\sf
    \langle D_{i-1}.C_{I}[q(\ol v)],c_{I}\rangle \rightarrow
    \langle D_{i-1}.C_{I}[C[ask(true) \rightarrow H]
    \Par tell(\ol{v}=\ol{r})],c_{I}\rangle \rightarrow^*
    \\ & \sf
    \langle D_{i-1}.C_k[ask(true) \rightarrow H], c''_{F}\rangle,
\end{array}\]
rule ${\bf R2}$ is not used and therefore each derivation step is done
in $\sf D_{i-1}$. Moreover, observe that since $\sf t \in \{ss,dd,ff\}$,
if
$\sf c''_{F}$ is satisfiable, then $\sf C_k$ is a guarding context.
Then, it is easy to check that
\[\begin{array}{ll}
    \sf \xi= &\sf
    \langle D_{i}.C_{I}[q(\ol v)],c_{I}\rangle \rightarrow
    \langle D_{i}.C_{I}[C[H]
    \Par tell(\ol{v}=\ol{r})],c_{I}\rangle \rightarrow^*
    \langle D_{i}.C_k[H], c''_{F}\rangle
\end{array}\]
is a split derivation in $\sf D_{i} \cup D_0$, such that
$\sf m(C_k[H], c''_{F})= m(C_k[ask(true) \rightarrow H], c''_{F})
\in \{dd,ff\}$ and then the thesis follows
by the previous observation.
\II

\NI \emph{Induction step}. Assume that
$\sf w_t(C_{I}[q(\ol v)],c_{I},c''_{F} )=n >0$ and
that $\chi$ has the form
\[\begin{array}{ll}
    \sf \chi= &\sf
    \langle D_{i-1}.C_{I}[q(\ol v)],c_{I}\rangle \rightarrow
    \langle D_{i-1}.C_{I}[C[ask(true) \rightarrow H]
    \Par tell(\ol{v}=\ol{r})],c_{I}\rangle \rightarrow^* \\
    & \sf \langle D_{i-1}.C_m[ask(true) \rightarrow H],
    c_m\rangle \rightarrow
    \langle D_{0}.C_m[H], c_m\rangle \rightarrow^*
    \langle D_0.B'',c''_{F} \rangle,
\end{array}\]
since the other case is immediate.
By the definition of $\sf D_{i}$ and since $\chi$ is a split derivation,
there exists a derivation
\[\begin{array}{ll}
\sf \xi_{0}= & \sf
\langle D_{i}.C_{I}[q(\ol v)],c_{I}\rangle \rightarrow
    \langle D_{i}.C_{I}[C[H]
    \Par tell(\ol{v}=\ol{r})],c_{I}\rangle \rightarrow^*
\langle D_{i}.C_{m}[H], c_{m}\rangle,
\end{array}\]
which does not use rule ${\bf R2}$. Moreover, by definition of
split derivation $$\sf w_t(C_m[H], c_m,c''_{F} ) <
w_t(C_{I}[q(\ol v)],c_{I},c''_{F} )$$ and therefore, by inductive
hypothesis there exists a split derivation in $\sf D_{i} \cup
D_0$,
\[\begin{array}{ll}
    \sf \xi_{1}= &\sf
    \langle D_{i}.C_m[H], c_m\rangle \rightarrow^*
    \langle D_0.B', c'_{F}\rangle,
\end{array}\]
such that
\begin{equation}
        {\sf m(B',c'_{F})}  = {\sf  m(B'',c''_{F}) = t}
        \mbox{ and }
        {\sf \exists_{-\Var(C_m[H], c_m)} c'_{F}=
        \exists_{-\Var(C_m[H], c_m)}  c_{F}'' }.
        \label{eq:app10nov1}
\end{equation}
Without loss of generality, we can assume that
$\sf \Var(\xi_{0}) \cap \Var(\xi_{1}) \subseteq
\Var(C_m[H], c_m)$.
Therefore, by (\ref{eq:app10nov1}) and by definition of $\sf c'_{F}$ and
$\sf c''_{F}$,
\begin{equation}
\label{eq:app21nov99}
\begin{array}{ll}
      \sf \exists_{-\Var(C_{I}[q(\ol v)],c_{I})} c'_{F} & =\\
      \sf \exists_{-\Var(C_{I}[q(\ol v)],c_{I})}(c_{m} \wedge
      \exists_{-\Var(C_m[H], c_m) } c'_{F} ) & = \\
      \sf \exists_{-\Var(C_{I}[q(\ol v)],c_{I})}(c_{m} \wedge
      \exists_{-\Var(C_m[H], c_m) } c''_{F} ) & = \\
      \sf \exists_{-\Var(C_{I}[q(\ol v)],c_{I})} c''_{F}.
\end{array}
\end{equation}
Then by definition of weight, since
$\sf w_t(C_m[H], c_m,c''_{F} )
< w_t(C_{I}[q(\ol v)],c_{I},c''_{F} )$ and by (\ref{eq:app10nov1})
\begin{equation}
\label{eq:app21nov299}
      \sf w_t(C_m[H], c_m,c'_{F} ) <
      w_t(C_{I}[q(\ol v)],c_{I},c'_{F} ).
\end{equation}
Moreover, by our hypothesis on the variables of $\xi_{0}$ and
of $\xi_{1}$, there exists a derivation $\sf D_{i} \cup D_0$,
\[\begin{array}{ll}
    \sf \xi= &\sf
    \langle D_{i}.C_{I}[q(\ol v)],c_{I}\rangle \rightarrow
    \langle D_{i}.C_{I}[C[H]
    \Par tell(\ol{v}=\ol{r})],c_{I}\rangle \rightarrow^* \\
    & \sf \langle D_{i}.C_m[H], c_m\rangle \rightarrow^*
    \langle D_0.B',c'_{F} \rangle.
\end{array}\]
By (\ref{eq:app10nov1}), (\ref{eq:app21nov299}),
since $\xi_{0}$ do not use Rule ${\bf R2}$ and  $\xi_{1}$ is
a split derivation in $\sf D_{i} \cup D_0$, we have that
$\xi$ is a split derivation in $\sf D_{i} \cup D_0$, such that
${\sf m(B',c'_{F})}$  = ${\sf  m(B'',c''_{F})}$.
Now, the thesis follows by (\ref{eq:app21nov99}).
\II

\NI {\bf Case 6:} An ask guard in $\sf d$ is distributed. Let

- $\sf d:\ q(\ol r) \la C[H \Par \sum_{j=1}^{n} ask(c_j)
   \rightarrow B_j] \in D_{i-1}$

- $\sf d':\   q(\ol r) \la  C[ \sum_{j=1}^{n} ask(c_j) \rightarrow
  (H \Par B_j)]\in D_{i}$,

where, for every constraint $\sf e'$ such that $\sf \Var(e')
\cap \Var(d) \subseteq \Var(q(\ol r),C)$, if
$\sf  \langle D_{i-1}.H, e' \wedge pc(C[\ ]) \rangle$ is productive
then there exists at least one $j\in [1,n]$ such that
$\sf {\cal D}\models (e' \wedge pc(C[\ ])) \rightarrow  c_j$
and  for each $j\in [1,n]$, either
$\sf {\cal D}\models (e' \wedge pc(C[\ ]))  \rightarrow  c_j$
or $\sf {\cal D}\models (e' \wedge pc(C[\ ]))
\rightarrow  \neg c_j$.

By the definition of split derivation, $\chi$ has the form
\[\begin{array}{ll}
    \sf \chi= &\sf
    \langle D_{i-1}.C_{I}[q(\ol v)],c_{I}\rangle \rightarrow
    \langle D_{i-1}.C_{I}[C[H \Par \sum_{j=1}^{n}
    ask(c_j) \rightarrow B_j]
    \Par tell(\ol{v}=\ol{r})],c_{I}\rangle \rightarrow^* \\
    & \sf
    \langle D_{i-1}.C_m[\sum_{j=1}^{n} ask(c_j)
    \rightarrow B_j], c_m\rangle \rightarrow
    \langle D_{0}.A_{m+1},c_{m}\rangle \rightarrow^*
    \langle D_0.B'',c''_{F} \rangle.
\end{array}\]
If the first $m-1$ steps of $\chi$ do not evaluate the agent $\sf H$
then the proof is analogous to that one of Case 6 of
Lemma~\ref{lem:apppesiindi}.
Otherwise, let us assume that
\[\begin{array}{ll}
    \sf \chi= &\sf
    \langle D_{i-1}.C_{I}[q(\ol v)],c_{I}\rangle \rightarrow
    \langle D_{i-1}.C_{I}[C[H \Par \sum_{j=1}^{n}
    ask(c_j) \rightarrow B_j]
    \Par tell(\ol{v}=\ol{r})],c_{I}\rangle \rightarrow^* \\
    & \sf
    \langle D_{i-1}.C_m[H' \Par \sum_{j=1}^{n} ask(c_j)
    \rightarrow B_j], c_m\rangle \rightarrow
    \langle D_{0}.A_{m+1},c_{m}\rangle \rightarrow^*
    \langle D_0.B'',c''_{F} \rangle.
\end{array}\]

Since by the inductive hypothesis for any agent $\sf A$,
$\sf {\cal O}(D_0.A) = {\cal O}(D_{i-1}.A)$
there exists a derivation
\[\begin{array}{ll}
    \sf \chi'= &\sf
    \langle D_{i-1}.C_{I}[q(\ol v)],c_{I}\rangle \rightarrow
    \langle D_{i-1}.C_{I}[C[H \Par \sum_{j=1}^{n}
    ask(c_j) \rightarrow B_j]
    \Par tell(\ol{v}=\ol{r})],c_{I}\rangle \rightarrow^* \\
    & \sf
    \langle D_{i-1}.C_k[H \Par \sum_{j=1}^{n} ask(c_j)
    \rightarrow B_j], c_k\rangle \rightarrow^*
    \langle D_{i-1}.\bar B,\bar c_{F} \rangle,
\end{array}\]
where $\sf \exists_{-\Var(C_{I}[q(\ol v)],c_{I})}\bar c_{F}=
\exists_{-\Var(C_{I}[q(\ol v)],c_{I})} c''_{F}$ and
${\sf m(\bar B,\bar c_{F})}= {\sf m(B'',c''_{F})}$.
By (\ref{eq:app7ott6}),
\begin{equation}
        {\sf \exists_{-\Var(C_{I}[q(\ol v)],c_{I})}\bar c_{F}=
        \exists_{-\Var(C_{I}[q(\ol v)],c_{I})} c_{F}}
        \mbox{ and }
        {\sf m(\bar B,\bar c_{F})} = {\sf m(B,c_{F})}.
        \label{eq:app26nov199}
\end{equation}

Without loss of generality we can assume that the first $k$ steps of $\sf
\chi'$ neither use rule ${\bf R2}$
nor contain the  evaluation of any (renamed) occurrence
$\sf {\bar H}$ of the agent $\sf H $, where
$\sf  q({\ol r}') \la  \bar C[\bar H \Par \sum_{j=1}^{n}
ask(\bar c_j) \rightarrow \bar B_j] $
is a renamed version of the declaration $\sf d$ and
$\sf \bar C[\bar H \Par \sum_{j=1}^{n}
ask(\bar c_j) \rightarrow \bar B_j] $ has
been introduced by the evaluation of a
procedure call of the form $\sf q(\ol s)$.
Moreover, we can assume that $k$ is maximal, in the
sense that either $\sf c_{k}$ is not satisfiable or
the $k+1$-th step can only either use rule ${\bf R2}$
or evaluate a (renamed) occurrence of $\sf H$ introduced
by a procedure call of the form $\sf q(\ol s)$.
If $\sf c_{k}$ is not satisfiable, then the proof is
analogous to that one of the previous Case 4.

Assume then that $\sf c_{k}$ is satisfiable.
By Lemma~\ref{lem:apppesideri} and (\ref{eq:app26nov199}), there exists a
constraint $\sf \bar d$, such that
$\sf \Var(\bar d) \subseteq
\Var(C_k[H \Par \sum_{j=1}^{n} ask(c_j)
    \rightarrow B_j], c_k)$ and
\begin{equation}
        \sf w_t(C_k[H \Par \sum_{j=1}^{n} ask(c_j)
        \rightarrow B_j], c_k,\bar d) \leq
        w_t(C_{I}[q(\ol v)],c_{I},c_{F}),
        \label{eq:app116nov4}
\end{equation}
where
\begin{equation}
        {\sf \exists_{-\Var(C_{I}[q(\ol v)],c_{I})}\bar d=
        \exists_{-\Var(C_{I}[q(\ol v)],c_{I})}c_{F}}
        \mbox{ and } {\sf t} = {\sf m(B,c_{F})}.
        \label{eq:app116nov5}
\end{equation}
By definition of weight, by (\ref{eq:app116nov4}) and since
$\sf \Var(\bar d) \subseteq
\Var(C_k[H \Par \sum_{j=1}^{n} ask(c_j)
\rightarrow B_j], c_k)$, there exists a derivation
\[
\sf \langle D_{0}.
C_k[H \Par \sum_{j=1}^{n} ask(c_j)
\rightarrow B_j], c_k\rangle
\rightarrow^* \langle D_{0}.\bar B',\bar d' \rangle
\]
such that $\sf \exists_{-\Var(C_k[H \Par \sum_{j=1}^{n} ask(c_j)
  \rightarrow B_j], c_k)} \bar d' = \bar d$ and $\sf m(\bar B',\bar
d') = t$. Then, by the definition of weight and by
(\ref{eq:app116nov4}),
\begin{equation}
        \sf w_t(C_k[H \Par \sum_{j=1}^{n} ask(c_j)
        \rightarrow B_j], c_k,\bar d') \leq
        w_t(C_{I}[q(\ol v)],c_{I},c_{F}).
        \label{eq:app118nov5}
\end{equation}

Without loss of generality, we can assume that
$ \sf \Var(\bar d') \cap \Var(C_{I}[q(\ol v)],c_{I})
\subseteq \Var(C_k, c_k)$.
Therefore from (\ref{eq:app116nov5}) it follows that
\begin{equation}
        {\sf \exists_{-\Var(C_{I}[q(\ol v)],c_{I})}\bar d' =
        \exists_{-\Var(C_{I}[q(\ol v)],c_{I})}c_{F}}
        \mbox{ and }
        {\sf m(\bar B',\bar d')} = {\sf m(B,c_{F})}.
         \nonumber
\end{equation}

Let
$\sf C'_k[ \sum_{j=1}^{n} ask(c_j) \rightarrow (H \Par B_j)]$
be the agent obtained from
$\sf C_k[H \Par \sum_{j=1}^{n} ask(c_j) \rightarrow B_j]$ as follows:
any (renamed) occurrence of the agent
$\sf H \Par \sum_{j=1}^{n} ask(c_j) \rightarrow B_j$
in $\sf C_{k}[\ ]$ which has been
introduced by a procedure call of the form $\sf q(\ol s)$
is replaced
by a (suitably) renamed occurrence of the agent
$\sf  \sum_{j=1}^{n} ask(c_j) \rightarrow (H \Par B_j)$.

By the definition of $\sf D_{i}$ and since
$\sf
\langle D_{i-1}.C_{I}[q(\ol v)],c_{I}\rangle \rightarrow^*
\langle D_{i-1}.C_k[H \Par \sum_{j=1}^{n} ask(c_j)
\rightarrow B_j], c_k\rangle$,
there exists a derivation
\[\sf \xi_{0}=
\langle D_{i}.C_{I}[q(\ol v)],c_{I}\rangle \rightarrow^*
\langle D_{i}.C'_{k}[\sum_{j=1}^{n} ask(c_j) \rightarrow
(H \Par B_j)],c_{k}\rangle
\]
which does not use rule ${\bf R2}$.

Now,  by construction,
$\sf C'_{k}[\sum_{j=1}^{n} ask(c_j) \rightarrow (H \Par B_j)]$
has the form $\sf A_{1}\Par \ldots \Par A_{l}$, where
$\sf A_{j}$ is either a choice agent or $\sf Stop$.

Moreover, since
$\sf D_{0}$ is weight complete,
$\sf \langle D_{0}.
C_k[H \Par \sum_{j=1}^{n} ask(c_j)
\rightarrow B_j], c_k\rangle
\rightarrow^* \langle D_{0}.\bar B',\bar d' \rangle$
and analogously to the Case 6 of Lemma~\ref{lem:apppesiindi},
there exists a split derivation
\[\begin{array}{ll}
    \sf \xi_{1}= & \sf
    \langle D_{0}.C'_k[ \sum_{j=1}^{n} ask(c_j)
    \rightarrow (H \Par B_j)], c_{k}\rangle \rightarrow^*
    \langle D_{0}.B' ,c'_{F} \rangle,
\end{array}\]
such that
$\sf \exists_{-\Var(C'_k[ \sum_{j=1}^{n} ask(c_j)
\rightarrow (H \Par B_j)], c_{k}) } c'_{F})=
\exists_{-\Var(C'_k[ \sum_{j=1}^{n} ask(c_j)
\rightarrow (H \Par B_j)], c_{k}) }\bar d')$ and
${\sf m(B', c'_{F})}$ = ${\sf m(\bar B',\bar d')}$.

Then, by using the same arguments as in Case 6 of
Lemma~\ref{lem:apppesiindi}, from the definition
of weight and (\ref{eq:app118nov5}) it follows that
\begin{eqnarray*}
\sf w_t(C'_k[ \sum_{j=1}^{n} ask(c_j)
\rightarrow (H \Par B_j)], c_{k},c'_{F}) & =\\
 \sf w_t(C'_k[ \sum_{j=1}^{n} ask(c_j)
    \rightarrow (H \Par B_j)], c_{k},\bar d')&  \leq\\
\sf w_t(C_k[H \Par \sum_{j=1}^{n} ask(c_j)
    \rightarrow B_j], c_k,\bar d')
    & \leq & \sf w_t(C_{I}[q(\ol v)],c_{I},c_{F}),
\end{eqnarray*}
where $\sf t=m(B ',c'_{F})$.

>From this point the proof proceeds exactly as in Case 4
by using Lemma~\ref{lem:appdasplitasplit}
and therefore it is omitted. \II

\NI
{\bf Case 7:} Finally assume that $\sf d$ is folded.\\
Let

- $\sf d: \ q(\ol r) \la C[H]$ be the folded declaration ($\in \sf
D_{i-1}$)

- $\sf f: \ p(\ol X) \la H$ be the folding declaration ($\in \sf D_{0}$),

- $\sf d': \ q(\ol r) \la C[p(\ol X)]$ be the result of the folding
operation $ (\in \sf D_{i})$,
\\
where, by definition of folding, $\sf \Var(d) \cap \Var(\ol X) \subseteq
\Var(H)$
and $\sf \Var(H) \cap (\Var(\ol r)\cup \Var(C))\subseteq \Var(\ol X)$.
Since $\sf C[\ ]$ is a guarding context, the agent $\sf H$ in $\sf
C[H]$ appears in the scope of an $\sf ask$ guard. By
definition of split derivation $\chi$ has the form
\[\begin{array}{ll}
\sf \langle D_{i-1}.C_I[q(\ol v)],c_I\rangle \rightarrow
\langle D_{i-1}.C_I[C[H]\Par tell(\ol{v}=\ol{r})],c_I\rangle
\rightarrow^* \langle
D_{i-1}.C_{m}[H],c_m\rangle \rightarrow \\
\sf \langle  D_0.C_{m+1}[H],c_{m}\rangle
\rightarrow^* \langle D_0.B'',c''_F\rangle ,
\end{array}\]
where $\sf C_{m}[\ ]$ is a guarding context.
Without loss of generality we can assume that
$\sf \Var (\chi)\cap \Var(\ol X)\subseteq \Var(H)$.
Then, from the definition of $\sf D_{i}$ it follows
that there exists a derivation
\[
\begin{array}[t]{ll}
\sf \xi_{0} = &\sf \langle D_{i}.C_I[q(\ol v)],c_I\rangle \rightarrow
\langle D_{i}.C_I[C[p(\ol X)]\Par tell(\ol{v}=\ol{r})],c_I\rangle
\rightarrow^*
\\
& \sf \langle  D_i.C_{m}[p(\ol X)],c_m\rangle \rightarrow \langle
D_0.C_{m+1}[p(\ol X)],c_{m}\rangle,
\end{array}
\]
which performs exactly the first $m$ steps as $\chi$. Since
$\sf \langle D_0.C_{m+1}[H],c_{m}\rangle \rightarrow^*
\langle D_0.B'',c''_F\rangle$,
the definition of weight implies that
$\sf w_{t}(C_{m+1}[H],c_{m},c''_F)$ is defined, where
$\sf t=m(B'',c''_F)$. Then, by (\ref{eq:app7ott6}), we have that
\begin{equation}
        {\sf t= m(B, c_{F})}.
        \label{eq:app29ott5}
\end{equation}
The definitions of derivation and folding  imply that
$\sf \Var(H ) \cap \Var(C_{m+1},c_{m}) \subseteq
\sf \Var(H ) \cap (\Var(C,\ol r)) \subseteq
\Var(\ol X)$ holds.  Moreover, from the assumptions on the
variables, we obtain that
$\sf \Var(c''_{F} ) \cap \Var(\ol X)\subseteq \Var(H)$.
Thus, from part 2 of Lemma~\ref{lem:apppesiind0} it follows that there exists
a constraint
$\sf d'$ such that
\begin{eqnarray}
 {\sf w_{t}(C_{m+1}[p(\ol X)],c_{m},d')}
        & \leq & {\sf w_{t}(C_{m+1}[H],c_{m},c''_F)}
        \mbox{ and } \nonumber
        \\
        {\sf \exists _{-\Var(C_{m+1}[p(\ol X)],c_{m})}d'} & =
        & {\sf \exists _{-\Var(C_{m+1}[p(\ol X)],c_{m})}c''_{F}.}
        \label{eq:app29ott2}
\end{eqnarray}
>From the definition of weight and the fact that
$\sf w_{t}(C_{m+1}[H],c_{m},c''_F)$ is defined it follows
that there exists a derivation
$\sf \xi_{1}=
\langle D_0.C_{m+1}[p(\ol X)],c_{m}\rangle \rightarrow^*
\langle D_0.B',c'_F\rangle$, where
$\sf m(B',c'_F)=t$ and
$\sf \exists _{-\Var(C_{m+1}[p(\ol X)],c_{m})}c'_{F}=
\exists _{-\Var(C_{m+1}[p(\ol X)],c_{m})}d'$.
Then, by the definition of weight,
$\sf w_{t}(C_{m+1}[p(\ol X)],c_{m},c'_{F})=
w_{t}(C_{m+1}[p(\ol X)],c_{m},d')$
and therefore, by (\ref{eq:app29ott2}),
\begin{eqnarray}
        {\sf \exists _{-\Var(C_{m+1}[p(\ol X)],c_{m})}c'_{F}} & =
       & {\sf \exists _{-\Var(C_{m+1}[p(\ol X)],c_{m})}c''_{F}}
        \mbox{ and }\nonumber
        \\
        {\sf w_{t}(C_{m+1}[p(\ol X)],c_{m},c'_{F})} & \leq
        &{\sf w_{t}(C_{m+1}[H],c_{m},c''_F)}
        \label{eq:app29ott1}
\end{eqnarray}
holds. Moreover, from (\ref{eq:app29ott5}) we obtain
\begin{equation}
            {\sf  m(B',c'_F)} = {\sf m(B, c_{F})}.
        \label{eq:app29ott51}
\end{equation}
Without loss of generality, we can now assume that
\[\sf \Var(\xi_{0}) \cap \Var(\xi_{1}) =
\Var(C_{m+1}[p(\ol X)],c_{m}).\]
Then, by (\ref{eq:app29ott1}) and
(\ref{eq:app7ott6}) it follows that
\begin{eqnarray}
         &  & \sf
         \exists _{-\Var(C_I[q(\ol v)],c_I)}c'_{F}=
        \exists _{-\Var(C_I[q(\ol v)],c_I)}(c_{m} \wedge
        \exists _{-\Var(C_{m+1}[p(\ol X)],c_{m})} c'_{F})=
        \nonumber \\
         &  & \sf \exists _{-\Var(C_I[q(\ol v)],c_I)}(c_{m} \wedge
        \exists _{-\Var(C_{m+1}[p(\ol X)],c_{m})} c''_{F})=
        \exists _{-\Var(C_I[q(\ol v)],c_I)}c''_{F} =
        \nonumber
        \\
         &  & \sf  \exists _{-\Var(C_I[q(\ol v)],c_I)}c_{F}.
        \label{eq:app29ott3}
\end{eqnarray}
>From the definition of weight $\sf w_{t}(C_I[q(\ol v)],c_I,c'_{F})=
w_{t}(C_I[q(\ol v)],c_I,c''_{F})$
and since $\chi$ is a split derivation we obtain
$\sf w_{t}(C_I[q(\ol v)],c_I,c''_{F}) >
w_{t}(C_{m+1}[H],c_{m},c''_F)$.
Then, from  (\ref{eq:app29ott1}) and (\ref{eq:app29ott3}) it follows that
\begin{equation}
        \sf \sf w_{t}(C_I[q(\ol v)],c_I,c'_{F})>
        w_{t}(C_{m+1}[p(\ol X)],c_{m},c'_{F})
        \label{eq:app29ott4}
\end{equation}
and therefore, by construction,
\begin{eqnarray*}
        \xi & = &
        \sf \langle D_{i}.C_I[q(\ol v)],c_I\rangle \rightarrow
        \langle D_{i}.C_I[C[p(\ol X)]\Par
        tell(\ol{v}=\ol{r})],c_I\rangle \rightarrow^*
        \langle  D_i.C_{m}[p(\ol X)],c_m\rangle \rightarrow  \\
        &  & \sf \langle D_0.C_{m+1}[p(\ol X)],c_{m}\rangle
        \rightarrow^* \langle D_0.B',c'_F\rangle
\end{eqnarray*}
is a derivation in $\sf D_{i} \cup D_{0}$ such that: (a) rule
${\bf R2}$ is not used in the first $m-1$ steps;
(b) rule ${\bf R2}$ is used in the $m$-th step.
The thesis then follows from (\ref{eq:app29ott3}), (\ref{eq:app29ott51})
and (\ref{eq:app29ott4}) thus concluding the proof.
\end{proof}

\subsection{Proof of correctness for intermediate results and traces}

In this subsection we show how the previous proofs can be adapted when
considering intermediate results and traces as observables. We first
consider Theorem~\ref{thm:correctness2}.  Since its proof is
essentially the same of that one already given for the total
correctness theorem, here we provide only the intuition illustrating
the (minor) modifications needed.

\begin{theorem} [\ref{thm:correctness2} (Total Correctness 2)]
Let
$\sf D_0, \ldots, D_n$ be a transformation sequence, and $\sf A$
be an agent.\begin{itemize}
  \item  If there exists a derivation $\sf \langle
  D_0.A,c\rangle \rightarrow^* \langle D_0.B,d\rangle$ then there
  exists a derivation $\sf \langle D_n.A,c\rangle \rightarrow^*
  \langle D_n.B',d'\rangle$ such that $\sf {\cal D}\models
  \exists_{-Var(A,c)}d'\rightarrow \exists_{-Var(A,c)}d$.
\item Conversely, if there exists a derivation $\sf \langle
  D_n.A,c\rangle \rightarrow^* \langle D_n.B,d \rangle$ then there
  exists a derivation $\sf \langle D_0.A,c\rangle \rightarrow^*
  \langle D_0.B',d'\rangle$ with $\sf {\cal D}\models
  \exists_{-Var(A,c)}d'\rightarrow \exists_{-Var(A,c)}d$.
\end{itemize}
\end{theorem}

\begin{proof}  The proof of this result is essentially the same as
that one of the total correctness Theorem~\ref{thm:correctness}
provided that in such a proof, as well as in the proofs of the
related preliminary results, we perform the following changes:
\begin{enumerate}
\item Rather than considering terminating derivations, we consider any
(possibly non-maximal) finite derivation.
\item Whenever in a proof we write that, given a derivation $\xi$,
a derivation $\xi'$ is constructed which performs the same steps of
$\xi$, possibly in a different order, we now write that
a derivation $\xi''$ is constructed which performs the same
steps as $\xi$ (possibly in a different order) plus some
other additional steps. Since the store grows monotonically
in ccp derivations, clearly if
a constraint $\sf c$ is the result of the derivation $\xi$, then
a constraint $\sf c''$ is the result of  $\xi''$ such that
${\cal D}\models {\sf c''\rightarrow c}$ holds. For example, for case
2 in the proof of Proposition~\ref{pro:partial} (in the Appendix),
when considering a (non-maximal) derivation $\xi$ which uses the
declaration
$\sf H\leftarrow C[tell(\ol{s} = \ol{t})]\parallel B]$
we can always construct a derivation $\xi''$ which performs all the steps
of
$\xi$ (possibly plus others) and such that the $\sf tell(\ol{s} = \ol{t})$
agent is evaluated before $\sf B$. Differently from the previous proof,
now we are not ensured that the result of $\xi$ is the same as that one
of $\xi''$, since $\xi$ is non-maximal (thus, $\xi$
could also avoid the evaluation of $\sf tell(\ol{s} = \ol{t})$).
However, we are ensured that the result of $\xi''$ is
stronger (i.e. implies) that one of $\xi$.
\end{enumerate} \end{proof}

We now consider the correctness results given for the restricted
transformation system \wrt the traces. Also in this case, the proofs
follow the guidelines of that one already presented in Section
\ref{sec:correctness} and in the previous part of this Appendix.  We
then sketch the proofs by showing which are the relevant new notions
and differences with respect to the previous ones.

In the remainder of this section we will always refer to the
restricted transformation system and to a given restricted
transformation sequence $\sf D_0, \ldots, D_n$.

We start with the following definition.

\begin{definition}
        \label{def:appdaderaseq}
        Let $\sf D$ be a set of declarations and let $\sf \xi$ be the
        derivation $$\sf \langle D.A_1,c_1\rangle \rightarrow^* \langle
        D.A_m,c_m\rangle \rightarrow^* \langle D.A_n,c_n\rangle.$$ We
        define $\sf tr(\xi)$ =
        $$\sf \exists
        _{-\Var(A_1,c_1)}(c_{1};c_{2};\ldots;c_{n}) = \sf
        (c_{1};(\exists _{-\Var(A_1,c_1)}c_{2});\ldots;(\exists
        _{-\Var(A_1,c_1)}c_{n})).$$
\end{definition}

The function {\em mode} ($\sf m(A,d)$) is extended to consider
also non-terminated derivations in the obvious way. We then
extend the notion of weight, split derivation and weight complete
programs to the case of traces. Here and in the following the
subscript $\sf t$ will denote a generic termination mode, that
is, we assume $\sf t \in \{ss,dd,pp,ff\}$. We also say that a
trace starts with $\sf c$ in case $\sf c$ is the first constraint
appearing in that trace.

\begin{definition}[\thetheorem\ (Weight for traces)]
  \label{def:appweighttr}
  Given an agent $\sf A$,  a
  satisfiable constraint $\sf c$ and a trace $\sf s$ starting with $\sf c$,
  we define the \emph{weight} of the agent $\sf A$
  w.r.t. the trace $\sf s$, notation $\sf w_{t}(A,s)$, as follows:
  \II

  ${\sf w_{t}(A,s)} = min\{ n \mid \begin{array}[t]{ll} n = wh(\xi)
  \hbox{ and }
  \xi \hbox{ is a derivation }
  \sf \langle D_0.A,c\rangle \rightarrow^* \langle D_0.B,d\rangle \\
  \hbox{ such that } {\sf \exists _{-\Var(A,c)} s \preceq  tr(\xi)}
  \hbox{ and }  \sf t=m(B,d) \}.
  \end{array}
  $
\end{definition}

\begin{definition}[\thetheorem\ (Split derivation for traces)]
\label{def:appsplitdertr} Let $\sf D_0,\ldots,D_i$ be a
transformation sequence. We call a derivation in $\sf D_i\cup
D_0$ a \emph{split derivation for traces} if it has the form

\II $\sf \langle D_i.A_1,c_1\rangle \rightarrow^* \langle
D_i.A_m,c_m\rangle \rightarrow \langle D_0.A_{m+1},c_{m+1}\rangle
\rightarrow^* \langle D_0.A_{n},c_n\rangle $ \II

\NI where $m\in[1,n]$ and the following conditions hold:
\begin{enumerate}
\parentalphi
\item the first $m-1$ derivation steps do not use rule ${\bf R2}$;
\item the $m$-th derivation step $\sf \langle D_i.A_m,c_m\rangle
\rightarrow \langle D_0.A_{m+1},c_{m+1}\rangle$ uses rule ${\bf R2}$;
\item $\sf w_{t}(A_1,(c_1;c_{2};\ldots;c_n)) >
w_{t}(A_{m+1},(c_{m+1};\ldots;c_n))$, where $\sf t=m(A_{n},c_n)$.
\end{enumerate}
\end{definition}

\begin{definition}
  \label{def:appprogcompltr}
  We call the program $\sf D_i$ \emph{weight complete for traces}
  iff, for any agent $\sf A$ and any satisfiable constraint $\sf c$ the
  following hold: If there exists a derivation
\II

$\sf \chi=\langle D_0.A,c\rangle
\rightarrow^* \langle D_0.B,d\rangle$
\II

\NI
such that $\sf  m(B,d)\in \{ss, dd, pp, ff\}$ then there
exists a split derivation in $\sf D_i \cup D_0$
\II

$\sf \xi=
\langle D_i.A,c\rangle \rightarrow^* \langle D_0.B',d' \rangle
$
\II

\NI where $\sf tr(\chi) \preceq  tr(\xi) $ and $\sf  m(B',d') =
\sf m(B,d)$.
\end{definition}

Proposition~\ref{pro:total1} holds also when considering as
observables $\sf {\cal O}_t$ rather than $\sf {\cal O}$ and its proof
is essentially the same, thus we omit it.

The following Lemma is obtained from Lemma~\ref{lem:proppc} by
considering the weakest produced constraint $\sf wpc$ rather than the
produced constraint. The proof is analogous to that one given for
Lemma~\ref{lem:proppc} and hence it is omitted.

\begin{lemma}
        \label{lem:apppropwpc}
        Assume that there exists a derivation
        $\sf \langle D.C[A],c\rangle \rightarrow^* \langle
        D.C'[A],c'\rangle $ where $\sf c$ is a satisfiable constraint
        and the context $\sf C'[\ ]$ has the form $$\sf A_{1}\Par
        \ldots\Par \bar C[\ ]\Par \ldots \Par A_{n}.$$  Then ${\cal D}
        \models \sf \ (wpc(\bar C[\ ]) \And c') \rightarrow wpc(C[\ ])
        $ holds and in case $\sf \bar C[\ ]$ is the empty context also
        ${\cal D} \models \sf \ c' \rightarrow wpc(C[\ ]) $ holds.
\end{lemma}

In the following we extend to set of observables the
(pre-order) relation $\preceq$ in the expected way:
Given two sets of observables $\sf {\cal O}_t(D_i.A)$ and
$\sf {\cal O}_t(D_j.A)$, we say that
$\sf {\cal O}_t(D_i.A) \preceq {\cal O}_t(D_j.A)$ iff,
for any $\sf \langle s,x\rangle \in \sf {\cal O}_t(D_i.A)$,
(with $\sf x\in \{ss,dd,pp,ff\}$), there exists
$\sf \langle s',x\rangle \in \sf {\cal O}_t(D_j.A)$
such that $\sf s\preceq s'$. We denote by $\equiv$ the
equivalence relation induced by $\preceq$ on sets of observables, that is,
$\sf {\cal O}_t(D_i.A) \equiv \sf {\cal O}_t(D_j.A)$ iff
$\sf {\cal O}_t(D_i.A) \preceq \sf {\cal O}_t(D_j.A)$ and
$\sf {\cal O}_t(D_j.A) \preceq \sf {\cal O}_t(D_i.A)$.

The following is analogous of Proposition~\ref{pro:partial} for
traces.

\begin{proposition}[\thetheorem\ (Partial Correctness for traces)]
        If, for each agent $\sf A$,
        $\sf {\cal O}_t(D_0.A) \equiv {\cal O}_t(D_i.A)$
        then, for each agent $\sf A$,
        $\sf {\cal O}_t (D_{i+1}.A) \preceq {\cal O}_t (D_i.A) $.
        \label{prp:apppartialtr}
\end{proposition}
\begin{proof}
We have to show that, given an agent $\sf A$ and a satisfiable
constraint $\sf c_{I}$, if there exists a derivation
$\sf \xi = \langle D_{i+1}.A,c_{I}\rangle \rightarrow^*
\langle D_{i+1}.B,c_{F}\rangle$,
then there exists also a derivation
$\sf \xi' = \langle D_{i}.A,c_{I}\rangle
\rightarrow^* \langle D_{i}.B',c_{F}'\rangle$
such that $\sf tr(\xi) \preceq  tr(\xi')$
and ${\sf m(B',c'_{F})}={\sf m(B,c_{F})}$.

The proof is analogous to that one given for
Proposition~\ref{pro:partial},
therefore we illustrate only the modifications needed to adapt such a
proof.

Assume that the first step of derivation $\sf \xi$ uses rule
${\bf R4}$ and let $\sf d' \in D_{i+1}$ be the declaration used
in the first step of $\sf \xi$. Assume also
that $\sf d' \not\in D_{i}$ and that $\sf d'$ is the result of the
transformation operation applied to obtain $\sf D_{i+1}$. As usual, we
distinguish various cases according to the kind of operation performed.
Here we consider only those cases whose proof is different from
that one of Proposition~\ref{pro:partial}, due to the fact that
here we consider traces (consisting of intermediate results) rather than
the final constraints.

\begin{description}
\item[Case 2] In this case $\sf d: \sf H \la C[tell(\ol{s}=\ol{t})
  \Par B] \in D_{i}$, $\sf d': \sf H \la C[B\sigma]\in D_{i+1}$, where
  $\sigma$ is a relevant most general unifier of $\sf \ol{s}$ and $\sf \ol{t}$
  (or a renaming, in case of $\sf \ol{s}$ and $\sf \ol{t}$ consist of distinct
  variables). From the definition of the operation we know that the
  variables in the domain of $\sigma$ do not occur neither in $\sf C[\
  ]$ nor in $\sf H$ and, differently from the case of
  Proposition~\ref{pro:partial},
  that $\sf Var(B) \I Var(H,C) = \emptyset$.\\
  For any derivation which uses a declaration $\sf \sf H \la
  C[tell(\ol{s}=\ol{t}) \Par B]$, if the agent $\sf
  tell(\ol{s}=\ol{t}) $ is evaluated before $\sf B$ then the proof is
  analogous to that one given for Case 2 of
  Proposition~\ref{pro:partial}. Otherwise, if the agent $\sf
  tell(\ol{s}=\ol{t}) $ is not evaluated before $\sf B$, then by using
  the condition $\sf Var(B) \I Var(H,C) = \emptyset$ we obtain that
  the evaluation of the agent $\sf B$ can add to the store only
  constraints on variables which do not occur neither in the global
  store (before the evaluation of $\sf B$) nor in $\sf \Var(A,c_{I})$.
  Therefore the contribution to the global store of the agent $\sf B$
  (before the evaluation of the agent $\sf tell(\ol{s}=\ol{t})$) when
  restricted to $\sf \Var(A,c_{I})$ is equivalent either to the
  constraint $\sf true$ or to the constraint $\sf false$.

        In the first case the global store is the same as
        that one existing before the evaluation of $\sf B$.
        In the second case we can obtain the
        constraint $\sf false$ by evaluating the same agents
        evaluated in $\sf B$ also in $\sf B\sigma$.

    \item[Case 3]  In this case the proof is analogous to that one given
    for Case~3  of Proposition~\ref{pro:partial}
    by observing the following: If
    in the derivation $\chi$ in $\sf D_{i}$ either the agent
    $\sf tell(b)$ or the agent $\sf tell(\ol t = \ol s)$
    are evaluated, then in the derivation $\chi'$ the agent
    $\sf p(\ol t)$ can be evaluated and then one
    performs exactly the same steps of $\chi$, except for
    the evaluation of a renamed version of the agents
    $\sf tell(b)$ and $\sf tell(\ol t = \ol s)$.

        \item[Cases 4]  For the ask simplification the proof of Case~4 of
        Proposition~\ref{pro:partial} is simplified by using
        Lemma~\ref{lem:apppropwpc} and by observing that, for any derivation,
        when the choice agent inside $\sf C[\ ]$ is evaluated the current
        store certainly implies $\sf wpc(C[\ ])$.
        Therefore we do not need to construct
        the new derivation $\chi'$. The same holds for the tell
simplification.

    \item[Case 7]  In this case the proof is analogous to that given for
    the previous Case 2, by observing that in the derivation
    \[\sf \beta=\langle D_{0}.
    C_{I}[ C [H'\Par tell(\ol X = \ol X') ]
    \Par tell(\ol v = \ol r)],c_{I}\rangle \rightarrow^*
    \langle D_{0}. B'_{0}, c_{0}\rangle,
    \]
    $\sf \Var(H')\cap \Var(C_{I}, C, c_{I},\ol X,\ol v, \ol r ) = \emptyset$.
    Therefore we can construct a derivation
    \[\sf \chi_{0}= \sf \langle D_{0}.
    C_{I}[C [H \Par tell(\ol X' = \ol X) ]
    \Par tell(\ol v = \ol r)],c_{I}\rangle \rightarrow^*
    \langle D_{0}. B''_{0}, c_{0}'\rangle
    \]
    where $\sf tr(\beta) \preceq tr(\chi_{0})$ and ${\sf m(B''_{0},
      c'_{0})} = {\sf m(B'_{0}, c_{0})}$.  Moreover, we can drop the
    constraint $\sf tell(\ol X' = \ol X)$, since the declarations used
    in the derivation are renamed apart and, by construction, $\sf
    \Var(C_{I}[C [H] \Par tell(\ol r = \ol v)], c_{I}) \cap \Var(\ol
    X') =\emptyset$.  We then obtain that there exists a derivation
    $\sf \beta'=\langle D_{0}.  C_{I}[C [H] \Par tell(\ol v = \ol
    r)],c_{I}\rangle \rightarrow^* \langle D_{0}.\bar B_{0},\bar
    c_{0}\rangle$ which performs exactly the same steps of $\sf
    \chi_{0}$ except for (possibly) the evaluation of $\sf tell(\ol X'
    = \ol X)$ and such that $\sf \exists_{-\Var(C_{I}[C [H] \Par
      tell(\ol v = \ol r)],c_{I})} tr(\chi_{0}) \preceq tr(\beta')$
    and ${\sf m(\bar B_{0},\bar c_{0})}= {\sf m(B''_{0}, c'_{0})}$.
    Now, the proof is the same to that given for Case~7 of
    Proposition~\ref{pro:partial}, since the evaluation of $\sf
    tell(\ol X' = \ol X)$ does not modify the current store with
    respect to the variables not in $\sf \Var(\ol X')$.
\end{description}
\end{proof}

The following Lemmata are the counterpart of previous
Lemma~\ref{lem:apppesoclausola} and Lemma~\ref{lem:apppesiind0},
when considering the observable
$\sf {\cal O}_t(D_i.A)$.

\begin{lemma}
    \label{lem:apppesoclausolatr}
    Let $ \sf q (\ol{r})\la H \in D_{0}$ and
    let $\sf C[\ ]$ be context. For any satisfiable
    constraint $\sf c$ and for any trace $\sf s$ starting with $\sf c$,
    such that
    $\sf \Var(C[q(\ol{t})],c) \cap \Var(\ol{r}) =\emptyset$ and
    $\sf w_t( C[q (\ol{t}) ],s)$ is defined,
    there exists a trace $\sf s'$ such that
    $\sf w_t( C[q (\ol{r}) \Par tell(\ol{t}=\ol{r})],s')\leq
    w_t( C[q (\ol{t}) ],s)$ and
    $\sf \exists_{-\Var( C[q (\ol{t}) ],c)}s \preceq
    \exists_{-\Var( C[q (\ol{t}) ],c)}s'$.
\end{lemma}
\begin{proof}
Immediate.  \end{proof}

\begin{lemma}
    \label{lem:apppesiind0tr}
    Let $\sf q (\ol{r})\la H \in D_{0}$.
    For any context $\sf C_{I}[\ ]$, any satisfiable
    constraint $\sf c$ and for any sequence $\sf s$ starting in $\sf c$,
    the following holds:
    \begin{enumerate}
        \item\label{pt:apppesiind01tr}
        If $\sf \Var(H) \cap \Var(C_{I},c) \subseteq \Var(\ol r)$
        and $\sf w_t( C_{I}[q (\ol{r}) ],s)$ is defined, then there
        exists a sequence $\sf s'$, such that
        $\sf \Var(s') \subseteq \Var(C_{I}[H],c)$,
        $\sf w_t( C_{I}[H],s') \leq  w_t( C_{I}[q (\ol{r}) ],s)$
        and $\sf \exists_{-\Var( C_{I}[q (\ol{r}) ],c)}s \preceq
        \exists_{-\Var( C_{I}[q (\ol{r}) ],c)}s'$.

        \item\label{pt:apppesiind02tr}
        If $\sf \Var(H) \cap \Var(C_{I},c) \subseteq \Var(\ol r)$,
        $\sf \Var(s) \cap \Var(\ol r) \subseteq \Var(C_{I}[H],c)$
        and $\sf w_t( C_{I}[H],s)$ is defined, then there exists a
        sequence $\sf s'$, such that
        $\sf w_t( C_{I}[q (\ol{r}) ],s') \leq  w_t( C_{I}[H],s)$
        and $\sf \exists_{-\Var( C_{I}[q (\ol{r}) ],c)}s \preceq
        \exists_{-\Var( C_{I}[q (\ol{r}) ],c)}s'$.
    \end{enumerate}
\end{lemma}
\begin{proof}
Immediate.
\end{proof}

Analogously to the case of the previous results, the following Lemma is
crucial in the proof of completeness for traces.

\begin{lemma}
    \label{lem:apppesiinditr}
    Let $0 \leq i \leq n$,
    $ \sf cl:\  q (\ol{r})\la H$ be a declaration in $D_{i}$
    and let $ \sf cl':\  q (\ol{r})\la H' $ be the corresponding
    declaration in $\sf D_{i+1}$ (in case $i<n$).
    For any context $\sf C_{I}[\ ]$, any satisfiable
    constraint $\sf c$ and for any sequence $\sf s$ starting in $\sf c$
    the following holds:
    \begin{enumerate}
       \item\label{pt:apppesiindi1tr}
       If $\sf \Var(H) \cap \Var(C_{I},c) \subseteq \Var(\ol r)$
       and $\sf w_t( C_{I}[q (\ol{r}) ],s)$ is defined,
       then there exists a sequence $\sf s'$, such that
       $\sf \Var(s') \subseteq \Var(C_{I}[H],c)$,
       $\sf w_t( C_{I}[H],s') \leq
       w_t( C_{I}[q (\ol{r}) ],s)$
       and $\sf \exists_{-\Var( C_{I}[q (\ol{r}) ],c)}s \preceq
       \exists_{-\Var( C_{I}[q (\ol{r}) ],c)}s'$;
       \item\label{pt:apppesiindi2tr}
       If $\sf \Var(H,H') \cap \Var(C_{I},c) \subseteq \Var(\ol r)$,
       $\sf \Var(c') \cap \Var(\ol r)\subseteq \Var(C_{I}[H],c)$ and
       $\sf w_t( C_{I}[H],s)$ is defined, then there exists a
       sequence $\sf s'$, such that $\sf \Var(s') \subseteq
       \Var(C_{I}[H'],c)$,
       $\sf w_t( C_{I}[H'],s') \leq  w_t( C_{I}[H],s)$
       and $\sf \exists_{-\Var( C_{I}[q (\ol{r}) ],c)}s \preceq
       \exists_{-\Var( C_{I}[q (\ol{r}) ],c)}s'$.
    \end{enumerate}
\end{lemma}
\begin{proof}
The proof is analogous to that given for Lemma~\ref{lem:apppesiindi}, by
using
Lemma~\ref{lem:apppesiind0tr} and \ref{lem:apppesoclausolatr} instead of
Lemma~\ref{lem:apppesiind0} and \ref{lem:apppesoclausola}, respectively.
We have only to observe the following facts:

For Case 3, Point {\em (1)} we can evaluate the agent $\sf
tell(b)$ after the global store implies ${\sf \exists_{-\Var(\ol{s})}
b}$. In this way the new derivation has the same sequence of
intermediate results.

For Case 6, Point {\em (2)}, by using Lemma~\ref{lem:apppropwpc}, if there
exists a derivation
\[
\begin{array}{ll}
   \sf \xi =
   & \sf \langle D_{0}.C_{I}[C[H \Par \sum_{j=1}^{n} ask(c_j)
   \rightarrow B_j]], c\rangle \rightarrow^*
   \langle D_{0}.C_{m}[H \Par \sum_{j=1}^{n}  ask(c_j)
   \rightarrow B_j], d_{m}\rangle \rightarrow \\
   & \sf \langle D_{0}.C_{m}[H' \Par \sum_{j=1}^{n}  ask(c_j)
   \rightarrow B_j], d_{m+1}\rangle \rightarrow^*
   \langle D_{0}.B,d\rangle,
\end{array}
\]
then $\sf {\cal D} \models d_{m} \ra e (= wpc(C[\ ]))$.
If  $\sf  \langle D_{0}.H, d_{m} \rangle$ is not productive then
the proof is straightforward.
Otherwise, assume that $\sf  \langle D_{0}.H, d_{m} \rangle$ is productive.
By definition of distribution there exists at least one $j\in [1,n]$
such that $\sf {\cal D}\models d_{m} \rightarrow  c_j$ and,
for each $j\in [1,n]$, either
$\sf {\cal D}\models d_{m} \rightarrow  c_j$
or $\sf {\cal D}\models d_{m} \rightarrow  \neg c_j$.
Then, by definition, there exists a derivation
$\sf \xi_{1}=\langle D_{0}.C_m[\sum_{j=1}^{n} ask(c_j)
\rightarrow (H  \Par B_j)], d_m\rangle\rightarrow^*
\langle D_{0}.B,d\rangle$
which performs the same steps of $\chi_{1}$ in the same order, except for
one step of evaluation of the agent
$\sf \sum_{j=1}^{n}  ask(c_j)
\rightarrow B_j$ which is performed before evaluating the agent $\sf H$.
Then the thesis follows
by definition of the relation $\preceq$.
\end{proof}

Also the proof of the following Lemma is analogous to that of its
previous counterpart (Lemma~\ref{lem:apppesideri}) and hence it is
omitted.

\begin{lemma}
   \label{lem:apppesideritr}
   Let $0 \leq i \leq n$, $\sf c_1$ be a satisfiable constraint and
   assume that there exists a derivation $\xi:\ \sf \langle
   D_i.A_1,c_1\rangle \rightarrow^{*} \langle D_i.A_m,c_m\rangle
   \rightarrow ^{*} \langle D_i.A_{k},c_{k}\rangle$, such that $\sf
   c_m$ is satisfiable. If
   \begin{description}
      \item{i)} in the first $m-1$ steps of $\xi$ rule ${\bf R2}$
      is used only for evaluating agents of the
      form $\sf ask(c)\rightarrow B$,
      \item{ii)} $\sf w_t(A_{1},tr(\xi))$ is defined
      (for ${\sf t} ={\sf m(A_{k},c_{k})\in \{ss,dd,pp,ff\}}$).
   \end{description}
   then there exists a sequence $\sf s'$ starting in $\sf c_{m}$, such
   that $\sf \Var(s') \subseteq \Var(A_{m},c_{m})$, $\sf
   \exists_{-\Var(A_{1},c_{1})}(c_{m};\ldots;c_{k}) \preceq
   \exists_{-\Var(A_{1},c_{1})}s'$ and $\sf w_t( A_{m},s') \leq
   w_t(A_{1},tr(\xi))$.
\end{lemma}

Finally we have the following.
\medskip


\begin{theorem}[\ref{thm:correctnessstrong} (Strong Total
  Correctness)] Let $\sf D_0, \ldots, D_n$ be a restricted
transformation sequence, and $\sf A$ be an agent.
\begin{itemize}
\item If $\sf \langle s,x\rangle \in \sf {\cal O}_t(D_0.A)$ (with $\sf
  x \in \{ss,dd,pp,ff\}$) then there exists $\sf \langle s',x\rangle
  \in \sf {\cal O}_t(D_n.A)$ such that $\sf s\preceq s'$.
\item  Conversely, if $\sf \langle
s,x\rangle \in \sf {\cal O}_t(D_n.A)$ then there exists $\sf \langle
s',x\rangle \in \sf {\cal O}_t(D_0.A)$ such that $\sf s\preceq s'$.
\end{itemize}
\end{theorem}
\begin{proof} The proof is analogous to that given for
Theorem~\ref{thm:correctness} and proceeds by showing
simultaneously, by induction on $i$, that for $i \in [0,n]$ and
for any agent $\sf A$:
\begin{enumerate}
    \item $\sf {\cal O}_t(D_0.A) \equiv {\cal O}_t(D_i.A)$;
    \item $\sf D_i$ is weight complete for the traces.
\end{enumerate}
The proof of the base case is analogous to that given for the
base case of Theorem~\ref{thm:correctness} and hence it is
omitted.  For the induction step we have that, by induction
hypothesis, for any agent $\sf A$, $\sf {\cal O}_t(D_0.A) \equiv
{\cal O}_t(D_{i-1}.A)$ and $\sf D_{i-1}$ is weight complete for
the traces. Proposition~\ref{pro:total1} holds also when
considering $\sf {\cal O}_t$ rather than $\sf {\cal O}$.   From
Proposition~\ref{prp:apppartialtr} and (the counterpart for traces
of) Proposition~\ref{pro:total1} then it follows that if $\sf
D_{i}$ is weight complete for traces then, for any agent $\sf A$,
$\sf {\cal
  O}_t(D_0.A) = {\cal O}_t(D_{i}.A)$.  So, in order to prove parts 1
and 2, we have only to show that, for any derivation $\sf \beta=\langle
D_0.A,c_{I}\rangle \rightarrow^* \langle D_0.B,c_{F}\rangle$ such that
$\sf c_{I}$ is a satisfiable constraint and ${\sf m(B,c_{F})} \in {\sf
  \{ss,dd,pp,ff\}}$, there exists a split derivation in $\sf D_{i}\cup
D_{0}$, $\sf \xi=\langle D_{i}.A,c_{I}\rangle \rightarrow^* \langle
D_0.B',c'_{F} \rangle$, such that $\sf tr(\beta) \preceq tr(\xi)$ and
${\sf m(B',c'_{F})}$ = ${\sf m(B,c_{F})}$.
\II

>From the inductive hypothesis it follows that there exists a split
derivation
\[\sf \chi=
\langle D_{i-1}.A,c_{I}\rangle \rightarrow^*
\langle D_0.B'',c''_{F} \rangle
\]
where ${\sf tr(\beta) \preceq tr(\chi)}$ and ${\sf
m(B'',c''_{F})} = {\sf m(B,c_{F})}$. Now, let $\sf d \in D_{i-1}
\backslash D_{i}$ be the modified clause in the transformation
step from $\sf D_{i-1}$ to $\sf D_{i}$.  The rest of the proof is
essentially analogous to that given for
Theorem~\ref{thm:correctness}. The only points which require some
case are the following:
\begin{description}
      \item[Case 2]  In this case, the proof
        is analogous to that given for Case 2 of
        Proposition~\ref{prp:apppartialtr}.

      \item[Case 3] In this case the proof is analogous to that
        given for Case~3 of Proposition~\ref{prp:apppartialtr}, provided
        we observe the following fact for case {\em 2a)}
        in such a proof:
        Given any set of declarations, if there exists a derivation
        $\sf \chi'$ for the configuration
        $\sf \langle C'[tell(\bar b)\Par \bar B \Par
        tell(\ol t = \ol s')], c' \rangle$
        where $\sf c'$ is satisfiable and
        $\sf \Var(C'[tell(\bar b)\Par \bar B
        \Par tell(\ol t = \ol s')], c') \cap  \Var(b, \ol s) =
        \emptyset$, then there exists a derivation for
        $\sf \langle C'[tell(\bar b)\Par \bar B \Par
        tell(\ol t = \ol s') \Par tell(b)\Par
        tell(\ol t = \ol s)], c'\rangle$
        which performs the same steps of  $\chi'$ plus (possibly)
        two steps corresponding to the evaluation of
        $\sf tell(\ol t = \ol s)$ and $\sf tell(b)$, after
        the evaluation of $\sf tell(\bar b)$ and
        $\sf tell(\ol t = \ol s')$.

     \item[Case 4]  Analogously to the proof of Case~4 of
        Proposition~\ref{prp:apppartialtr}, it is sufficient
        to observe the following.
        From Lemma~\ref{lem:apppropwpc} it follows that, for any
        derivation, when the choice agent inside a context
        $\sf C[\ ]$ is evaluated the current store implies
        $\sf wpc(C[\ ])$. Then, by definition of ask simplification,
        the constraint $\sf c_{j}$ and $\sf c_{j}'$ are equivalent
        with respect to the current store (and therefore we do not
        need to construct the new derivation $\chi'$).
        The same reasoning applies to the case
        of tell simplification.

     \item[Case 6]  The proof is analogous to that of Case~6 of
        Lemma~\ref{lem:apppesiinditr}.
\end{description}
\end{proof}

\end{document}